\documentclass[final,3p,times]{elsarticle}
\usepackage{mwe}
\usepackage{tikz}
\usepackage{structmech}
\usepackage{float}
\usepackage{caption}
\usepackage{subcaption}
\usepackage{amsmath}

\usepackage{comment}
\usepackage{algorithm2e}
\SetKwComment{Comment}{/* }{ */}
\usepackage{amsthm}
\usepackage{amssymb}

\usepackage{expl3}
\ExplSyntaxOn
\newcommand\latinabbrev[1]{
  \peek_meaning:NTF . {
    #1\@}%
  { \peek_catcode:NTF a {
      #1.\@ }%
    {#1.\@}}}
\ExplSyntaxOff
\usepackage{mathrsfs}

\usepackage{amsmath,amssymb,amsthm,color,xfrac,mathrsfs}
\usepackage{bbm}
\usepackage{makecell}
\usepackage{epsfig,amsfonts}
\usepackage{graphicx,epsfig,epstopdf} 
\usepackage{array} 
\usepackage{color,soul}
\usepackage{xcolor}
\soulregister\ref{7}
\soulregister\pageref{7}
\sethlcolor{yellow}
\usepackage{appendix}
     
\usepackage{pmat} 
\usepackage{indentfirst} 
\usepackage{enumerate} 
\usepackage{lineno}
 \usepackage{placeins}
 \usepackage{silence}
\usepackage{booktabs,ctable,multirow,longtable} 
 \usepackage{rotating}
\usepackage{hyphenat}

\usepackage{changepage}

\NewDocumentCommand{\evalat}{sO{\big}mm}{%
  \IfBooleanTF{#1}
   {\mleft. #3 \mright|_{#4}}
   {#3#2|_{#4}}%
}

\DeclareMathAlphabet{\mathpzc}{OT1}{pzc}{m}{it}
\DeclareMathAlphabet{\mathcalligra}{OT1}{calligra}{m}{it}

\AtBeginDocument{

}

\newcolumntype{P}[1]{>{\centering\arraybackslash}m{#1}}

\graphicspath{{./}}

\journal{}
\makeatletter
\def\ps@pprintTitle{%
  \let\@oddhead\@empty
  \let\@evenhead\@empty
  \let\@oddfoot\@empty
  \let\@evenfoot\@oddfoot
}
\makeatother

\theoremstyle{definition}
\newtheorem{rmk}{Remark}

\graphicspath{{Figures/}}
\begin{document}
\begin{frontmatter}

\title{A variationally consistent and asymptotically convergent phase-field model for solute precipitation and dissolution}
\author[1]{A. Lamperti}
\author[1]{L. De Lorenzis\corref{cor1}}
\ead{ldelorenzis@ethz.ch}

\address[1]{Eidgen\"{o}ssische Technische Hochschule Z\"{u}rich, Computational Mechanics Group, Tannenstrasse 3, 8092 Z\"{u}rich, Switzerland}
\cortext[cor1]{Corresponding author}


\begin{abstract}
We propose a novel phase-field model for solute precipitation and dissolution in liquid solutions. Unlike in previous studies with similar scope, in our model the two non-linear coupled governing equations of the problem, which deliver the solute ion concentration and the phase-field variable, are derived in a variationally consistent way starting from a free energy functional of Modica-Mortola type. The phase-field variable is assumed to follow the non-conservative Allen-Cahn evolution law, whereas the solute ion concentration obeys the conservative Cahn-Hilliard equation. We also assess the convergence of the new model to the corresponding sharp-interface model via the method of matched asymptotic expansions, and derive a novel expression of the reaction rate of the sharp-interface model. Through a finite element discretization, we present several numerical examples in two and three dimensions.
\end{abstract}

\begin{keyword}
Phase-field model \sep Precipitation and dissolution \sep Variationally consistent model \sep Allen-Cahn \sep Cahn-Hilliard \sep Matched asymptotic expansions
\end{keyword}

\end{frontmatter}



\section{Introduction}
\label{sct:intro}
Phase-field modeling is a powerful approach for solving problems which involve complex and evolving interfaces. Instead of explicitly tracking sharp boundaries, the method introduces one or more continuous field variables—called phase fields—that smoothly vary across interfaces, allowing for a diffuse, yet physically consistent, description of transitions such as fracture, solidification, or phase separation. In the discretized setting, this approach leads to highly flexible computational tools which can seamlessly handle topology and shape changes of the moving interfaces. Among the countless contributions in the literature, we limit ourselves to mentioning the pioneering work of Allen, Cahn and Hilliard on phase transformations and spinodal decomposition \cite{cahn1958free, cahn1965phase, allen1979microscopic}, and the comprehensive reviews in \cite{Boettinger2002, Chen2002, Brener2009, plapp2015phase, Elder2016}.
\par 
In this paper, we focus on phase-field models used to formulate and solve solute precipitation and dissolution problems in liquid solutions, a topic that has received some attention in the literature \cite{xu2008phase, van2011phase, redeker2016upscaling, bringedal2020phase, rohde2021ternary,von2021investigation}. 
All these studies consider a liquid and a solid phase, and assume the solute concentration to be variable in the liquid phase and equal to its (constant) molar density in the solid. 
The effect of liquid flow is sometimes neglected \cite{xu2008phase, van2011phase, redeker2016upscaling} and sometimes accounted for \cite{bringedal2020phase, rohde2021ternary,von2021investigation}. Another difference among available models concerns the Gibbs-Thomson effect related to the interface curvature, which is neglected in \cite{xu2008phase, von2021investigation} and considered in \cite{van2011phase, redeker2016upscaling, bringedal2020phase,  rohde2021ternary}. Beside the aforementioned studies which are very close in scope to the present work, we mention at this point the more remotely related research on solid-state precipitation in binary alloys, see e.g. \cite{amirouche2009phase}; these studies typically consider three phases, and the solute concentration is variable in both the depleted and precipitate phases. Also, a more recent phase-field study \cite{ji2022phase} offers a general framework for stoichiometric precipitation and dissolution. 


Beside the mentioned differences in the detailed assumptions of the available models, two fundamental aspects are also handled differently, possibly because of different priorities in the respective scientific communities: the availability or not of a variational principle, from which the governing equations are derived, and the attention paid to asymptotic convergence of the phase-field model to a sharp-interface limit. With the exception of \cite{von2021investigation, amirouche2009phase, ji2022phase}, all cited models are constructed in such a way that asymptotic convergence to a target sharp-interface model is achieved, which is proved by the method of matched asymptotic expansions. On the other hand, all these models lack a variational structure. In other words, their governing equations are not derived starting from an underlying free energy functional, but rather constructed directly and (to some extent) independently from each other in strong form.  Interestingly, the contrary is true for the models in \cite{amirouche2009phase, ji2022phase}; they do possess a variational structure, but are not proved convergent to a sharp-interface limit. 
\par
In this paper, our goal is to develop a novel phase-field model for precipitation and dissolution which reconciles both requirements: variational consistency, i.e. existence of an underlying free energy potential, and asymptotic convergence to a target sharp-interface model. While the importance of asymptotic convergence has been recognized in the most relevant literature \cite{xu2008phase, van2011phase, redeker2016upscaling, bringedal2020phase, rohde2021ternary}, we here briefly mention the advantages of a variational structure. The variational approach leverages the calculus of variations to analyze the existence of solutions. It also provides a natural framework to interpret quasi-static equilibrium states and their evolution as energy minimization or incremental minimization problems, in line with principles common in many areas of physics. From a computational perspective, a variational formulation ensures a symmetric tangent stiffness matrix in the discrete setting, facilitates the use of optimization algorithms, and offers error bounds for the finite element discretization that can effectively guide mesh adaptation \cite{ORTIZ1999419}. Finally, the presence of an underlying free energy functional may enable the future incorporation of nucleation mechanisms, which are not included in any of the referenced models nor in our current formulation. 

Our new phase-field model for solute precipitation and dissolution assumes fixed solute concentration in the solid phase, neglects the liquid flow and takes into account the Gibbs-Thomson effect. In addition to being asymptotically convergent to the target sharp-interface model like the models in \cite{xu2008phase, van2011phase, 
 redeker2016upscaling,  bringedal2020phase,  rohde2021ternary}, our model is also variationally consistent, i.e. the governing equations are derived starting from a free energy potential based on the Modica-Mortola functional \cite{modica1977esempio, modica1987gradient, ambrosio2000variational} through suitable out-of-equilibrium thermodynamically consistent evolution laws. Furthermore, a new expression of the reaction rate is derived from the asymptotic analysis.
Our intended application is the simulation of Microbially Induced Calcium carbonate Precipitation (MICP) \cite{castro2019microbially}, a biochemical process that leverages microbes to promote calcium carbonate precipitation. MICP can be exploited, e.g., to induce self-healing in concrete \cite{lee2018current, yang2020review, feng2021microbial, chang2024application, zhang2024application}, carbon capture \cite{ chang2024application, fang2024enhancing, ma2024biomimetic} and soil strengthening/stabilization \cite{wang2017review,   mujah2017state, fu2023microbially}. However, the same model may be used for other applications as well. \par
 The paper is structured as follows: Section \ref{sectionsharp} introduces the basic quantities involved and the target sharp-interface model. Section \ref{sectionphasefield} formulates the proposed variationally consistent phase-field model for solute precipitation and dissolution, including the detailed derivation of the governing equations and a proof of thermodynamic consistency. Section \ref{sectionasymptotic} is devoted to the asymptotic analysis of the phase-field model in Section \ref{sectionphasefield}, which proves its convergence to the sharp-interface model in Section \ref{sectionsharp}. Section \ref{numerical examples} illustrates the results obtained from the proposed model upon finite element discretization, with several numerical tests in two and three dimensions. Finally, the main conclusions are drawn in Section \ref{conclusions}.\par
 We now summarize some of the notation used throughout the paper. $V \subset \mathbb{R}^{n_d}$ is the domain, $n_d$ is the number of space dimensions, with $n_d=2$ in two dimensions (2D) and $n_d=3$ in three dimensions (3D), and $\mathbf{x}\in V$ denotes a point in $V$. The letter $t$ denotes time. For vectors we use boldface letters, like $\mathbf{n}$, $\mathbf{s}$ and $\mathbf{y}_\varepsilon$. The symbols $\nabla (\cdot)$, $\nabla^2(\cdot)$ and $\nabla \cdot (\cdot)$ respectively denote the gradient, the Laplace and the  divergence operators applied to $(\cdot)$. The symbol $|(\cdot)|$ denotes the Euclidean norm of $(\cdot)$. The variational derivative of a functional $(\cdot)$ with respect to a field $(*)$ is indicated as $\delta (\cdot)/\delta(*)$.  We typically refer to liquid and solid phase by using the subscripts $l$ and $s$, respectively. 
 For clarity on the dimensions of the quantities introduced throughout the paper, Tab. \ref{Dataex} summarizes a possible set of consistent units.


\section{Sharp-interface model for solute precipitation and dissolution}
\label{sectionsharp}
As follows, we introduce some basic notions on the problem of solute precipitation and dissolution and formulate the target sharp-interface model. We conclude the section with some simple results on the critical radius of an initial precipitate of circular or spherical shape, which are useful for the later comparison with numerical results.

\subsection{Solute precipitation and dissolution reaction}
\label{subsct:reaction}
Solute \textit{precipitation} occurs when cations $c^+$ and anions $a^-$ react to form a crystalline solid $s$. The opposite reaction is also possible and is referred to as \textit{dissolution}. The generic precipitation/dissolution reaction can be written as
\begin{equation}
    n_c c^+ + n_a a^- \rightleftarrows s,
\label{reaction}
\end{equation}
where $n_c$ and $n_a$ denote appropriate stoichiometric coefficients.
As an example, the calcium carbonate precipitation/dissolution reaction reads
 \begin{equation}
     \mathrm{Ca^{2+}}+\mathrm{CO_3^{2-}}\rightleftarrows \mathrm{CaCO_3},
 \end{equation}
where $\mathrm{Ca^{2+}}$, $\mathrm{CO_3^{2-}}$ and $\mathrm{CaCO_3}$  denote calcium ions, carbonate ions and calcium carbonate, respectively. In this specific example $n_c = n_a = 1$.\par
\begin{figure}
     \centering
     \includegraphics[width=0.5\linewidth]{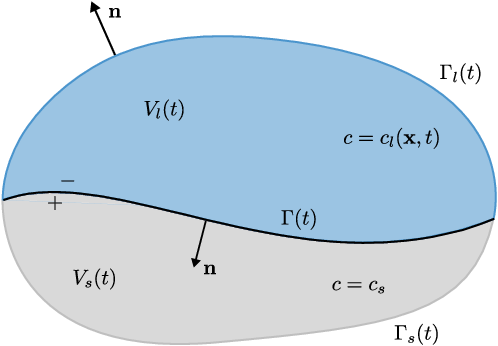}
     \caption{Scheme of the precipitation/dissolution problem.}
     \label{prepdiss}
 \end{figure}

Consider the fixed volume $V$ composed of the time-dependent liquid solution volume $V_l(t)$ and the crystalline solid portion $V_s(t)$, as shown in Fig. \ref{prepdiss}. 
A generic point $\mathbf{x}$ at time $t$ can be associated to the infinitesimal volume $dV$, composed of a liquid portion $dV_l$ and a solid portion $dV_s$, such that $dV = dV_l+dV_s$\footnote{This is equivalent to saying that both phases can potentially be present at a certain point.}. The local volume fractions of liquid and solid are defined as
 \begin{equation}
     \varphi_l(\mathbf{x},t) := \frac{dV_l}{dV}, \qquad \varphi_s(\mathbf{x},t)  := \frac{dV_s}{dV},
 \end{equation}
 and they satisfy $\varphi_l+\varphi_s=1$. From now onwards the relation $\varphi_l = 1-\varphi_s$ is used to express everything in terms of the solid local volume fraction $\varphi_s$. The local ion molar concentrations in the liquid solution are defined as
 \begin{equation}
     c^l_{c^+}(\mathbf{x},t) := \frac{dN^l_{c^+}}{dV_l}, \qquad c^l_{a^-}(\mathbf{x},t)  := \frac{dN^l_{a^-}}{dV_l},
 \end{equation}
 where $dN^l_{c^+}$ and $dN^l_{a^-}$ respectively denote the numbers of moles of cations and anions contained by the infinitesimal liquid volume $dV_l$.
 The (constant) crystalline solid molar density is defined as
  \begin{equation}
     c_s := \frac{dN_s}{dV_s},
 \end{equation}
where $dN_s$ denotes the number of moles of crystalline solid contained by the infinitesimal solid volume $dV_s$; its value can be found in literature, depending on the specific crystalline solid. As an example, its value for $\mathrm{CaCO_3}$ is $c_s = 2.71\times10^4 \ \mathrm{mol/m^3}$ \cite{bringedal2020phase, von2021investigation}.
 The total number of moles of cations in the infinitesimal volume $dV$ is then given by
\begin{equation}
     dN_{c^+} = dN^l_{c^+}+dN^s_{c^+} = dN^l_{c^+}+n_cdN_s,
 \end{equation}
 where $dN^s_{c^+}$ is the number of moles of cations contained by the solid infinitesimal volume $dV_s$. Due to stoichiometric considerations, one has $dN^s_{c^+} = n_cdN_s$ since one dissolved mole of crystalline solid gives $n_c$ moles of cations and $n_a$ moles of anions -- see (\ref{reaction}). Similarly, it is
 \begin{equation}
     dN_{a^-} = dN^l_{a^-}+dN^s_{a^-} = dN^l_{a^-}+n_adN_s.
 \end{equation}
 The local total molar concentration of cations is defined as
  \begin{equation}
     c_{c^+}(\mathbf{x},t) := \frac{dN_{c^+}}{dV} = \left[1-\varphi_s(\mathbf{x},t)\right]c^l_{c^+}(\mathbf{x},t) +n_c\varphi_s(\mathbf{x},t) c_s.
 \end{equation}
 Similarly, the local total molar concentration of anions is given by
  \begin{equation}
     c_{a^-}(\mathbf{x},t) := \frac{dN_{a^-}}{dV} = \left[1-\varphi_s(\mathbf{x},t)\right]c^l_{a^-}(\mathbf{x},t) +n_a\varphi_s(\mathbf{x},t) c_s.
 \end{equation}
 We assume that the ratio $dN^l_{a^-}/dN^l_{c^+}$ in the infinitesimal liquid solution volume $dV_l$ is the same as $\textbf{$dN^s_{a^-}/dN^s_{c^+}$}=n_a/n_c$ in the infinitesimal solid volume $dV_s$. It follows $c^l_{a^-}(\mathbf{x},t)=n_ac^l_{c^+}(\mathbf{x},t)/n_c$ and $c_{a^-}(\mathbf{x},t)=n_ac_{c^+}(\mathbf{x},t)/n_c$. This assumption is reasonable in the case of equal diffusivity of anions and cations and considering electroneutrality conditions, as pointed out in \cite{van2011phase, bringedal2020phase}. As a consequence, it is possible to model a single concentration field, e.g. by defining
 \begin{equation}\label{mixtureruleconc}
      c(\mathbf{x},t) := \frac{c_{c^+}(\mathbf{x},t)}{n_c} = \frac{c_{a^-}(\mathbf{x},t)}{n_a} = \left[1-\varphi_s(\mathbf{x},t)\right]c_l(\mathbf{x},t)+\varphi_s(\mathbf{x},t)c_s,
 \end{equation}
 which we refer to as the \textit{total concentration}, where $c_l(\mathbf{x},t):=c^l_{c^+}(\mathbf{x},t)/n_c = c^l_{a^-}(\mathbf{x},t)/n_a$.
 
  \begin{table}[h!]
\centering
\begin{tabular}{lccccccccccccccccccccc}
\toprule
  $\varphi_s$ & $c_l$ & $c_s$ &  $c$ & $D_l$ & $v_n$ & $\gamma$ & $\kappa$ & $r$  & $k$  & $c_l^{eq}$ & $R_{0c}$ & $h$ & $\delta$ & $\psi_l$ & $A$ & $B$ & $c_w$& $W$& $\varepsilon$ & $M_\phi$\\
\midrule
    /            & $\mathrm{\frac{mol}{m^3}}$     &  $\mathrm{\frac{mol}{m^3}}$       & $\mathrm{\frac{mol}{m^3}}$            & $\mathrm{\frac{m^2}{s}}$           & $\mathrm{\frac{m}{s}}$     & $\mathrm{\frac{m^2}{s}}$       & $\mathrm{\frac{1}{m}}$         & $\mathrm{\frac{m}{s}}$         & $\mathrm{\frac{m}{s}}$                     &  $\mathrm{\frac{mol}{m^3}}$    & $ \mathrm{m}    $  & / & /  & $\mathrm{\frac{J}{m^3}}$ &   $\mathrm{\frac{J}{m^3}}$ & $\mathrm{\frac{J}{m^2}}$ & /&/ & $\mathrm{m}$ & $\mathrm{\frac{m^3}{Js}}$    \\
\bottomrule
\end{tabular}
\caption{Set of possible consistent units for the quantities used throughout the paper.}
\label{Dataex}
\end{table}

\subsection{Sharp-interface model} 
\label{subsct:sharp}
In this section, we present a sharp-interface model which describes the free-boundary problem of solute precipitation and dissolution, neglecting liquid flow and, therefore, considering diffusion as the only ion transport mechanism in the liquid phase.
In the sharp-interface model, the moving liquid-solid interface $\Gamma(t)$ is assumed to have zero thickness; the total concentration $c(\mathbf{x},t)$ is discontinuous at the interface, passing from $c|_{-} $ on the liquid side to the constant value $c|_{+}= c_s$ on the solid side (see Fig. \ref{prepdiss}), and the mixture rule in Eq. \eqref{mixtureruleconc} applies only in the two bulk phases (not on the interface). The external boundary of the liquid volume $V_l(t)$ is $\Gamma_l(t)=\partial V \bigcap \partial V_l(t)$, $\partial V$ and $\partial V_l(t)$ being the fixed total external domain boundary and the total liquid boundary (including the moving interface $\Gamma(t)$), respectively. Similarly, $\Gamma_s(t)=\partial V \bigcap \partial V_s(t)$ is the external boundary of the solid domain, $\partial V_s(t)$ being the total solid boundary. The unit outward normal vector to $\partial V_l(t)$ is denoted as $\mathbf{n}(\mathbf{x},t)$; at the interface $\Gamma(t)$, the vector $\mathbf{n}$ points into the solid. The general form of the sharp-interface model for solute precipitation and dissolution with homogeneous Neumann boundary conditions and without liquid flow \cite{van2011phase}  is given by
\begin{equation}\label{sharpinterfacemodel}
\begin{cases}
    \frac{\partial c}{\partial t}=D_l\nabla^2c\quad\mathrm{in}\quad V_l(t)\\
    c=c_s \quad\mathrm{in}\quad V_s(t)\\
    \nabla c\cdot \mathbf{n}=0\quad \mathrm{on}\quad  \Gamma_l(t)\\
    v_n\left(c_s-c|_{-}\right)=D_l\nabla c|_{-}\cdot\mathbf{n}\quad \mathrm{on}\quad \Gamma(t)\\
    v_n = -\gamma \kappa -r(c|_{-}) \quad \mathrm{on} \quad \Gamma(t)\\
    c(\mathbf{x},0)=c_0(\mathbf{x})\quad \mathrm{in}\quad V_l(0).
\end{cases}
\end{equation}
The equations in (\ref{sharpinterfacemodel}) in the given order represent
\begin{itemize}
\item{Fick's diffusion law for the ions in the liquid domain, where $D_l$ is the (constant) diffusivity of the ions in the liquid solution, assumed to be the same for anions and cations;}

\item{the condition that the total ion concentration equals the constant molar density in the solid phase;}
\item{homogeneous Neumann boundary conditions on the external boundary of the liquid domain;}
\item{mass balance at the interface, where $v_n$ is the (unknown) normal component of the interface velocity;}
\item{the kinetic law (also known as the \emph{Gibbs-Thomson} equation) which expresses the interface motion due to curvature effects and precipitation/dissolution. Here $\gamma$ is the interface diffusivity; $\kappa=\nabla\cdot\mathbf{n}$ vanishes in $1$D and is equal to the interface curvature in $2$D, whereas in 3D it is the sum of the two interface principal curvatures (as stated also in \cite{allen1979microscopic, caginalp1994phase}) and not the mean curvature. 
 This aspect is particularly important for the computation of the critical radius, as discussed in Section \ref{criticalradiusprep}. The meaning of $\kappa$ will become clearer from the asymptotic analysis carried out in Section \ref{sectionasymptotic}. Finally, the term $r(c|_{-})$ is the precipitation/dissolution reaction rate. In this paper, the following new expression is proposed
\begin{equation}\label{reactionrate}
    r(c|_{-})=\frac{k}{c_l^{eq2}}\left[\left(c|_{-} - c_l^{eq}\right)^2-2\left(c|_{-}-c_l^{eq}\right)\left(c|_{-}-c_s\right)\right],
\end{equation}
which is valid for $0\le c|_{-}\le c_s$. The quantity $k$ is a rate coefficient and $c_l^{eq}$ is the (constant) equilibrium concentration of the ions in the liquid solution. The reaction rate (Fig. \ref{rate}) is negative for $0\le\ c|_{-} < c_l^{eq}$, positive for $c_l^{eq}<c|_{-}\le c_s$ and vanishes for $c|_{-}=c_l^{eq}$. The proposed expression for $r(c|_{-})$ in \eqref{reactionrate} is different from the ones introduced in other works \cite{xu2008phase, van2011phase, redeker2016upscaling, bringedal2020phase, rohde2021ternary, von2021investigation} and is derived as a result of the asymptotic analysis of the phase-field model, as discussed in Section \ref{sectionasymptotic};}
\item{the initial conditions in the initial liquid domain.}

\end{itemize}

\par
The unknowns of the sharp-interface model \eqref{sharpinterfacemodel} are the total ion concentration $c$ and the normal interface velocity $v_n$. 
This model typically cannot be solved analytically, but needs numerical approaches such as level-set \cite{xu2012phase} or Arbitrary Lagrangian Eulerian (ALE) methods \cite{van2011phase}. The former  tracks the sharp interface implicitly by means of a level-set function conveniently chosen as a signed distance function, whereas the latter performs explicit interface tracking. 


In this paper, we do not solve the sharp-interface model \eqref{sharpinterfacemodel} but propose a phase-field model, which implicitly tracks the moving interface by means of a phase indicator variable. The phase field smoothly changes across a diffuse regularization of the sharp interface characterized by a small regularization length scale. The discretized diffuse problem can be solved using a fixed mesh of standard Lagrangian finite elements, thereby circumventing many complications of level-set and ALE methods. The drawback of the phase-field model in the computational setting is that very fine meshes are required to resolve the width of the regularized interface, which (in absence of adaptive local refinement) leads to a  high computational cost.
 
\subsection{Critical radius of a circular/spherical initial precipitate}
\label{criticalradiusprep}
\begin{figure}[htbp]
    \centering
    \begin{subfigure}[b]{0.48\textwidth}
        \centering
        \includegraphics[width=\linewidth]{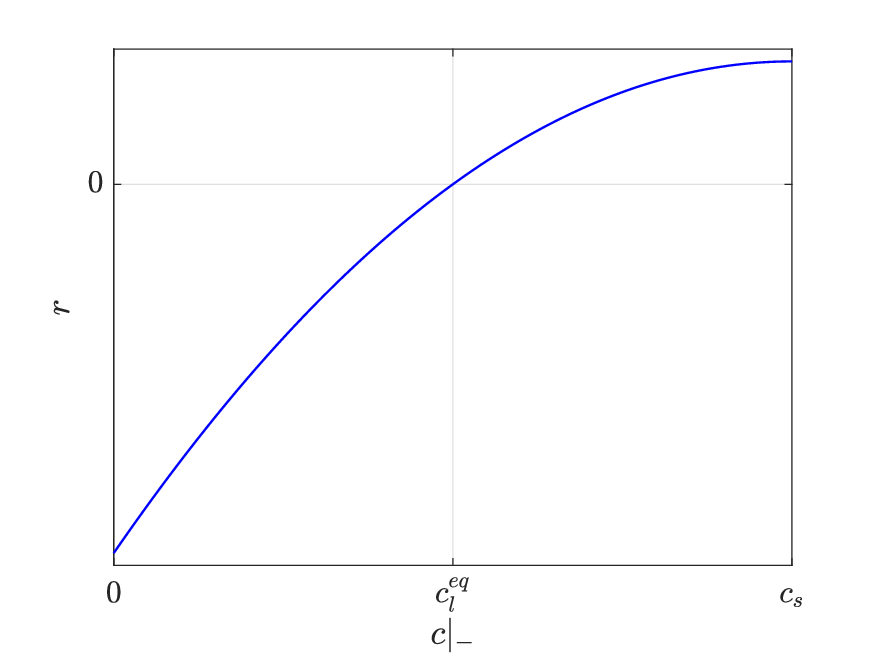}
        \caption{}
        \label{rate}
    \end{subfigure}
    \hfill
    \begin{subfigure}[b]{0.48\textwidth}
        \centering
        \includegraphics[width=\linewidth]{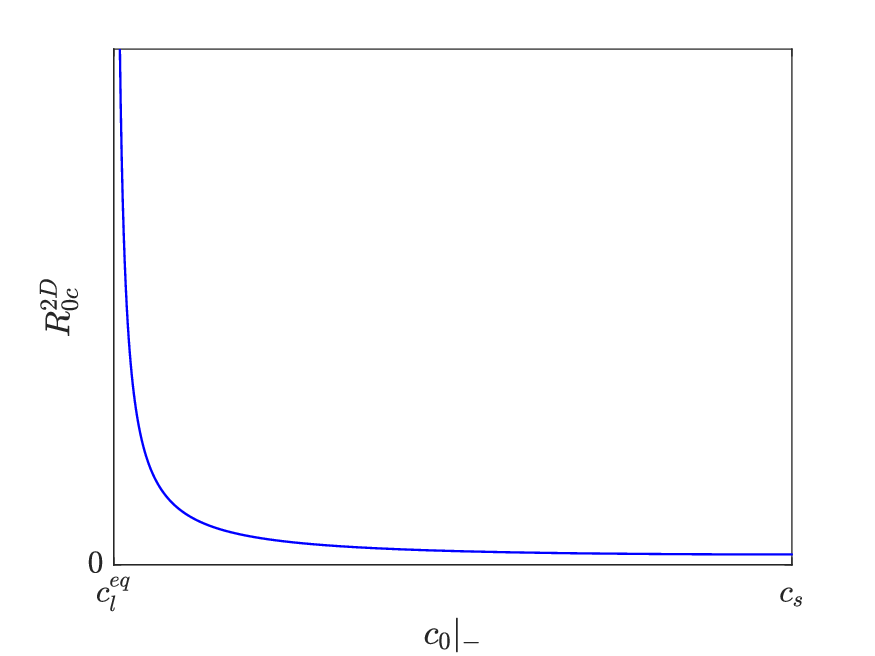}
        \caption{}
        \label{criticalradius}
    \end{subfigure}
    \caption{(a) Precipitation/dissolution reaction rate and (b) initial critical radius in 2D as functions of the total concentration at the interface on the liquid side.}
    \label{combined}
\end{figure}

In the case of a circular (in $2$D) or spherical (in $3$D) initial precipitate inside a liquid domain with $c_0|_{-}>c_l^{eq}$, where $c_0|_{-}$ is the initial concentration at the liquid side of the interface, we can solve (\ref{sharpinterfacemodel}) to find a closed-form expression of the critical initial radius $R_{0c}$, i.e. the minimum radius for which further precipitation rather than dissolution of the initial precipitate is observed. 
In the $2$D case, $\kappa$ is the interface curvature and can be expressed as $\kappa = -1/R$, $R$ being the radius of the circular precipitate\footnote{The minus sign comes from the assumed  convention on the vector $\mathbf{n}$.}. Evaluating the kinetic law of the sharp-interface model \eqref{sharpinterfacemodel} at initial time and setting $v_n=v_{n0}=0$, we find the following expression for the initial critical radius for precipitation
\begin{equation}\label{criticalradiusformula2D}
    R_{0c}^{2D} = \frac{\gamma c_l^{eq2}}{k\left[\left(c_0|_{-} - c_l^{eq}\right)^2-2\left(c_0|_{-}-c_l^{eq}\right)\left(c_0|_{-}-c_s\right)\right]} \qquad \mathrm{for}\quad c_0|_{-}>c_l^{eq}.
\end{equation}
$R_{0c}^{2D}$ from Eq. \eqref{criticalradiusformula2D} is plotted in Fig. \ref{criticalradius} as a function of $c_0|_{-}$.
 The initial critical radius tends to $\infty$ as $c_0|_{-}\to c_l^{eq}$, which means that it is impossible to observe the growth of a finite-radius initial precipitate starting from equilibrium conditions. The initial critical radius decreases as the initial concentration at the liquid side increases, which means that precipitates in more supersaturated solutions require a smaller initial radius to grow (Gibbs-Thomson effect).\par
 In the $3$D case, $\kappa$ is the sum of the two principal curvatures of the interface and can be expressed as $\kappa = -(1/R+1/R)=-2/R$, $R$ being the radius of the spherical precipitate. Again from the kinetic law of the sharp-interface model \eqref{sharpinterfacemodel} at initial time and setting $v_n=v_{n0}=0$, the initial critical radius for precipitation follows as
\begin{equation}\label{criticalradiusformula3D}
    R_{0c}^{3D} = \frac{2\gamma c_l^{eq2}}{k\left[\left(c_0|_{-} - c_l^{eq}\right)^2-2\left(c_0|_{-}-c_l^{eq}\right)\left(c_0|_{-}-c_s\right)\right]}  \qquad \mathrm{for}\quad c_0|_{-}>c_l^{eq},
\end{equation}
which is twice  $R_{0c}^{2D}$ in \eqref{criticalradiusformula2D}.

\section{Variationally consistent phase-field model for solute precipitation and dissolution}
\label{sectionphasefield}
In this section, we present a novel phase-field model for solute precipitation and dissolution, under the same assumptions of the sharp-interface model in Section \ref{sectionsharp}, i.e. fixed concentration in the solid, negligible liquid flow and equal concentration and diffusivity of cations and anions.
\subsection{Relevant quantities and free energy functional} 
\label{subsct:energy}
\begin{figure}
     \centering
     \includegraphics[width=0.9\linewidth]{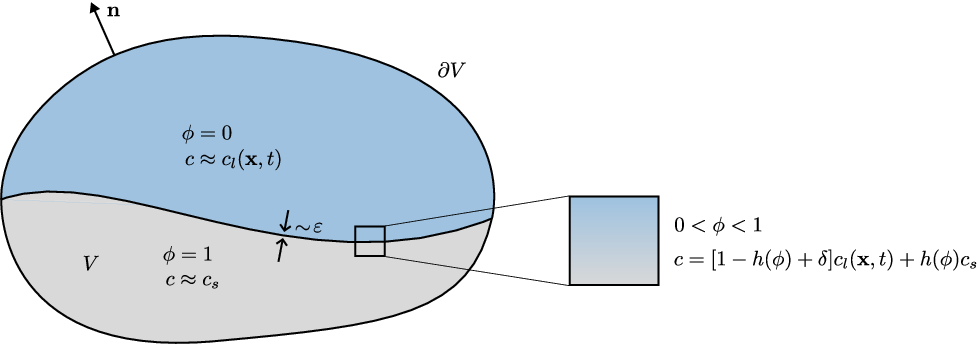}
     \caption{Scheme of the precipitation/dissolution problem with a diffuse interface approximation as in the phase-field model.}
     \label{pf}
 \end{figure}

\begin{figure}[htbp]
    \centering
    \begin{subfigure}[b]{0.48\textwidth}
        \centering
        \includegraphics[width=\linewidth]{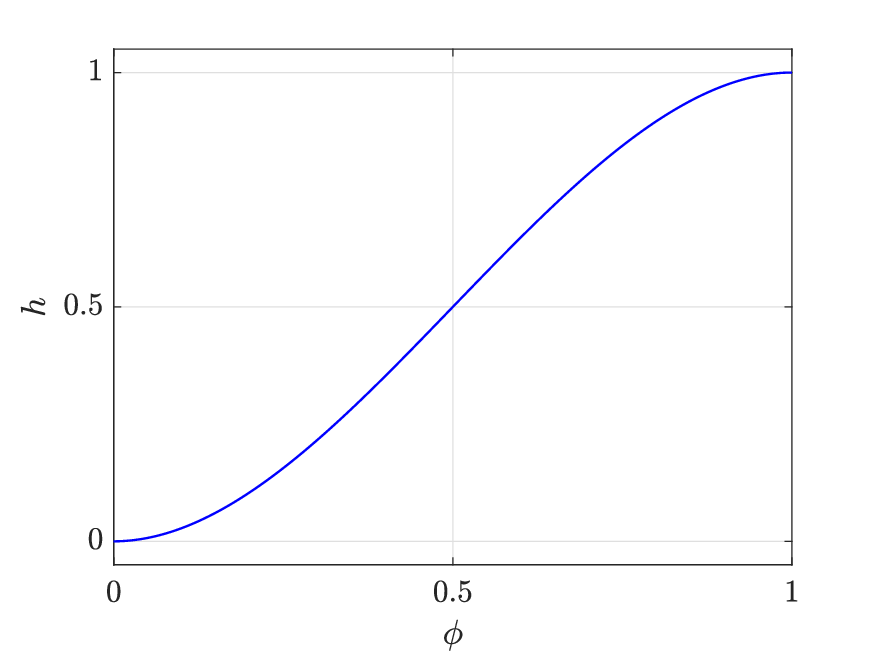}
        \caption{}
        \label{interpolation}
    \end{subfigure}
    \hfill
    \begin{subfigure}[b]{0.48\textwidth}
        \centering
        \includegraphics[width=\linewidth]{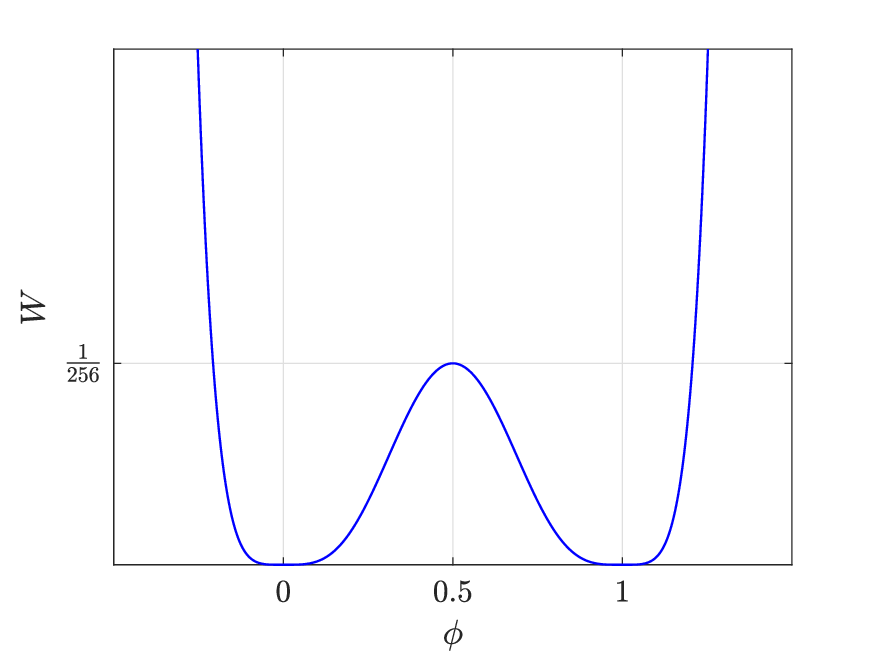}
        \caption{}
        \label{doublewell}
    \end{subfigure}
    \caption{(a) Interpolation function and (b) double-well function adopted in the phase-field model.}
    \label{combined2}
\end{figure}

In the phase-field model, 
the moving interface is implicitly tracked by means of a new variable referred to as \emph{phase indicator}. The phase indicator $\phi$ is assumed to be equal to the local volume fraction of crystalline solid, i.e. $\phi(\mathbf{x},t)\equiv\varphi_s(\mathbf{x},t)$. It takes the values $\phi=0$ in the liquid solution,  $\phi=1$ in the solid region, and intermediate values $0<\phi<1$ inside the diffuse interface region (Fig. \ref{pf}). 
The variables of the phase-field model are the phase indicator $\phi(\mathbf{x},t)$ and the total ion concentration $c(\mathbf{x},t)$. The latter is defined according to the following mixture rule
\begin{equation}\label{ctildePF}
    c(\mathbf{x},t):= \left[1-h(\phi)+\delta\right]c_l(\mathbf{x},t)+h(\phi)c_s,
\end{equation}
which is a non-linear interpolation of the ion concentration in the liquid $c_l(\mathbf{x},t)$ and the crystalline solid molar density $c_s$, based on the interpolation function $h(\phi)$. A similar interpolation of the concentration field is considered in \cite{kim1999phase, plapp2011unified,mai2016phase}. Eq. (\ref{ctildePF}) is the counterpart of Eq. \eqref{mixtureruleconc} for the phase-field model, and is valid in the whole domain. It implies that in the liquid phase  $c=(1+\delta)c_l(\mathbf{x},t)\approx c_l(\mathbf{x},t)$  and in the solid phase $c=\delta c_l(\mathbf{x},t)+c_s\approx c_s$. The constant $\delta$ is a small parameter required to prevent numerical issues, as will be clear later.
The function $h(\phi)$ has to satisfy
\begin{equation}
    \begin{cases}
        h(0)=0\\
        h(1)=1\\
        h'(0)=h'(1)=0.
    \end{cases}
\end{equation}
The reason for the conditions on the first derivative will become clear in Section \ref{subsct:eqs_dim}. The simplest polynomial function that satisfies the previous requirements is
\begin{equation}\label{interpolationeq}
h(\phi)=3\phi^2-2\phi^3,
\end{equation}
see Fig. \ref{interpolation}, which we adopt in the remainder of this paper.

The proposed free energy functional of the \emph{Modica-Mortola} type \cite{modica1977esempio, modica1987gradient, ambrosio2000variational} reads
\begin{equation}\label{free energy}
    \Psi[\phi, c]=\underbrace{\int_V \left \{ \left[1-h(\phi)+\delta\right]\psi_l(c_l(\phi,c))+h(\phi)\psi_s \right \} \,\textrm{d}V}_{\Psi_{bulk}\left[\phi,c\right]}+\underbrace{\frac{B}{c_w}\int_V   \left ( \frac{W(\phi)}{\varepsilon}+\varepsilon|\nabla\phi|^2\right )\,\textrm{d}V}_{\Psi_{int}[\phi]},
\end{equation}
and is the sum of the bulk free energy $\Psi_{bulk}\left[\phi, c \right]$ and the interfacial free energy $\Psi_{int}[\phi]$. The integrand of the first term is a non-linear interpolation of the liquid  free energy density $\psi_l$ and the solid free energy density $\psi_s$, based on the previously introduced interpolation function $h(\phi)$ and on the same mixture rule used in Eq. \eqref{ctildePF}. The liquid free energy density is expressed using a parabolic approximation
\begin{equation}\label{psiL}
    \psi_l(c_l)=\frac{A \left(c_l-c_l^{eq} \right)^2}{2 c_l^{eq2}},
\end{equation}
where $A$ is a coefficient which has units of an energy per unit volume and $c_l$ is expressed as a function of $\phi$ and $c$ by inverting Eq. \eqref{ctildePF}
\begin{equation}\label{cLPF}
    c_l(\phi, c) = \frac{c-h(\phi)c_s}{1-h(\phi)+\delta}.
\end{equation}
The solid free energy density, using the same approximation used for $\psi_l$ (in general with a different constant $A_s$), reads
\begin{equation}\label{psiS}
    \psi_s = \frac{A_s \left(c_s-c_s^{eq}\right)^2}{2 c_s^{eq2}}=0,
\end{equation}
since at equilibrium $c_s^{eq}=c_s$.
Parabolic approximations analogous to \eqref{psiL} and \eqref{psiS} are used in \cite{amirouche2009phase} within a phase-field model of solute precipitation in alloys, and in \cite{abubakar2015phase}  and  \cite{mai2016phase, gao2020efficient} for phase-field models of hot corrosion and pitting corrosion, respectively.
\par
In the interfacial free energy, the constant $B$ is a coefficient which has units of an energy per unit surface and can be interpreted as the \emph{liquid-solid interfacial tension} (energy required to create a unit area of new liquid-solid surface). The term $W(\phi)$ is a double-well function expressed by
\begin{equation}\label{W}
    W(\phi)=\phi^4(1-\phi)^4,
\end{equation}
which models an energy barrier between the two bulk phases described by $\phi=0$ and $\phi=1$ (Fig. \ref{doublewell}).
The reason for such a high-degree double well is the requirement for the phase-field model to converge to the sharp-interface model when the regularization parameter $\varepsilon \to 0$, as discussed in Section \ref{sectionasymptotic} and Appendix \ref{appx:asymptotic}.
The coefficient $c_w$ is defined as
\begin{equation}
    c_w := 2\int_0^1 \sqrt{W(\phi)}\,\textrm{d}\phi=\frac{1}{15}.
\end{equation}
The regularization parameter $\varepsilon$ is proportional to the diffuse interface length.
The last term in the integrand of \eqref{free energy} is the gradient free energy density. This term penalizes sharp interfaces and promotes the creation of diffuse interfaces. The energy constants $A$ and $B$ must satisfy
\begin{equation}\label{constraint}
    \frac{A}{B} = \frac{2k}{\gamma},
\end{equation}
where $k$ and $\gamma$ are the reaction rate and the interface diffusivity, previously introduced in the sharp-interface model. The constraint in Eq. \eqref{constraint} is again needed for the phase-field model to converge to the corresponding sharp-interface model \eqref{sharpinterfacemodel} in the limit $\varepsilon\to 0$, as shown in Section \ref{sectionasymptotic}.\par

\subsection{Governing equations} 
\label{subsct:eqs_dim}
The governing equations are  derived in a variationally consistent way starting from the free energy functional \eqref{free energy}. 
The non-conserved phase indicator $\phi(\mathbf{x},t)$ is assumed to follow the Allen-Cahn evolution law \cite{allen1979microscopic}
\begin{equation}\label{allen-cahn}
    \frac{\partial \phi}{\partial t} = -M_\phi \frac{\delta \Psi}{\delta \phi},
\end{equation}
where $M_\phi>0$ is the mobility related to interface kinetics, assumed constant, and $\frac{\delta \Psi}{\delta \phi}$ denotes the variational derivative of the functional $\Psi$ with respect to $\phi$. Explicitly, the first governing equation reads\begin{equation}\label{eqphi}
    \frac{\partial \phi}{\partial t} = -M_\phi \left \{ h'(\phi)\left[-\psi_l(c_l(\phi, c))+ \frac{c-c_s(1+\delta)}{1-h(\phi)+\delta}\frac{\partial \psi_l}{\partial c_l}(c_l(\phi, c))\right]+\frac{B}{c_w}\left[\frac{W'(\phi)}{\varepsilon}-2\varepsilon\nabla^2\phi \right] \right \},
\end{equation}
where
\begin{equation}
   \frac{\partial \psi_l}{\partial c_l}(c_l) = A\frac{c_l-c_l^{eq}}{c_l^{eq2}}.
\end{equation}
The term $2M_\phi B\varepsilon/c_w$ has units of a diffusivity and can be interpreted as the interface diffusivity $\gamma$. In the rest of the paper we assume $M_\phi = \gamma c_w/(2B\varepsilon)$ and this choice allows to converge to the sharp-interface model \eqref{sharpinterfacemodel}, as shown in Section \ref{sectionasymptotic}. The first term on the right-hand side of Eq. \eqref{eqphi} models the precipitation/dissolution reaction and is proportional to $h'(\phi)$. Thus, it becomes clear that the requirement $h'(0)=h'(1)=0$ is needed to restrict the precipitation/dissolution reaction exclusively within the diffuse interface where $h'(\phi)\neq 0$, whereas anywhere else $h'(\phi)= 0$ with a smooth transition between interface and bulk. \par
The conserved total ion concentration $c$ obeys the Cahn-Hilliard equation \cite{cahn1965phase}
\begin{equation}\label{cahn-hilliard}
    \frac{\partial c}{\partial t} = \nabla \cdot \left[ M_c\nabla \left( \frac{\delta \Psi}{\delta c} \right) \right],
\end{equation}
where $M_c>0$ is the mobility related to the solute diffusivity and $\frac{\delta \Psi}{\delta c}$ denotes the variational derivative of the functional $\Psi$ with respect to $c$. The latter is given by
\begin{equation}\label{variationalderivativectilde}
    \frac{\delta \Psi}{\delta c} = A\frac{c_l(\phi, c)-c_l^{eq}}{c_l^{eq2}}.
\end{equation}
The gradient of Eq. \eqref{variationalderivativectilde} 
combined with
Eq. \eqref{cLPF} 
yields
\begin{equation}\label{Mnablavarderctilde}
    M_c\nabla \left( \frac{\delta \Psi}{\delta c} \right) = \frac{M_cA}{c_l^{eq2}}\left \{  \frac{\nabla c}{1-h(\phi)+\delta}+\frac{\left[ c-c_s(1+\delta)\right]h'(\phi)\nabla\phi}{\left[1-h(\phi)+\delta\right]^2} \right \}.
\end{equation}
As discussed in \cite{ji2022phase} in the context of stoichiometric reactions, $M_c$ is related to the total diffusivity as follows
\begin{equation}\label{MDrelation}
    M_c (\phi)= \frac{D(\phi)}{\frac{\partial^2 \psi_l}{\partial c_l^2}} = \frac{D(\phi)c_l^{eq2}}{A},
\end{equation}
with $D(\phi)$ as the total diffusivity, defined as
\begin{equation}\label{totaldiffusivity}
    D(\phi) := \left[1-h(\phi)+\delta\right]D_l+h(\phi)\underbrace{D_s}_{\approx 0} \approx \left[1-h(\phi)+\delta\right]D_l.
\end{equation}
Here we have assumed the diffusivity in the solid $D_s$ to be negligible, since it is typically a few orders of magnitude smaller than the one in the liquid.
With Eqs. \eqref{Mnablavarderctilde}-\eqref{totaldiffusivity}, the final expression of Eq. \eqref{cahn-hilliard} reads
\begin{equation}\label{eqctilde}
    \frac{\partial c}{\partial t} = D_l\nabla \cdot \left \{  \nabla c+\frac{\left[ c-c_s(1+\delta)\right]h'(\phi)\nabla\phi}{1-h(\phi)+\delta} \right \}.
\end{equation}

From the two governing  equations (\ref{eqphi}) and (\ref{eqctilde}), the reason behind the introduction of the small parameter $\delta$ becomes clear. The term $1-h(\phi)$ tends to zero in the solid region, where $\phi=1$, hence this parameter prevents divisions by zero. In \cite{van2011phase}, the parameter $\delta$ is introduced at the very end in the evolution equation of the concentration field. Consistently with our variational framework, we prefer to introduce it directly in the free energy functional \eqref{free energy} and in the definitions \eqref{ctildePF} and \eqref{totaldiffusivity}. This choice leads to the presence of $\delta$ also at the numerator on the right-hand side, which is the only difference between (\ref{eqctilde}) and  the  concentration evolution equation in \cite{van2011phase} (apart from the different convention for the phase indicator and the different expression for the interpolation function).

The phase-field model given by Eqs. \eqref{eqphi} and \eqref{eqctilde} is completed by assuming homogeneous Neumann boundary conditions for both $\phi$ and $c$
\begin{equation}\label{boundaryconditionsPFmodel}
    \begin{cases}
        \nabla \phi \cdot \mathbf{n} = 0 \quad \mathrm{on}\quad \partial V \times[0,T]\\
        \nabla c \cdot \mathbf{n} = 0  \quad \mathrm{on}\quad \partial V \times[0,T],\\
    \end{cases}
\end{equation}
where $T$ denotes the final time, and initial conditions
\begin{equation}\label{initialconditionsPFmodel}
    \begin{cases}
        \phi(\mathbf{x},0)=\phi_0(\mathbf{x})\quad \mathrm{in}\quad  V\\
        c(\mathbf{x},0)=c_0(\mathbf{x})\quad \mathrm{in}\quad  V.   
        \end{cases}
\end{equation}
\par
The proposed variationally consistent phase-field model is summarized in Tab. \ref{phasefieldmodelsummary}.

\begin{table}[h!]
\centering
\renewcommand{\arraystretch}{2}
\begin{tabular}{@{} m{4cm} m{10cm} @{}}
\toprule
\textbf{Free energy functional} & 
$\Psi[\phi, c]=\int_V  \left[1-h(\phi)+\delta\right]\psi_l(c_l(\phi,c))  \,\textrm{d}V+\frac{B}{c_w}\int_V   \left ( \frac{W(\phi)}{\varepsilon}+\varepsilon|\nabla\phi|^2\right )\,\textrm{d}V $\\

\textbf{Evolution equations} & 
$ \frac{\partial \phi}{\partial t} = -M_\phi \left \{ h'(\phi)\left[-\psi_l(c_l(\phi, c))+ \frac{c-c_s(1+\delta)}{1-h(\phi)+\delta}\frac{\partial \psi_l}{\partial c_l}(c_l(\phi, c))\right]+\frac{B}{c_w}\left[\frac{W'(\phi)}{\varepsilon}-2\varepsilon\nabla^2\phi \right] \right \}$ \\

 & 
$\frac{\partial c}{\partial t} = D_l\nabla \cdot \left \{  \nabla c+\frac{\left[ c-c_s(1+\delta)\right]h'(\phi)\nabla\phi}{1-h(\phi)+\delta} \right \} \quad \mathrm{in} \quad V\times[0,T]$ \\

\textbf{Boundary conditions} & 
$ \nabla \phi \cdot \mathbf{n} = 0$ \\

 & 
$\nabla c \cdot \mathbf{n} = 0  \quad \mathrm{on} \quad \partial V\times[0,T]$ \\
\bottomrule
\end{tabular}
\caption{Summary of the proposed variationally consistent phase-field model.}
\label{phasefieldmodelsummary}
\end{table}

\subsection{Thermodynamic consistency} 
The proposed free energy \eqref{free energy} can be written as
\begin{equation}
    \Psi \left[ \phi, c \right] = \int_V \psi(\phi, \nabla\phi, c)\,\textrm{d}V,
\end{equation}
where
\begin{equation}
      \psi(\phi, \nabla\phi, c) = \left[1-h(\phi)+\delta\right]\frac{A \left[c_l(\phi,c)-c_l^{eq}\right]^2}{2 c_l^{eq2}}+\frac{B}{c_w}\left(\frac{W(\phi)}{\varepsilon}+\varepsilon|\nabla \phi |^2\right) , 
\end{equation}
with $c_l(\phi, c)$ given by Eq. \eqref{cLPF}. As follows, we aim to verify the energy dissipation property
\begin{equation}
\label{2P}
    \frac{d}{dt}\int_V \psi(\phi, \nabla \phi, c)\,\textrm{d}V =  \dot{\mathcal{W}}-\dot{\mathcal{D}},
\end{equation}
where $\dot{\mathcal{W}}$ is the \emph{power} associated with the external work exerted through the boundary $\partial V$ and $\dot{\mathcal{D}}\ge 0$ is the \emph{dissipation rate}. 
All the three quantities in brackets depend on time $t$, hence
\begin{equation}
    \frac{d}{dt}\int_V \psi(\phi, \nabla \phi, c)\,\textrm{d}V = \int_V \left(\frac{\partial \psi }{\partial \phi}\frac{\partial \phi}{\partial t} + \frac{\partial \psi }{\partial \nabla\phi}\cdot\frac{\partial \nabla\phi}{\partial t}+\frac{\partial \psi }{\partial c}\frac{\partial c}{\partial t}\right)\,\textrm{d}V.
\end{equation}
Using integration by parts, we obtain
\begin{equation}\label{ddt}
\begin{aligned}
    \frac{d}{dt}\int_V \psi(\phi, \nabla \phi, c)\,\textrm{d}V &= \int_V \left[\frac{\partial \psi}{\partial \phi}-\nabla \cdot \left(\frac{\partial \psi}{\partial \nabla \phi}\right)\right]\frac{\partial \phi}{\partial t}\,\textrm{d}V +\int_V \frac{\partial \psi }{\partial c}\frac{\partial c}{\partial t} \,\textrm{d}V +\int_{\partial V} \frac{\partial \psi}{\partial \nabla\phi}\cdot \mathbf{n} \frac{\partial \phi}{\partial t} \,\textrm{d}S\\
    &=\int_V \frac{\delta \Psi}{\delta \phi}\frac{\partial \phi}{\partial t}\,\textrm{d}V +\int_V \frac{\delta \Psi }{\delta c}\frac{\partial c}{\partial t} \,\textrm{d}V +\int_{\partial V} \frac{\partial \psi}{\partial \nabla\phi}\cdot \mathbf{n} \frac{\partial \phi}{\partial t} \,\textrm{d}S,
    \end{aligned}
\end{equation}
with
\begin{equation}
    \frac{\delta \Psi}{\delta \phi} = \frac{\partial \psi}{\partial \phi}-\nabla \cdot \left(\frac{\partial \psi}{\partial \nabla \phi}\right), \qquad 
    \frac{\delta \Psi}{\delta c} = \frac{\partial \psi}{\partial c}.
\end{equation}
Considering the Allen-Cahn and Cahn-Hilliard equations given by Eqs. \eqref{allen-cahn} and \eqref{cahn-hilliard}, with $M_\phi$ and $M_c$ positive mobilities, Eq. \eqref{ddt} gives
\begin{equation}
     \frac{d}{dt}\int_V \psi(\phi, \nabla \phi, c)\,\textrm{d}V = -\int_V  M_\phi \left(\frac{\delta\Psi}{\delta \phi}\right)^2\,\textrm{d}V +\int_V \frac{\delta \Psi }{\delta c}\nabla \cdot \left[M_c\nabla\left(\frac{\delta\Psi}{\delta c}\right)\right]  \,\textrm{d}V +\int_{\partial V} \frac{\partial \psi}{\partial \nabla\phi}\cdot \mathbf{n} \frac{\partial \phi}{\partial t} \,\textrm{d}S.
\end{equation}
Using again integration by parts yields
\begin{equation}
     \frac{d}{dt}\int_V \psi(\phi, \nabla \phi, c)\,\textrm{d}V = -\int_V  M_\phi \left(\frac{\delta\Psi}{\delta \phi}\right)^2\,\textrm{d}V -\int_V M_c\biggr\rvert\nabla\left(\frac{\delta\Psi}{\delta c}\right)\biggr\rvert^2  \,\textrm{d}V+\int_{\partial V} \frac{\partial \psi}{\partial \nabla\phi}\cdot \mathbf{n} \frac{\partial \phi}{\partial t} \,\textrm{d}S+\int_{\partial V} \frac{\delta \Psi }{\delta c} M_c\nabla\left(\frac{\delta\Psi}{\delta c}\right) \cdot \mathbf{n} \,\textrm{d}S.
\end{equation}
By defining the \emph{dissipation rate}
\begin{equation}
    \dot{\mathcal{D}} := \int_V  M_\phi \left(\frac{\delta\Psi}{\delta \phi}\right)^2\,\textrm{d}V +\int_V M_c\biggr\rvert\nabla\left(\frac{\delta\Psi}{\delta c}\right)\biggr\rvert^2  \,\textrm{d}V \ge 0
\end{equation}
and the \emph{power}
\begin{equation}
    \dot{\mathcal{W}} := \int_{\partial V} \frac{\partial \psi}{\partial \nabla\phi}\cdot \mathbf{n} \frac{\partial \phi}{\partial t} \,\textrm{d}S+\int_{\partial V} \frac{\delta \Psi }{\delta c}M_c\nabla\left(\frac{\delta\Psi}{\delta c}\right)\cdot \mathbf{n}  \,\textrm{d}S,
\end{equation}
we proved (\ref{2P})
with $\dot{\mathcal{D}}\ge0$.

\section{Asymptotic analysis}
\label{sectionasymptotic}

This section focuses on the asymptotic analysis of the phase-field model proposed in Section \ref{sectionphasefield} for $\varepsilon\to 0$. For conciseness, we report here the main results, whereas most of the details can be found in Appendix \ref{appx:asymptotic}. The analysis is carried out similarly as in  \cite{van2011phase, redeker2016upscaling, bringedal2020phase}, using the method of \emph{matched asymptotic expansions} (see \cite{holmes2012introduction} for an exhaustive explanation of the technique). This expansion consists of two parts: an
\textit{outer expansion} valid far away from the interface, and an \textit{inner expansion} valid close to the
interface. For the sake of simplicity and without loss of generality, we set $\delta=\varepsilon$ (this involves a slight abuse of notation since  $\delta$ is dimensionless and $\varepsilon$ is a length).
The objective of asymptotic analysis  is twofold: 1) prove the convergence of the proposed phase-field model to the  sharp-interface model \eqref{sharpinterfacemodel}; 2) determine the expression of the reaction rate in \eqref{sharpinterfacemodel} which stems from the assumed free energy functional \eqref{free energy}.
\subsection{Outer expansions}
We start by introducing the following outer expansions (valid far from the interface region) of $\phi$ and $c$
\begin{equation}\label{phiout}
    \phi^{o}(\mathbf{x},t)=\phi_0^{o}(\mathbf{x},t)+\varepsilon\phi_1^{o}(\mathbf{x},t)+\dots
\end{equation}
\begin{equation}\label{ctildeout}
    c^{o}(\mathbf{x},t)=c_0^{o}(\mathbf{x},t)+\varepsilon c_1^{o}(\mathbf{x},t)+\dots .
\end{equation}
 The functions $\phi_0^{o}$ and $c_0^{o}$ represent the limits of the solutions $\phi$ and $c$ of \eqref{eqphi} and \eqref{eqctilde} for $\varepsilon\rightarrow 0$.  
 
 Multiplying Eq. \eqref{eqphi} by $\varepsilon/M_\phi=2B\varepsilon^2/(\gamma c_w)$, inserting the outer expansions \eqref{phiout}-\eqref{ctildeout} and retaining only the $\mathcal{O}(1)$ terms (i.e. setting $\varepsilon=0$), we end up with
 \begin{equation}
 W'\left(\phi_0^{o}\right)=0,
 \end{equation}
 which gives the solutions $\phi_0^{o}= 0$, $\frac{1}{2}$ and $1$. The solution $\phi_0^{o}=\frac{1}{2}$ is discarded as it represents an unstable stationary state which cannot be reached by the limit for $\varepsilon\rightarrow 0$ \cite{van2011phase}. Hence, the two admissible solutions are $\phi_0^{o}=0$ and $\phi_0^{o}=1$.\par
 Let us now focus on Eq. \eqref{eqctilde}. It can be shown (see Appendix \ref{B1}) that for both $\phi^{o}_0=0$ and $\phi^{o}_0=1$, retaining only the $\mathcal{O}(1)$ terms in Eq. \eqref{eqctilde} leads to
\begin{equation}\label{diffusioneqouter}
\frac{\partial c_0^{o}}{\partial t} =D_l\nabla^2 c_0^{o}.
\end{equation}
For $\phi_0^{o}=0$, this is the diffusion equation in the liquid phase of the sharp-interface model \eqref{sharpinterfacemodel}. For $\phi_0^{o}=1$, (\ref{diffusioneqouter}) represents the starting point to prove the convergence to $c_0^{o}=c_s$ in the solid phase, as will be discussed in the following.\par

\subsection{Local curvilinear coordinates} \label{curvcoord}
\begin{figure}
     \centering
     \includegraphics[width=0.8\linewidth]{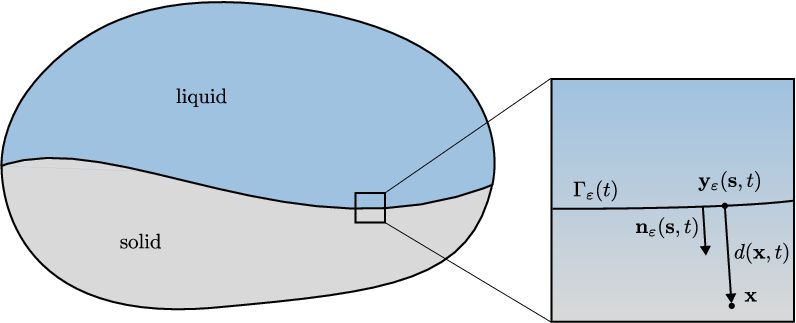}
     \caption{Local curvilinear coordinates.}
     \label{inner}
 \end{figure}

In order to perform the inner expansions (valid inside the interface region) of $\phi$ and $c$, we first need to introduce local curvilinear coordinates. Let us define $\Gamma_\varepsilon(t)$ as the centerline, i.e. the $0.5$-level set, of the diffuse interface in the phase-field model, i.e.
\begin{equation}
\Gamma_\varepsilon(t):=\left \{\mathbf{x}\in\mathbb{R}^{n_d}|\phi(\mathbf{x},t)=0.5\right \}.
\end{equation}
We assume that this level set is sufficiently smooth and smoothly approaches $\Gamma(t)$ for $\varepsilon\rightarrow 0$. We then introduce the arc length vector $\mathbf{s}\in \mathbb{R}^{n_d-1}$ along $\Gamma_\varepsilon(t)$, which induces a curvilinear parametrization of $\Gamma_\varepsilon (t)$, i.e. $\mathbf{s}\to \mathbf{y}_\varepsilon(\mathbf{s},t)\in \Gamma_\varepsilon(t)$, see Fig. \ref{inner}. The unit normal vector to $\Gamma_\varepsilon (t)$ (pointing into the solid region) at $\mathbf{y}_\varepsilon(\mathbf{s},t)$ is denoted as $\mathbf{n}_\varepsilon(\mathbf{s},t)$, see Fig. \ref{inner}. Given a point at position $\mathbf{x}$ close to $\Gamma_\varepsilon(t)$, we denote by $d(\mathbf{x},t)$ its signed distance with respect to $\Gamma_\varepsilon(t)$, such that $d>0$ when $\mathbf{x}$ lies in the solid region and $d<0$ when $\mathbf{x}$ lies in the liquid region. The coordinates $d$ and $\mathbf{s}$ form a set of local \emph{orthogonal} curvilinear coordinates.  As a result, the position $\mathbf{x}$ of any point close to $\Gamma_\varepsilon (t)$ can be expressed as
\begin{equation}\label{xclosetoGamma}
\mathbf{x}=\mathbf{y}_\varepsilon(\mathbf{s},t)+d\,\mathbf{n}_\varepsilon(\mathbf{s},t),
\end{equation}
as represented in Fig. \ref{inner}.
The functions $d(\mathbf{x},t)$ and $\mathbf{s}(\mathbf{x},t)$ have the following properties \cite{chen2006rapidly}
\begin{equation}
\label{signd}
|\nabla d|=1,
\qquad 
\nabla d\cdot \nabla s_i=0,
\qquad 
\frac{\partial d}{\partial t} = -v_n,
\qquad
\nabla^2 d=\kappa,
\end{equation}
where $s_i$ denotes the $i^{\mathrm{th}}$ component of $\mathbf{s}$, $i=1,...,n_d-1$.
Moreover, we assume that $v_n$ and $\kappa$ admit the following expansions
\begin{equation}
\label{expkappa}
    v_n = v_{n0}+\mathcal{O}(\varepsilon), \qquad
    \kappa = \kappa_0+\mathcal{O}(\varepsilon),
\end{equation}
where $v_{n0}$ and $\kappa_0$ are the normal velocity and the sum of the principal curvatures of the sharp interface $\Gamma(t)$, respectively. Considering \eqref{signd}$_4$ and \eqref{expkappa}, we can write
\begin{equation}
    \nabla^2 d = \kappa_0+\mathcal{O}(\varepsilon).
\end{equation}

With the local curvilinear coordinates at hand, we now perform the change of coordinates from $(\mathbf{x},t)$ to $(d,\mathbf{s},t)$. 
For the purpose of the inner expansion to be carried out next, we further substitute $d$ with the dimensionless coordinate
\begin{equation}
\xi(\mathbf{x},t):=\frac{d(\mathbf{x},t)}{\varepsilon}.
\end{equation}
This coordinate change is typically referred to as \emph{stretching transformation} and $\xi$ is denoted as \emph{boundary-layer coordinate} \cite{holmes2012introduction}. 

\subsection{Expansions close to the interface and matching conditions} \label{matching}
With the new coordinates in Section \ref{curvcoord}, we introduce the functions $\phi^{in}$ and $c^{in}$ as follows
\begin{equation}
    \phi(\mathbf{x},t) = \phi^{in}(\xi(\mathbf{x},t),\mathbf{s}(\mathbf{x},t),t), \qquad
    c(\mathbf{x},t) = c^{in}(\xi(\mathbf{x},t),\mathbf{s}(\mathbf{x},t),t).
\end{equation}
The inner expansions of $\phi$ and $c$ expressed in the new local coordinates read
\begin{equation}\label{innerphi}
    \phi^{in}(\xi,\mathbf{s},t)=\phi_0^{in}(\xi,\mathbf{s},t)+\varepsilon\phi_1^{in}(\xi,\mathbf{s},t)+\dots
\end{equation}
\begin{equation}\label{innerctilde}
    c^{in}(\xi,\mathbf{s},t)=c_0^{in}(\xi,\mathbf{s},t)+\varepsilon c_1^{in}(\xi,\mathbf{s},t)+\dots .
\end{equation}
Rewriting \eqref{eqphi} and \eqref{eqctilde} in the new coordinate system $(\xi,\mathbf{s},t)$ (see transformation formulae \eqref{asymptoticinner1}-\eqref{asymptoticinner4} in Appendix \ref{B2}) and substituting \eqref{innerphi} and \eqref{innerctilde}, we obtain
\eqref{eqphi_inner} and \eqref{inner2} in Appendix \ref{B2}.

We now expand the vectors $\mathbf{n}_\varepsilon(\mathbf{s},t)$ and $\mathbf{y}_\varepsilon (\mathbf{s},t)$ as follows
\begin{equation}
\label{neps}
\mathbf{n}_\varepsilon(\mathbf{s},t)=\mathbf{n}_{\varepsilon 0}(\mathbf{s},t)+\varepsilon \mathbf{n}_{\varepsilon 1}(\mathbf{s},t)+\dots
\end{equation}
\begin{equation}
\label{yeps}
\mathbf{y}_\varepsilon(\mathbf{s},t)=\mathbf{y}_{\varepsilon 0}(\mathbf{s},t)+\varepsilon \mathbf{y}_{\varepsilon1}(\mathbf{s},t)+\dots,
\end{equation}
where
\begin{equation}
\mathbf{y}_{\varepsilon 1}(\mathbf{s},t)=y_{\varepsilon1}(\mathbf{s},t)\mathbf{n}_{\varepsilon 0}(\mathbf{s},t), \qquad y_{\varepsilon1}(\mathbf{s},t)=|\mathbf{y}_{\varepsilon1}(\mathbf{s},t)|.
\end{equation}

Considering Eq. \eqref{xclosetoGamma}, we perform the following first-order Taylor expansions of $\phi^{o}(\mathbf{x},t)$ and $c^{o}(\mathbf{x},t)$ close to the interface
\begin{equation}\label{asymptoticinner5}
\phi^{o}(\mathbf{y}_\varepsilon+\varepsilon \xi\mathbf{n}_\varepsilon,t) = \phi^{o}\left(\mathbf{y}_{\varepsilon }^\pm,t\right)+\nabla \phi^{o}(\mathbf{y}_\varepsilon,t)\cdot \varepsilon \xi\mathbf{n}_\varepsilon+\dots
\end{equation}
\begin{equation}\label{asymptoticinner6}
    c^{o}(\mathbf{y}_\varepsilon+\varepsilon \xi \mathbf{n}_\varepsilon,t) = c^{o}\left(\mathbf{y}_{\varepsilon }^\pm,t\right)+\nabla c^{o}(\mathbf{y}_\varepsilon,t)\cdot \varepsilon \xi\mathbf{n}_\varepsilon+\dots,
\end{equation}
which, taking into account Eqs. \eqref{phiout}-\eqref{ctildeout} and \eqref{neps}-\eqref{yeps}, collecting common terms and considering that $\nabla\phi_0^{o}\left(\mathbf{y}_{\varepsilon0}^{\pm},t\right)=\mathbf{0}$, lead to\footnote{$\nabla\phi_0^{o}\left(\mathbf{y}_{\varepsilon0}^{\pm},t\right)=\mathbf{0}$ since $\phi_0^{o}\left(\mathbf{y}_{\varepsilon 0}^+, t\right) =  1$ and $\phi_0^{o}\left(\mathbf{y}_{\varepsilon 0}^-, t\right) =  0$.}
\begin{equation}\label{asymptoticinner7}
\phi^{o}(\mathbf{y}_\varepsilon+\varepsilon \xi\mathbf{n}_\varepsilon,t) = \phi_0^{o}\left(\mathbf{y}_{\varepsilon 0}^\pm,t\right)+\varepsilon \phi_1^{o}\left(\mathbf{y}_{\varepsilon 0}^\pm,t\right)+\dots
\end{equation}
\begin{equation}\label{asymptoticinner8}
c^{o}(\mathbf{y}_\varepsilon+\varepsilon \xi\mathbf{n}_\varepsilon,t) = c_0^{o}\left(\mathbf{y}_{\varepsilon 0}^\pm,t\right)+\varepsilon \left[c_1^{o}\left(\mathbf{y}_{\varepsilon 0}^\pm,t\right)+ (y_{\varepsilon 1}+\xi)\nabla c_0^{o}\left(\mathbf{y}_{\varepsilon 0}^\pm,t\right)\cdot  \mathbf{n}_{\varepsilon 0}\right]+\dots .
\end{equation}

We are now ready to write the matching conditions of inner and outer solutions and of their derivatives with respect to $\xi$, which have to hold separately for  the $\mathcal{O}(1)$ and the $\mathcal{O}(\varepsilon)$ terms, as follows
\begin{itemize}
\par
\item{Solutions, $\mathcal{O}(1)$ terms
\begin{equation}\label{phi0=1matching}
\lim_{\xi \to + \infty} \phi_0^{in}(\xi, \mathbf{s},t) = \phi_0^{o}\left(\mathbf{y}_{\varepsilon 0}^+, t\right) =  1,
\qquad
\lim_{\xi \to - \infty} \phi_0^{in}(\xi, \mathbf{s},t) = \phi_0^{o}\left(\mathbf{y}_{\varepsilon 0}^-, t\right) =  0,
\qquad
\lim_{\xi \to \pm \infty} c_0^{in}(\xi, \mathbf{s},t) = c_0^{o}\left(\mathbf{y}_{\varepsilon 0}^\pm, t\right).
\end{equation}} 
\item{Solutions, $\mathcal{O}(\varepsilon)$ terms \begin{equation}\label{matchingphi1}
\lim_{\xi \to \pm \infty} \phi_1^{in}(\xi, \mathbf{s},t) = \lim_{\xi \to \pm \infty} \phi_1^{o}\left(\mathbf{y}_{\varepsilon 0}^\pm,t\right),
\qquad
\lim_{\xi \to \pm \infty} c_1^{in}(\xi, \mathbf{s},t) = \lim_{\xi \to \pm \infty} \left[c_1^{o}\left(\mathbf{y}_{\varepsilon 0}^\pm,t\right)+ (y_{\varepsilon 1}+\xi)\nabla c_0^{o}\left(\mathbf{y}_{\varepsilon 0}^\pm,t\right)\cdot  \mathbf{n}_{\varepsilon 0}\right].
\end{equation}}
\item{Solution derivatives with respect to $\xi$, $\mathcal{O}(1)$ terms \begin{equation}\label{bcderivativephi}
\lim_{\xi \to \pm \infty} \frac{\partial }{\partial \xi}\phi_0^{in}(\xi, \mathbf{s},t) = 0,
\qquad
\lim_{\xi \to \pm \infty} \frac{\partial }{\partial \xi}c_0^{in}(\xi, \mathbf{s},t) = 0.
\end{equation}}
\item{Solution derivatives with respect to $\xi$, $\mathcal{O}(\varepsilon)$ terms \begin{equation}\label{bcderivativephi1}
\lim_{\xi \to \pm \infty} \frac{\partial }{\partial \xi}\phi_1^{in}(\xi, \mathbf{s},t) = 0,
\qquad
\lim_{\xi \to \pm \infty} \frac{\partial }{\partial \xi}c_1^{in}(\xi, \mathbf{s},t) = \nabla c_0^{o}\left(\mathbf{y}_{\varepsilon 0}^\pm,t\right)\cdot  \mathbf{n}_{\varepsilon 0}.
\end{equation}}
\end{itemize}

\subsection{Inner expansions}
As follows, we take as starting points \eqref{eqphi_inner} and \eqref{inner2} in Appendix \ref{B2}, which were obtained by inserting the inner expansions \eqref{innerphi} and \eqref{innerctilde} in the governing equations \eqref{eqphi} and \eqref{eqctilde}, and solve them with the boundary conditions delivered by the matching conditions in Section \ref{matching}. We consider separately the $\mathcal{O}(1)$ and the $\mathcal{O}(\varepsilon)$ terms of each equations. 

\subsubsection{Eq. \eqref{eqphi_inner}, $\mathcal{O}(1)$  terms}
As discussed in Appendix \ref{B4}, retaining only the $\mathcal{O}(1)$ terms in Eq. \eqref{eqphi_inner} leads to
\begin{equation}\label{phi0insecond}
2\frac{\partial^2 \phi_0^{in}}{\partial \xi^2}=W'\left(\phi_0^{in}\right),
\end{equation}
with boundary conditions given by Eqs. \eqref{phi0=1matching}$_{1,2}$.
Solving Eq. \eqref{phi0insecond} (see Eq. \eqref{integralw'} in Appendix \ref{appx:asymptotic}), we obtain
\begin{equation}\label{2W}
\left(\frac{\partial \phi_0^{in}}{\partial \xi}\right)^2=W\left(\phi_0^{in}\right),
\end{equation}
where we used Eq. \eqref{bcderivativephi} and the condition $W(0)=0$.

\subsubsection{Eq. \eqref{inner2}, $\mathcal{O}(1)$  terms}
For the sake of brevity, we use the notation $h$ to indicate $h\left(\phi_0^{in}\right)$ and similarly for its derivatives $h'$ and $h''$. From Eq. \eqref{inner2}, using Eq. \eqref{h'term} in Appendix \ref{appx:asymptotic}, multiplying the equation by $\varepsilon^2$, retaining only the terms of order $\mathcal{O}(1)$ and deleting the common term $D_l$, we obtain
\begin{equation}
\frac{\partial ^2 c_0^{in}}{\partial \xi^2}=-\frac{\partial}{\partial \xi}\left[\left(c_0^{in}-c_s\right)\frac{h'}{1-h}\frac{\partial \phi_0^{in}}{\partial \xi}\right],
\end{equation}
with boundary conditions given by Eq. \eqref{phi0=1matching}$_{3}$.
Integrating from $-\infty$ to $\xi$ yields
\begin{equation}\label{dcodxi}
\frac{\partial  c_0^{in}}{\partial \xi}=-\left(c_0^{in}-c_s\right)\frac{h'}{1-h}\frac{\partial \phi_0^{in}}{\partial \xi},
\end{equation}
where we made use of Eqs. \eqref{bcderivativephi}$_{1,2}$.
As derived in Appendix \ref{appx:asymptotic}, the solution of Eq. \eqref{dcodxi} is
\begin{equation}\label{c0in}
c_0^{in}(\xi, \mathbf{s},t)=c_0^{o}\left(\mathbf{y}_{\varepsilon 0}^-, t\right) +\left[c_s-c_0^{o}\left(\mathbf{y}_{\varepsilon 0}^-, t\right) \right]h\left(\phi_0^{in}(\xi)\right).
\end{equation}
In Appendix \ref{appx:asymptotic} we also discuss that it must be $c_0^{o}=c_s$ where $\phi_0^{o}=1$ (solid region).

\subsubsection{Eq. \eqref{eqphi_inner}, $\mathcal{O}(\varepsilon)$  terms}
Now we collect the terms of order $\mathcal{O}(\varepsilon)$ in Eq. \eqref{eqphi_inner}. We end up with
\begin{equation}
\mathcal{L}\phi_1^{in} = \mathcal{A}\left(\phi_0^{in}\right),
\end{equation}
where
\begin{equation}
\mathcal{L}\phi_1^{in} = -\frac{B}{c_w}\left[2\frac{\partial^2 \phi_1^{in}}{\partial \xi^2}-W''\left(\phi_0^{in}\right)\phi_1^{in}\right]
\end{equation}
\begin{equation}\label{rhsfredholm}
\mathcal{A}\left(\phi_0^{in}\right) = \frac{2B}{c_w}\left(\frac{v_{n0}}{\gamma}+\kappa_0\right)\frac{\partial \phi_0^{in}}{\partial \xi}-h'\left(\phi_0^{in}\right)\frac{A}{2c_l^{eq2}}\left \{-\left(c_l^{eq}-c_s\right)^2+\left[\frac{c_0^{in}-c_s}{1-h\left(\phi_0^{in}\right)}\right]^2\right \}.
\end{equation}
$\mathcal{L}$ is a Fredholm operator of index zero. Hence, the equation has a solution if and only if $\mathcal{A}\left(\phi_0^{in}\right)$ is orthogonal to the kernel of $\mathcal{L}$, indicated as $\mathrm{ker}(\mathcal{L})$. Since $\frac{\partial \phi_0^{in}}{\partial \xi}\in \mathrm{ker({\mathcal{L}})}$, assuming that the solution exists (solvability condition) implies
\begin{equation}\label{solvability}
\int_{-\infty}^{+\infty} \mathcal{A}\left(\phi_0^{in}\right) \frac{\partial \phi_0^{in}}{\partial \xi} \,\textrm{d}\xi= 0.
\end{equation}
As described in Appendix \ref{appx:asymptotic} (see Eqs. \eqref{rhsfredholm2}-\eqref{rhsfredholmfinal}), Eq. \eqref{solvability} leads to
\begin{equation}
    v_{n0}=-\gamma\kappa_0-\frac{A\gamma}{2B c_l^{eq2}} \left \{\left[c_0^{o}\left(\mathbf{y}_{\varepsilon0}^-,t\right)-c_l^{eq}\right]^2-2\left[c_0^{o}\left(\mathbf{y}_{\varepsilon0}^-,t\right)-c_l^{eq}\right]\left[ c_0^{o}\left(\mathbf{y}_{\varepsilon0}^-,t\right)-c_s\right] \right \},
\end{equation}
which corresponds to the following kinetic law
\begin{equation}
    v_{n0}=-\gamma\kappa_0-\frac{k}{c_l^{eq2}} \left \{ \left[c_0^{o}\left(\mathbf{y}_{\varepsilon0}^-,t\right)-c_l^{eq}\right]^2-2\left[c_0^{o}\left(\mathbf{y}_{\varepsilon0}^-,t\right)-c_l^{eq}\right]\left[ c_0^{o}\left(\mathbf{y}_{\varepsilon0}^-,t\right)-c_s\right] \right \}
\end{equation}
as long as Eq. \eqref{constraint} holds.

\subsubsection{Eq. \eqref{inner2}, $\mathcal{O}(\varepsilon)$  terms}
As shown in Appendix \ref{appx:asymptotic}, retaining the $\mathcal{O}(\varepsilon)$ terms in the inner expansion of Eq. \eqref{eqctilde} and integrating the resulting equation leads to
\begin{equation}\label{kineticb}
    -v_{n0}c_0^{in}(\xi) = D_l\left[\frac{\partial b(\xi)}{\partial \xi}+b(\xi)f\left(\phi_0^{in}(\xi)\right)\frac{\partial \phi_0^{in}(\xi)}{\partial \xi}\right]+ C,
\end{equation}
where $C$ is an integration constant and
\begin{equation}
    b(\xi):=c_1^{in}(\xi)-c_s+\left[c_0^{o}\left(\mathbf{y}_{\varepsilon 0}^-, t\right)-c_s\right]\left[ h'\left(\phi_0^{in}(\xi)\right)\phi_1^{in}-1\right].
\end{equation}
The limit as $\xi\to -\infty$ of Eq. \eqref{kineticb} (see Appendix \ref{appx:asymptotic}) gives
\begin{equation}\label{lim-}
    -v_{n0}c_0^{o}\left(\mathbf{y}_{\varepsilon 0}^-, t\right) = D_l\nabla c_0^{o}\left(\mathbf{y}_{\varepsilon 0}^-,t\right)\cdot  \mathbf{n}_{\varepsilon 0}+ C,
\end{equation}
whereas the limit as $\xi \to +\infty$ of Eq. \eqref{kineticb} gives
\begin{equation}\label{lim+}
    -v_{n0}c_s = C.  
\end{equation}
Subtracting Eq. \eqref{lim+} from Eq. \eqref{lim-} we obtain
\begin{equation}\label{asymptoticmassbalance}
    v_{n0}\left[c_s-c_0^{o}\left(\mathbf{y}_{\varepsilon 0}^-, t\right)\right] =  D_l\nabla c_0^{o}\left(\mathbf{y}_{\varepsilon 0}^-,t\right)\cdot  \mathbf{n}_{\varepsilon 0},
\end{equation}
that is the mass balance condition of the sharp-interface model \eqref{sharpinterfacemodel}.\par

\subsection{Summary}
We define the domains $V_s (t) = \left \{\mathbf{x} \in V | \phi_0^{o}(\mathbf{x},t) = 1 \right \}$ and $V_l (t) = \left \{\mathbf{x} \in V| \phi_0^{o}(\mathbf{x},t) = 0 \right \}$ with common boundary $\Gamma(t)$. The results of the asymptotic analysis, along with the constraint given by Eq. \eqref{constraint}, can be summarized as
\begin{equation}
\begin{cases}
    \frac{\partial c_0^{o}}{\partial t} = D_l \nabla^2 c_0^{o}\quad \mathrm{in}\quad V_l(t)\\
   c_0^{o} = c_s\quad \mathrm{in} \quad V_s(t)\\
   v_{n0}\left[c_s-c_0^{o}\left(\mathbf{y}_{\varepsilon 0}^-, t\right)\right] =  D_l\nabla c_0^{o}\left(\mathbf{y}_{\varepsilon 0}^-,t\right)\cdot  \mathbf{n}_{\varepsilon 0}\quad \mathrm{on}\quad \Gamma(t)\\
  v_{n0} = -\gamma \kappa_0 -r\left(c_0^{o}\left(\mathbf{y}_{\varepsilon 0}^-, t\right)\right)\quad \mathrm{on}\quad \Gamma(t),
  \end{cases}
  \end{equation}
  with the following expression of the reaction rate
\begin{equation}
    r(c_0^{o}(\mathbf{y}_{\varepsilon 0}^-, t))=\frac{k}{c_l^{eq2}}\left \{ \left[c_0^{o}\left(\mathbf{y}_{\varepsilon 0}^-, t\right)-c_l^{eq} \right]^2 -2\left[c_0^{o}\left(\mathbf{y}_{\varepsilon 0}^-, t\right)-c_l^{eq} \right]\left[c_0^{o}\left(\mathbf{y}_{\varepsilon 0}^-, t\right)-c_s \right] \right \}.
\end{equation}
By setting $c|_{-} := c_0^{o}\left(\mathbf{y}_{\varepsilon 0}^-, t\right)$ and assigning appropriate boundary and initial conditions, the convergence of the proposed phase-field model to \eqref{sharpinterfacemodel} has been proved.

\rmk{In Appendix \ref{appx:asymptotic} it is shown that Eq. \eqref{lim+} and, as a consequence, Eq. \eqref{asymptoticmassbalance} can be retrieved using the double-well function in \eqref{W}. Conversely, the standard choice $W(\phi)=\phi^2(1-\phi)^2$ does not allow to retrieve the mass balance at the interface. This result is also in agreement with what is discussed in \cite{van2011phase}, considering that our evolution equation for the concentration is very similar to that in \cite{van2011phase}.
We also note that the choice $W(\phi)=8\phi^2(1-\phi)^2$ was made in \cite{bringedal2020phase}. However, the governing equation of the concentration field is different in that case, without the enforcement of the condition $c=c_s$ in the solid domain.}

\section{Numerical examples}
\label{numerical examples}
In this section, we present several numerical results obtained with the proposed phase-field model for solute precipitation/dissolution, both in $2$D and in $3$D. The phase-field model is discretized with the finite element method and the numerical implementation is carried out using the deal.II library \cite{bangerth2007deal}, suitable for dimension-independent programming and parallelization. Time discretization is performed with a backward Euler scheme with constant time step $\Delta t=5\times 10^{-3}$ over the time interval $[0, T]$. As for space discretization, in $2$D simulations we adopt a finite element mesh of $160^2=25600$ bilinear quadrilateral elements, whereas a mesh of $160^3=4096000$ trilinear hexahedral elements is used in the $3$D case. In the $2$D case we adopt full Gauss quadrature with $5$ points per parametric direction (required by the high-degree double-well potential in Eq. \eqref{W}), whereas in the $3$D case we limit ourselves to $4$ points per parametric direction, which upon a preliminary study are found to lead to a quadrature error of the same order of the discretization error. 
In the following, the units of the involved quantities are not specified though they have to be assumed in a consistent manner, e.g. $\left[\mathrm{m}\right]$ for lengths, $\left[\mathrm{s}\right]$ for time, $\left[\mathrm{mol/m^3}\right]$ for molar concentrations, $\left[\mathrm{m^2/s}\right]$ for diffusivities, $\left [\mathrm{m/s}\right]$ for $k$, $\left [\mathrm{J/m^3}\right]$ for $A$ and $\left [\mathrm{J/m^2}\right]$ for $B$.\par

\subsection{Precipitation and dissolution of a $2\mathrm{D}$ precipitate}
We start by simulating the precipitation and the dissolution of a $2$D precipitate of initially circular or irregular shape. The domain is the square $V = \left[0, L\right]^2$ and the data used in the simulations are listed in Tab. \ref{Dataex1}, where $c_{0}$ denotes the initial ion concentration in the liquid phase. A concentration equal to $c_s$ is assigned to the initial solid region. 
\begin{table}[h!]
\centering
\begin{tabular}{lcccccccccccc}
\toprule
\textbf{Parameter} & $L$ & $\gamma$ & $\varepsilon$ &  $k$ & $c_l^{eq}$ & $c_s$ & $D_l$ &  $c_{0}$ & $A$  & $B$  & $\delta$  \\
\midrule
\textbf{Value}     & 1            & 0.1     &  $6.25\times10^{-3} $        & 1                  & 0.5           & 1     & 1       & 0.7         & 1         & $5\times10^{-2}  $                     &  $10^{-2}   $                   \\
\bottomrule
\end{tabular}
\caption{Data used in the $2$D simulations.}
\label{Dataex1}
\end{table}
\subsubsection{Circular precipitate}
\begin{figure}[htbp]
    \centering
    \subfloat[$\phi(\mathbf{x},t=0)$]{\includegraphics[width=0.2\textwidth]{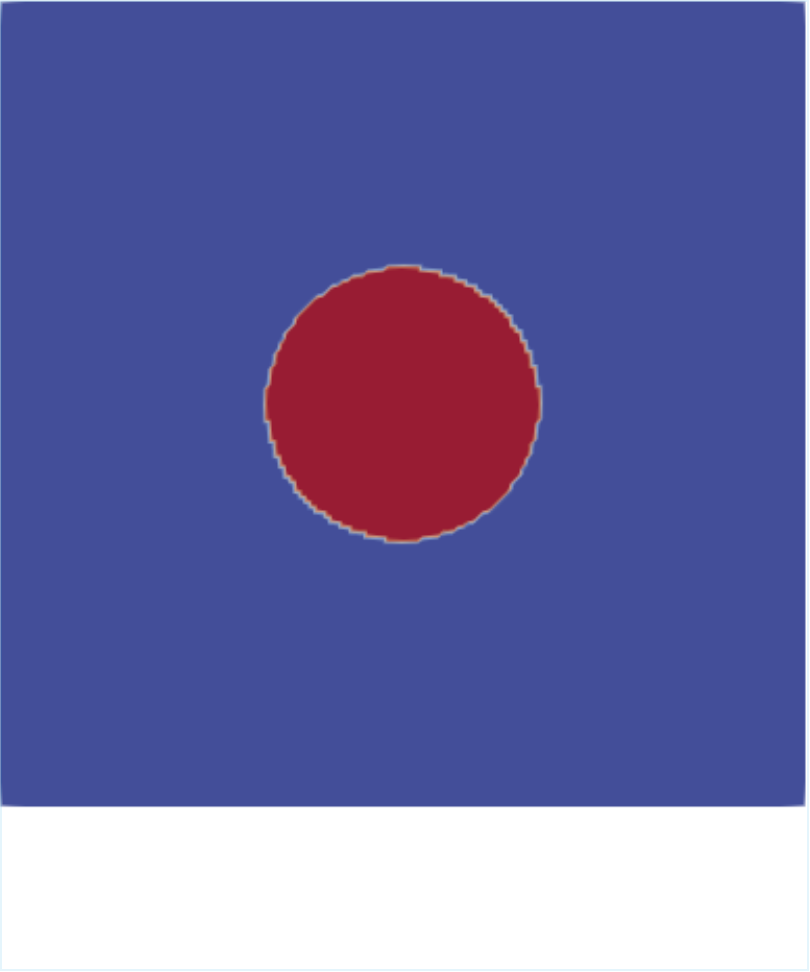}} \quad
    \subfloat[$\phi(\mathbf{x},t=2.5)$]{\includegraphics[width=0.2\textwidth]{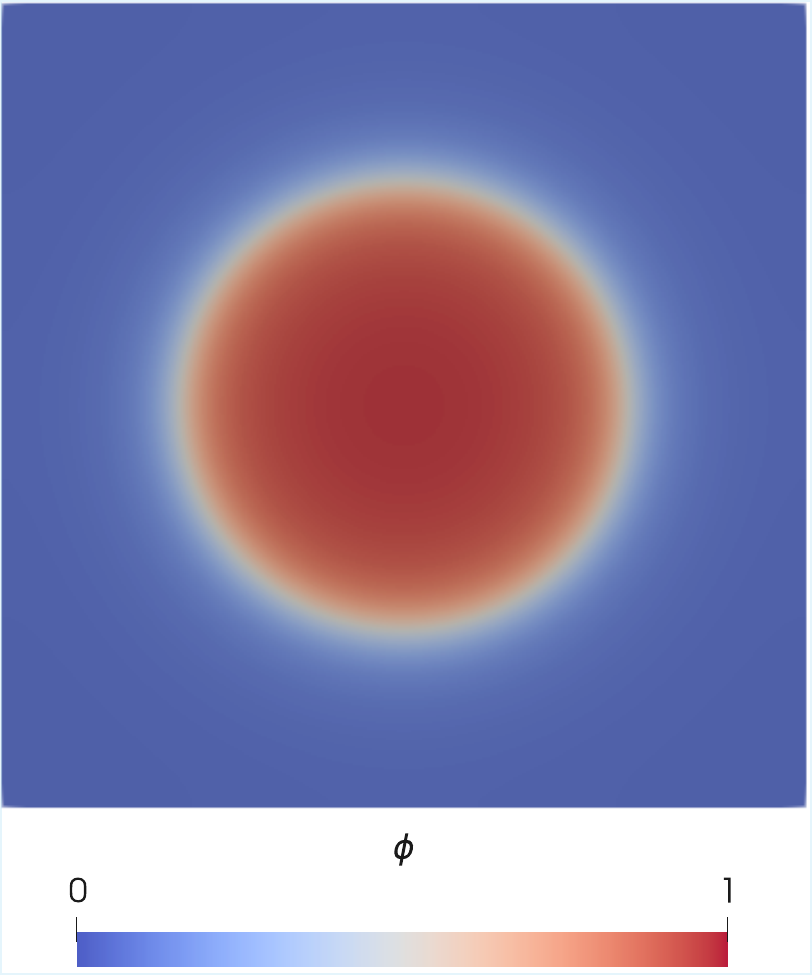}} \quad
    \subfloat[$\phi(\mathbf{x},t=7.5)$]{\includegraphics[width=0.2\textwidth]{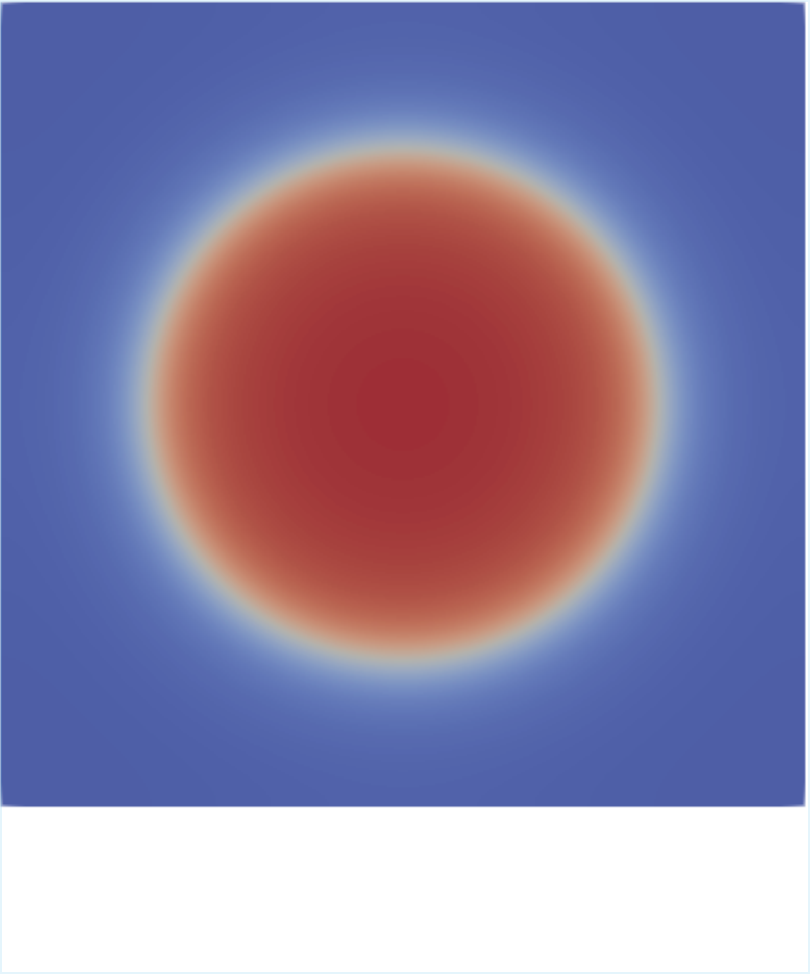}}\\
    \subfloat[$c(\mathbf{x},t=0)$]{\includegraphics[width=0.2\textwidth]{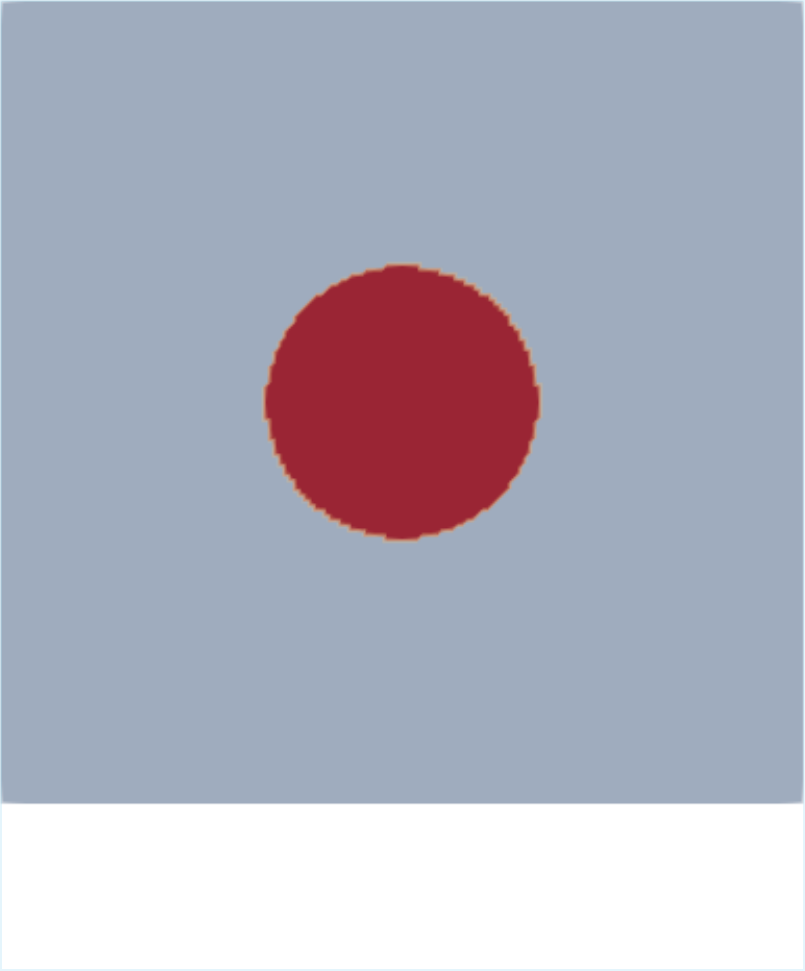}} \quad
    \subfloat[$c(\mathbf{x},t=2.5)$]{\includegraphics[width=0.2\textwidth]{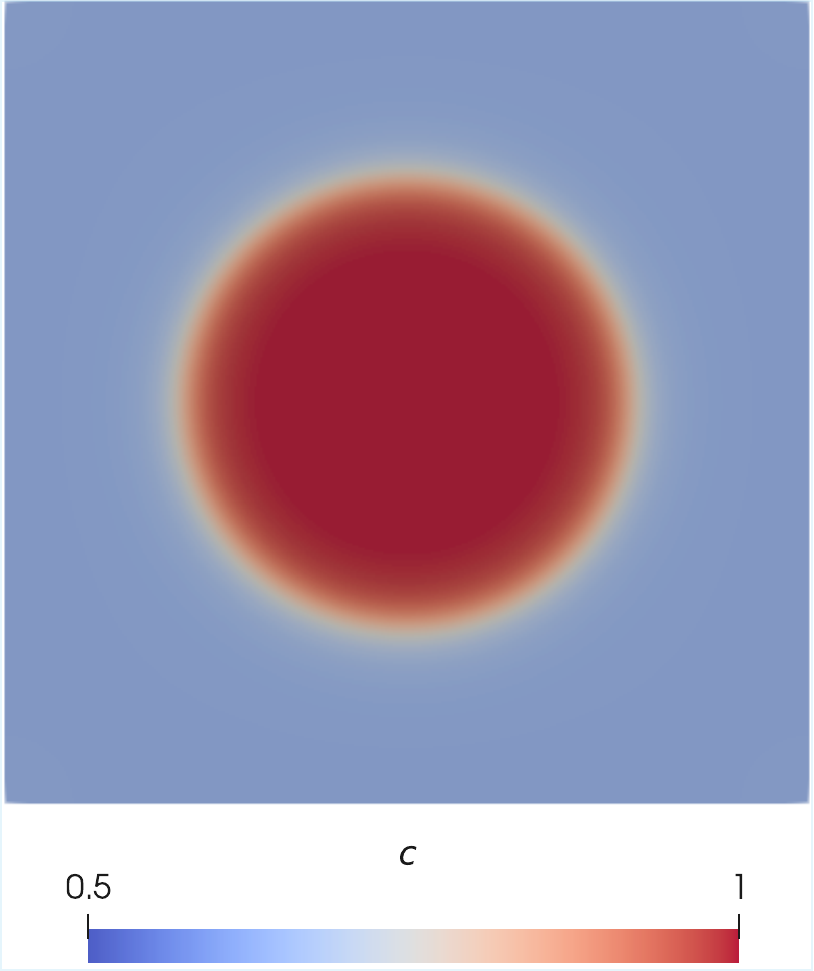}} \quad
    \subfloat[$c(\mathbf{x},t=7.5)$]{\includegraphics[width=0.2\textwidth]{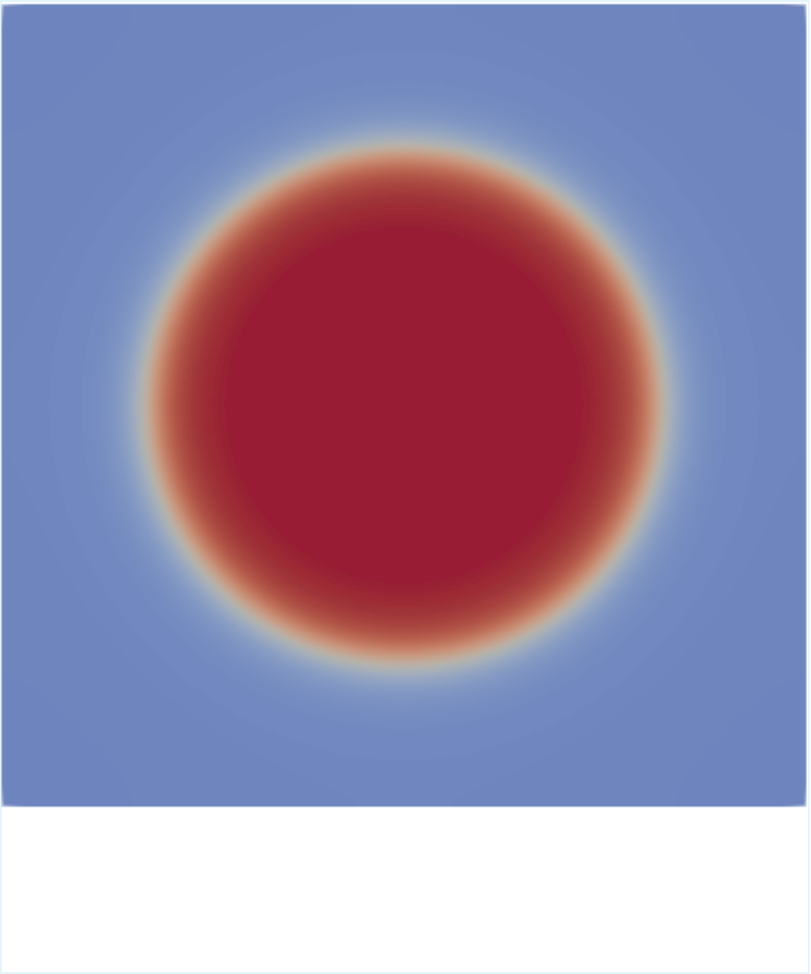}}
    \caption{Initial circular precipitate of radius $R_0=0.17$ (larger than the critical radius): evolution of phase indicator $\phi$ and total concentration $c$.}
    \label{phi_prep_circ_2D}
\end{figure}

\begin{figure}[htbp]
    \centering
    \subfloat[$\phi(\mathbf{x},t=0)$]{\includegraphics[width=0.2\textwidth]{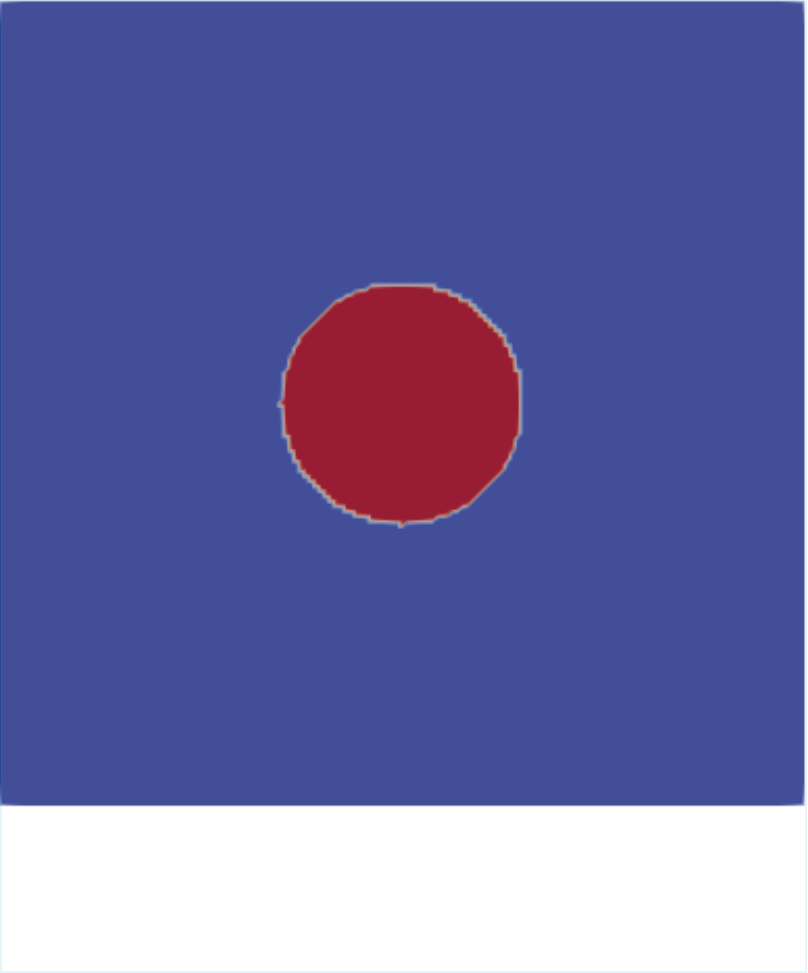}} \quad
    \subfloat[$\phi(\mathbf{x},t=0.15)$]{\includegraphics[width=0.2\textwidth]{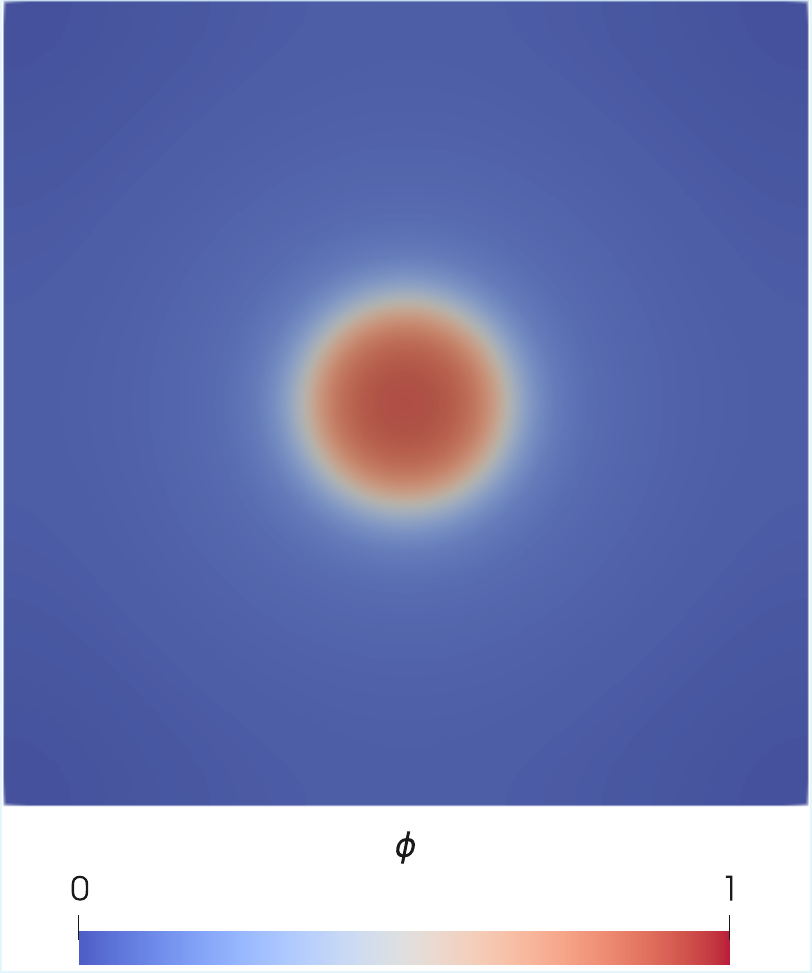}} \quad
    \subfloat[$\phi(\mathbf{x},t=0.3)$]{\includegraphics[width=0.2\textwidth]{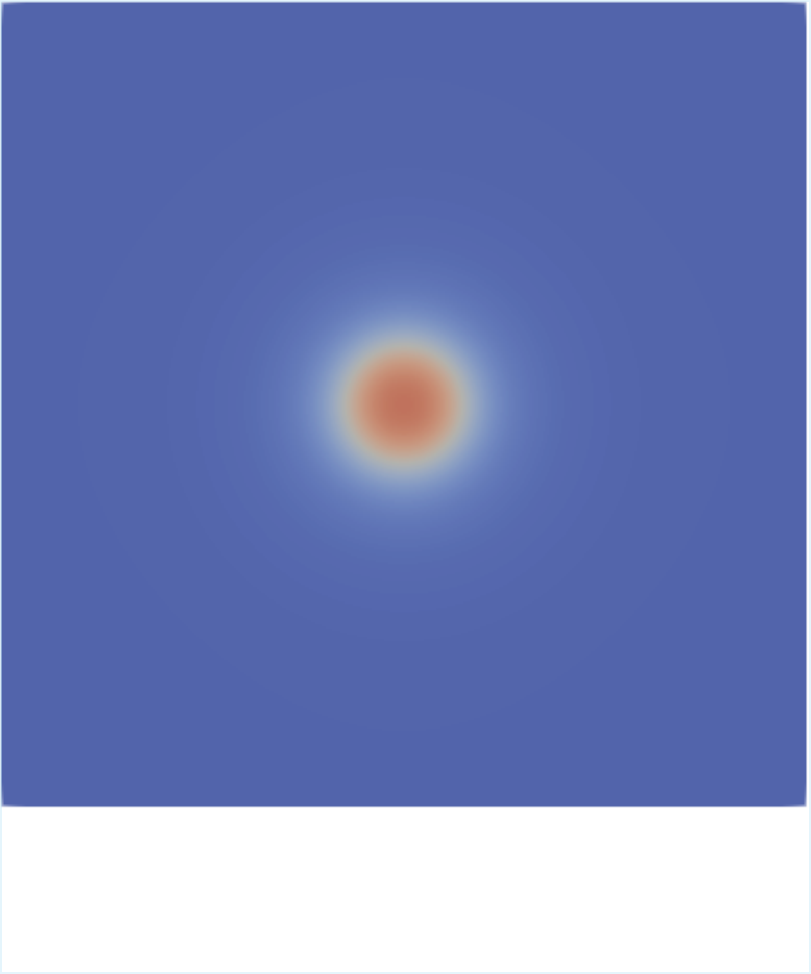}} \\
    \subfloat[$c(\mathbf{x},t=0)$]{\includegraphics[width=0.2\textwidth]{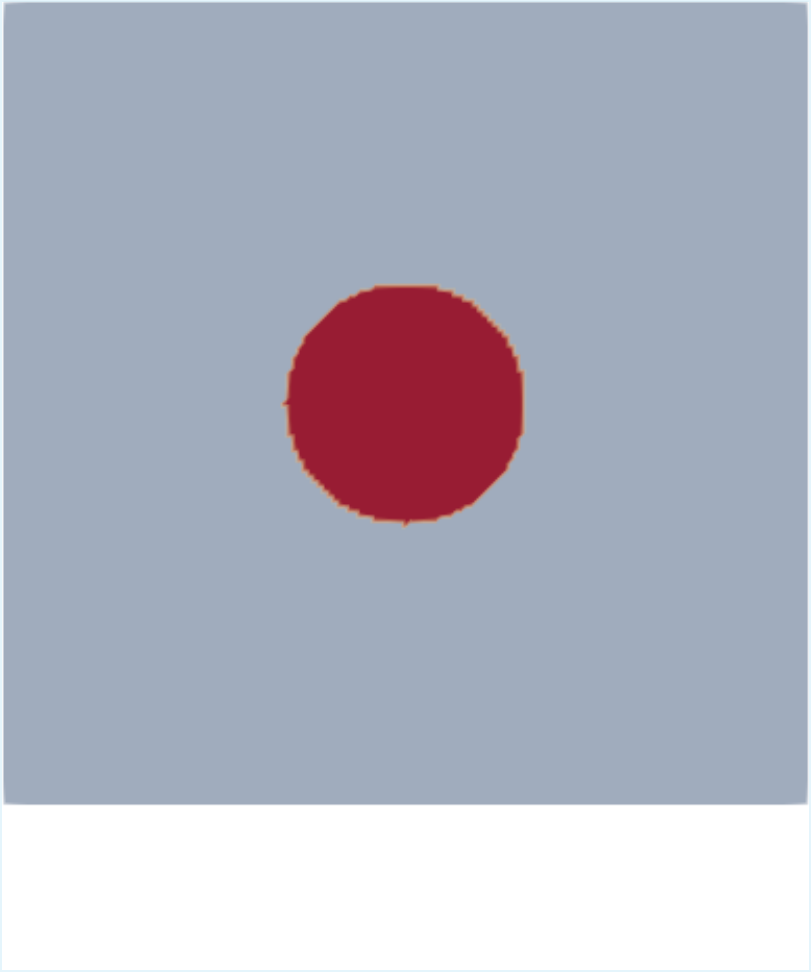}} \quad
    \subfloat[$c(\mathbf{x},t=0.15)$]{\includegraphics[width=0.2\textwidth]{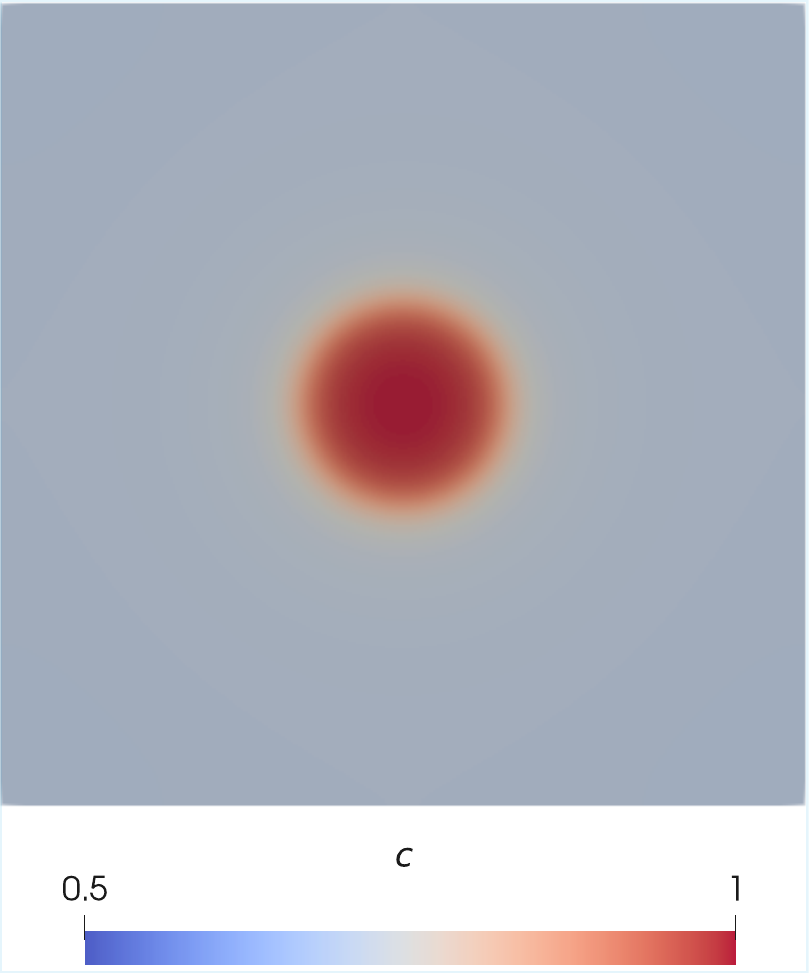}} \quad
    \subfloat[$c(\mathbf{x},t=0.3)$]{\includegraphics[width=0.2\textwidth]{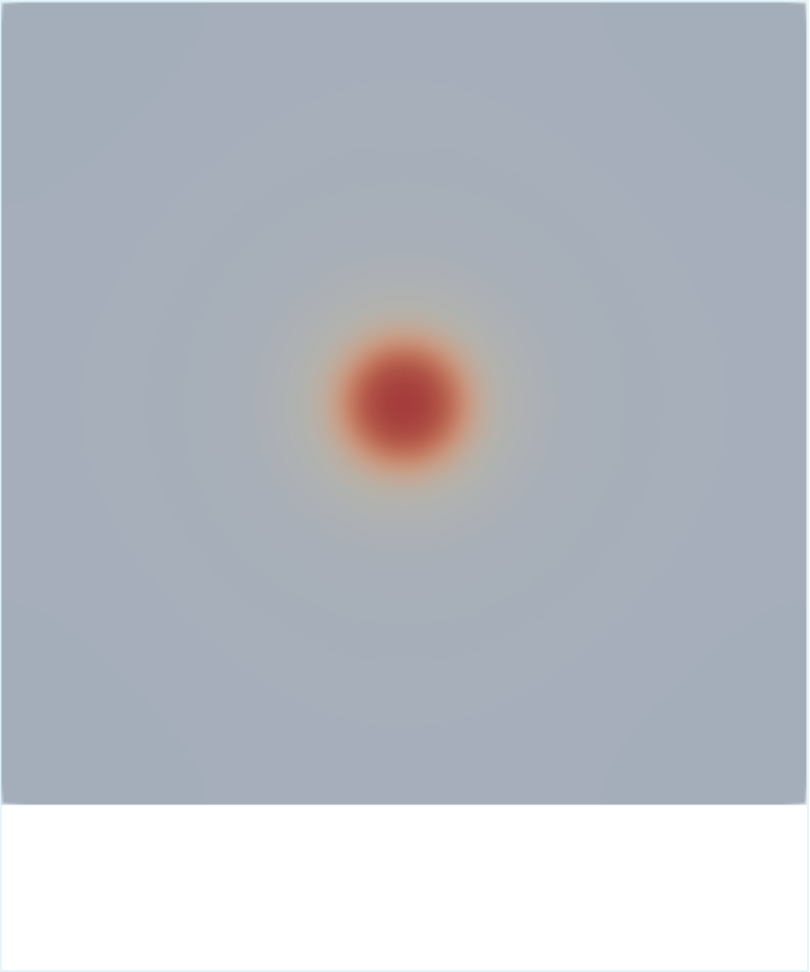}} 
    \caption{Initial circular precipitate of radius $R_0=0.15$ (smaller than the critical radius): evolution of phase indicator $\phi$ and total concentration $c$.}
    \label{phi_diss_circ_2D}
\end{figure}

First, we consider the case of an initial circular precipitate centered at $(L/2, L/2)$ surrounded by a liquid solution. Using the data from Tab. \ref{Dataex1} and Eq. \eqref{criticalradiusformula2D}, the critical radius of an initial circular precipitate from the sharp-interface model is $R_{0c}^{2D}=0.156$. In the finite element simulations we consider two different initial radii: $R_0=0.17$ and $R_0=0.15$, respectively larger and smaller than the critical radius, to verify whether the numerical simulations give, as expected from theory, precipitation (growth of the initial precipitate) in the first case and dissolution in the second case. The final times for the two simulations are chosen as $T=7.5$ and $T=0.45$, respectively, which are sufficient to reach steady-state conditions. The contour plots of the phase indicator and of the total ion concentration are reported for three different time instants in Fig. \ref{phi_prep_circ_2D} for the case with $R_0=0.17$ and in Fig. \ref{phi_diss_circ_2D} for the case with $R_0=0.15$. In both cases, the evolution obtained numerically matches the expectations from theory. In the precipitation case, the final size of the precipitate is dictated by mass balance. The steady state in the dissolution case, not reported, corresponds to a fully liquid domain, as a result of the complete dissolution of the circular precipitate, with a constant concentration again dictated by mass balance.

\subsubsection{Irregularly shaped precipitate}
\begin{figure}[htbp]
    \centering
    \subfloat[$\phi(\mathbf{x},t=0)$]{\includegraphics[width=0.2\textwidth]{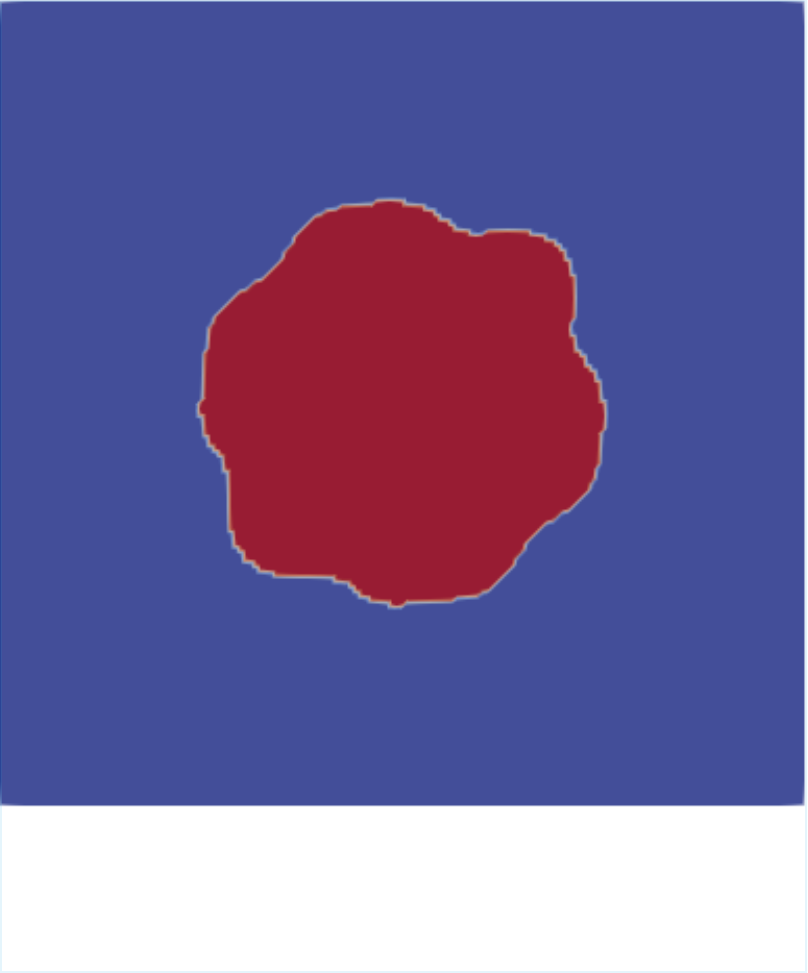}} \quad
    \subfloat[$\phi(\mathbf{x},t=0.025)$]{\includegraphics[width=0.2\textwidth]{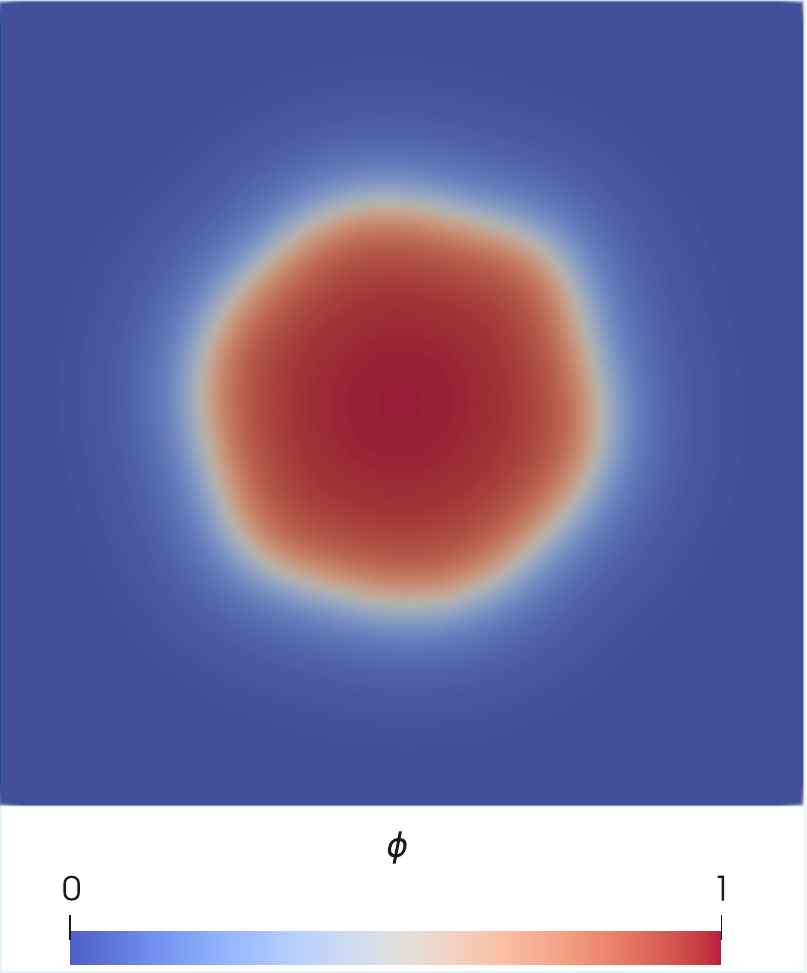}} \quad
    \subfloat[$\phi(\mathbf{x},t=7.5)$]{\includegraphics[width=0.2\textwidth]{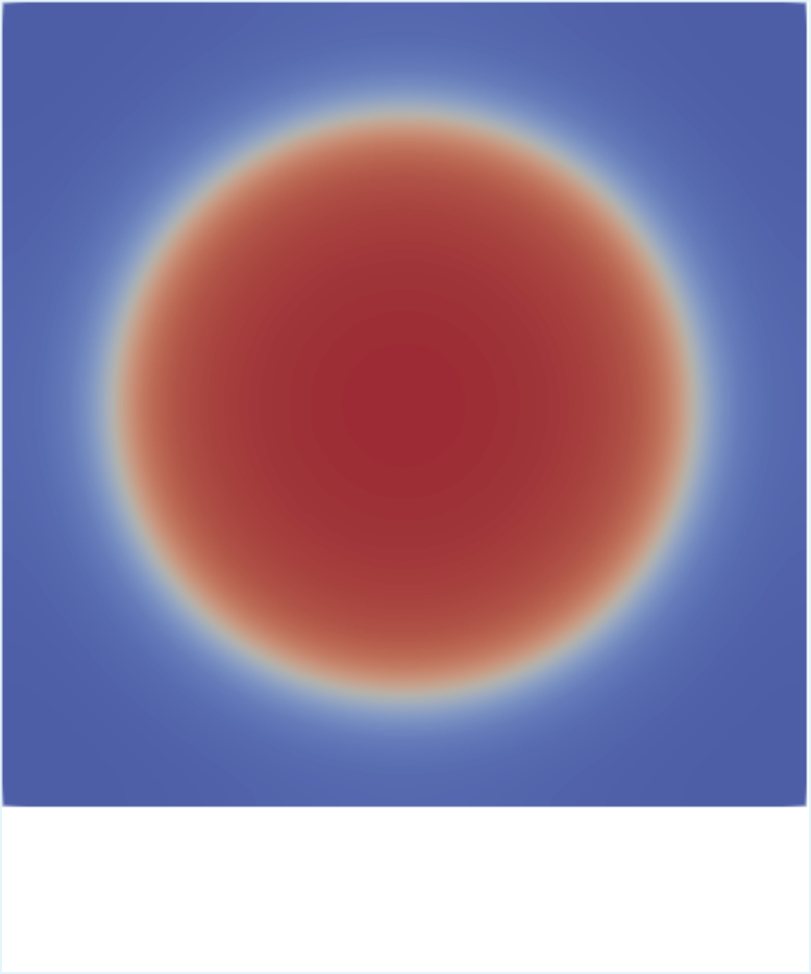}}\\
    \subfloat[$c(\mathbf{x},t=0)$]{\includegraphics[width=0.2\textwidth]{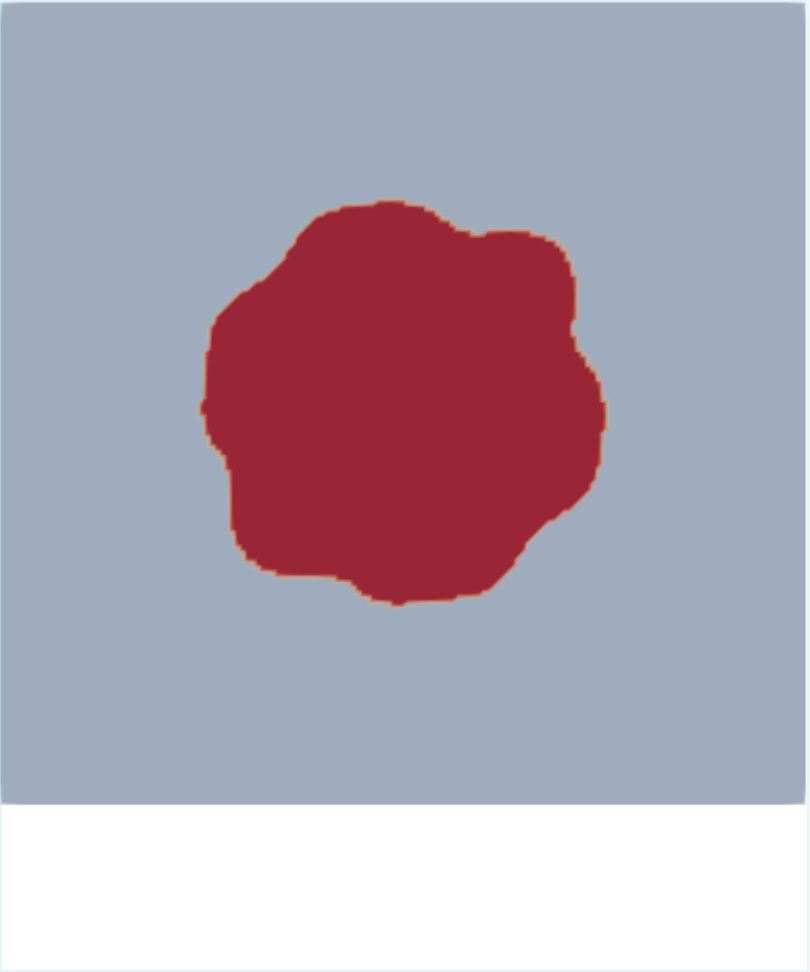}} \quad
    \subfloat[$c(\mathbf{x},t=0.025)$]{\includegraphics[width=0.2\textwidth]{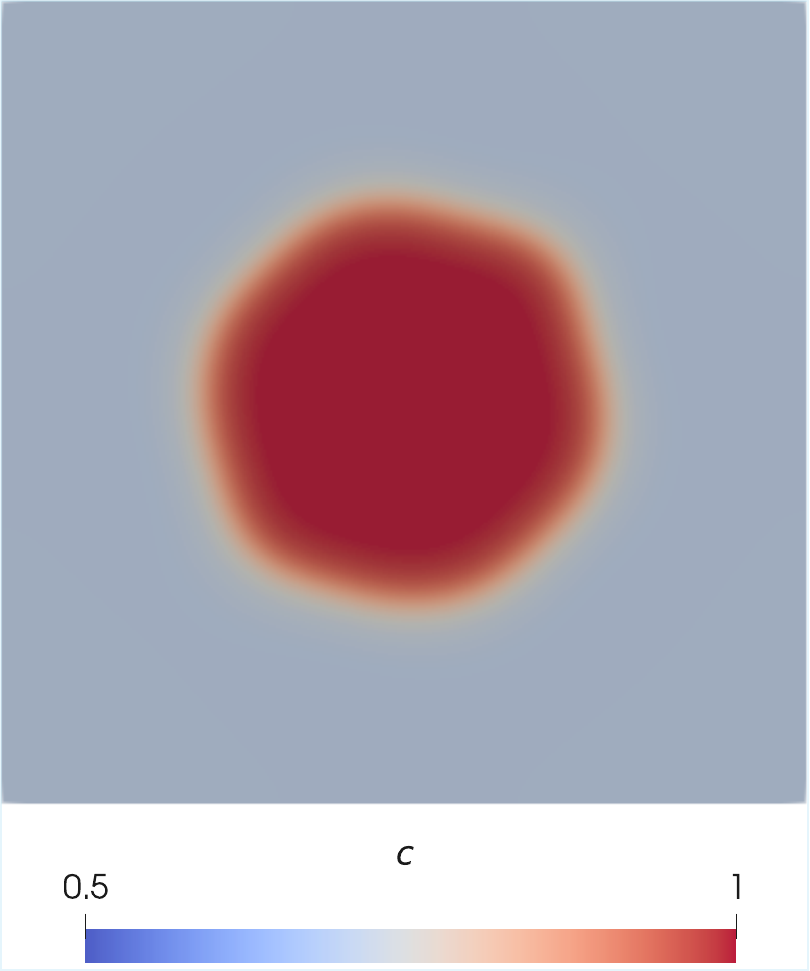}} \quad
    \subfloat[$c(\mathbf{x},t=7.5)$]{\includegraphics[width=0.2\textwidth]{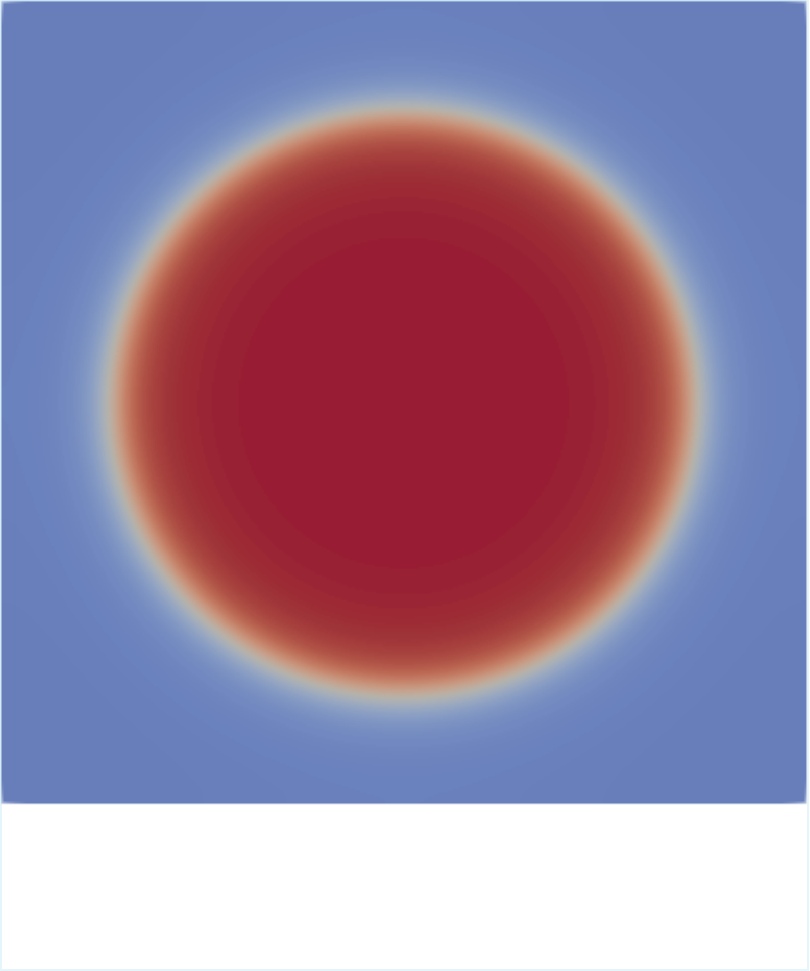}}
    \caption{Initial irregular precipitate: evolution of phase indicator $\phi$ and total concentration $c$.}
    \label{phi_prep_irr_2D}
\end{figure}

\begin{figure}[htbp]
    \centering
    \subfloat[$\phi(\mathbf{x},t=0)$]{\includegraphics[width=0.2\textwidth]{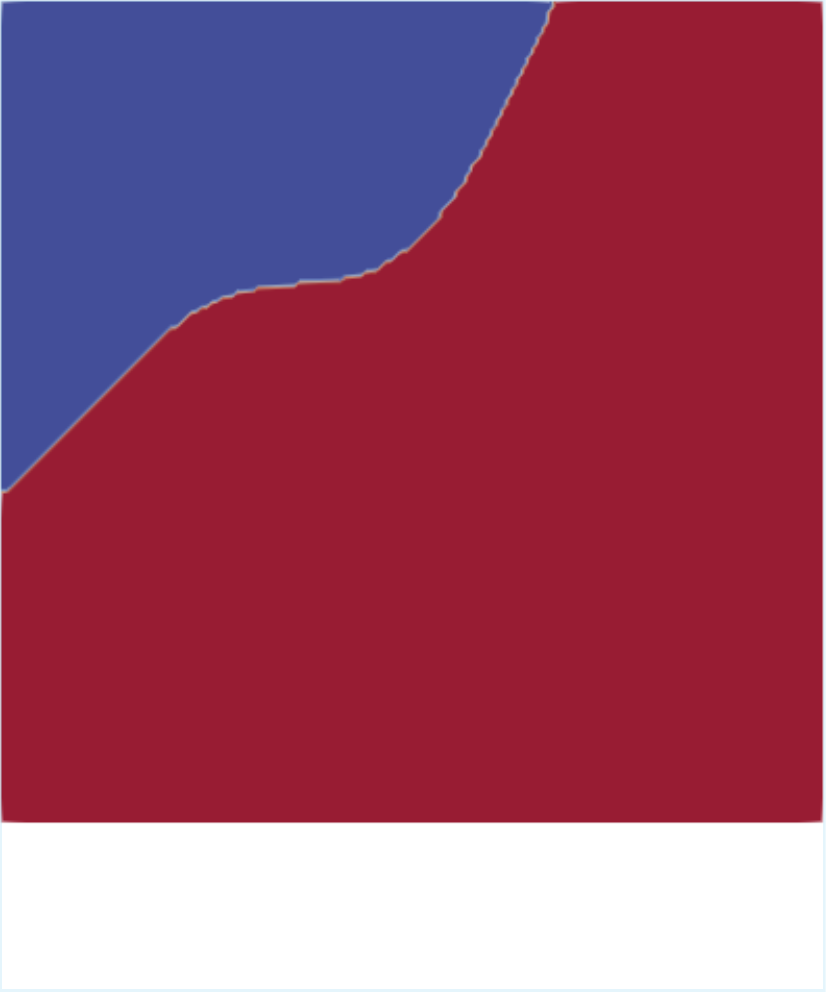}} \quad
    \subfloat[$\phi(\mathbf{x},t=0.1)$]{\includegraphics[width=0.2\textwidth]{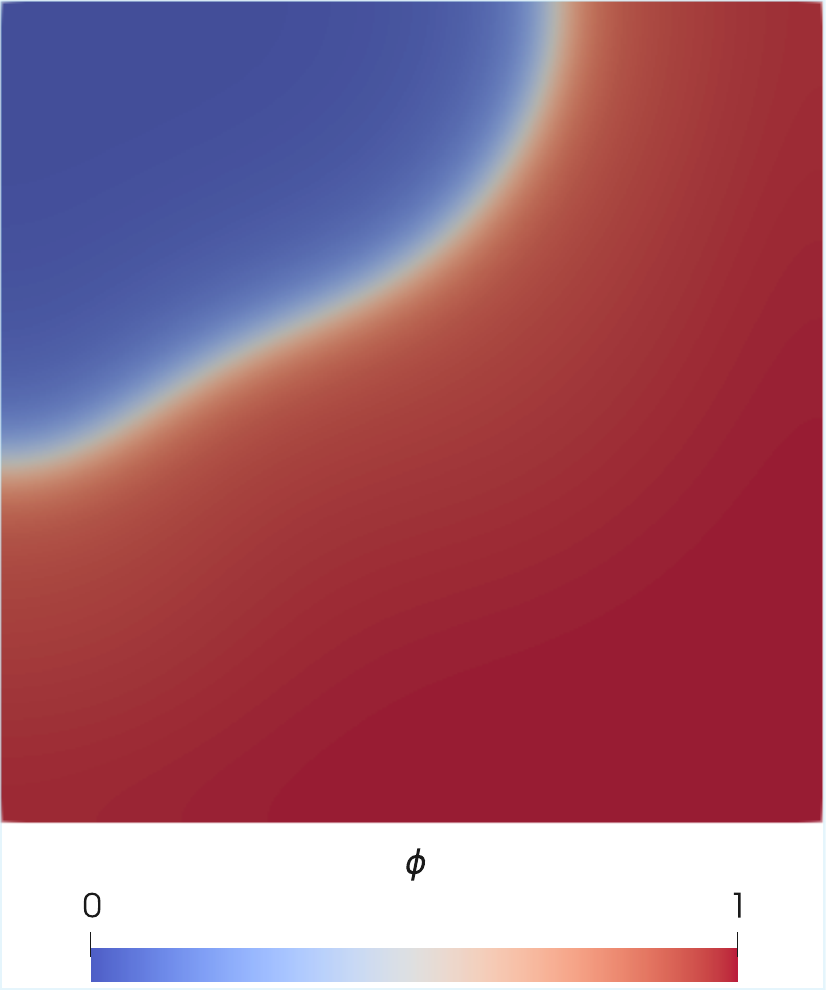}} \quad
    \subfloat[$\phi(\mathbf{x},t=1.5)$]{\includegraphics[width=0.2\textwidth]{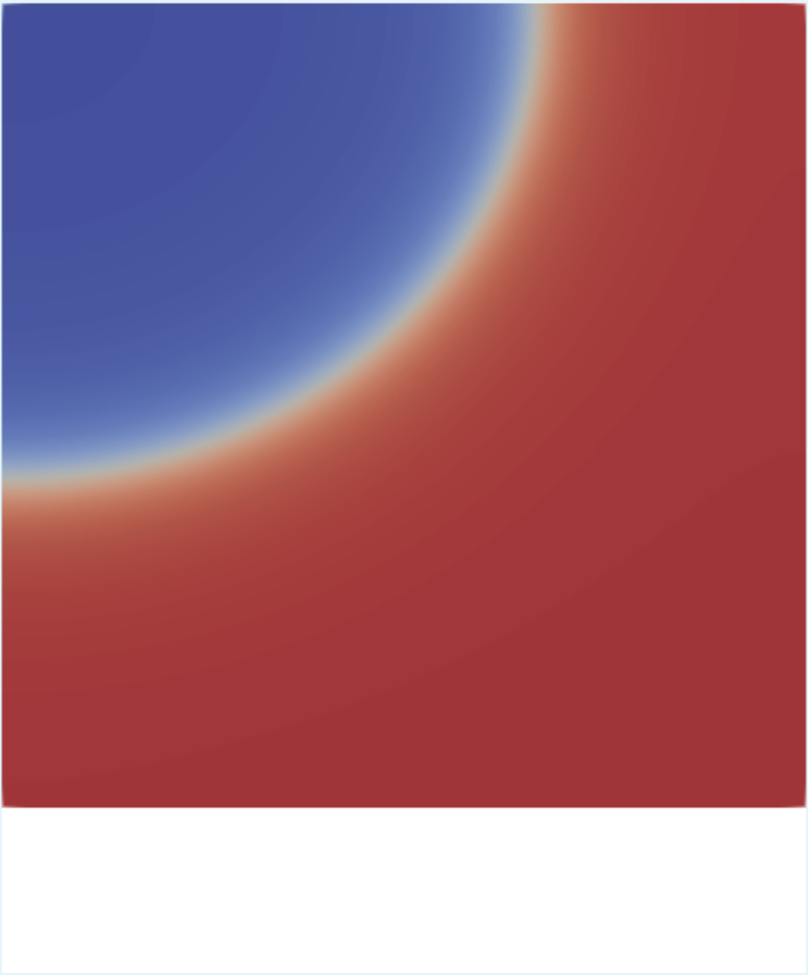}}\\
    \subfloat[$c(\mathbf{x},t=0)$]{\includegraphics[width=0.2\textwidth]{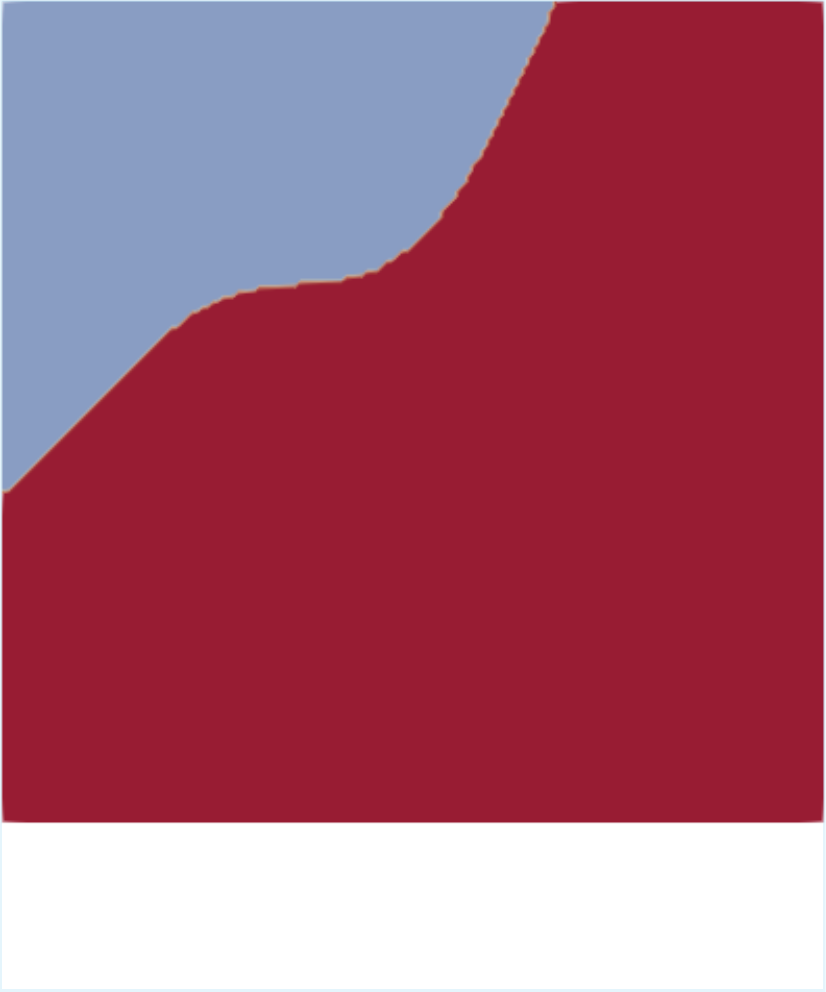}} \quad
    \subfloat[$c(\mathbf{x},t=0.1)$]{\includegraphics[width=0.2\textwidth]{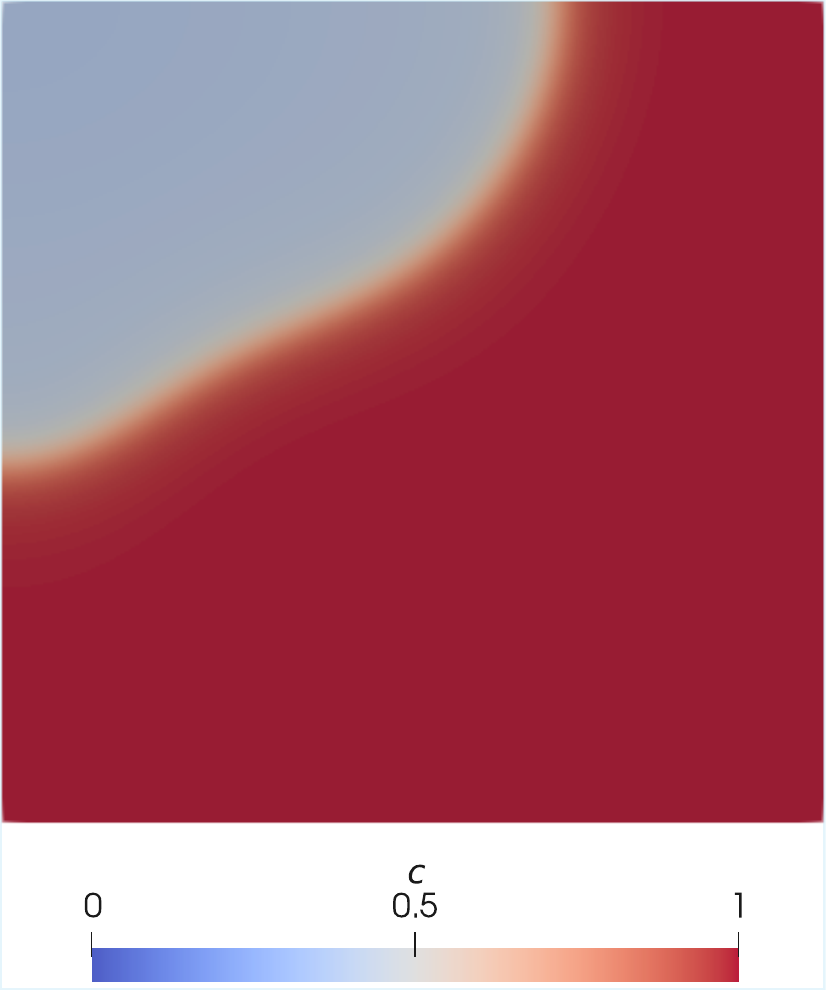}} \quad
    \subfloat[$c(\mathbf{x},t=1.5)$]{\includegraphics[width=0.2\textwidth]{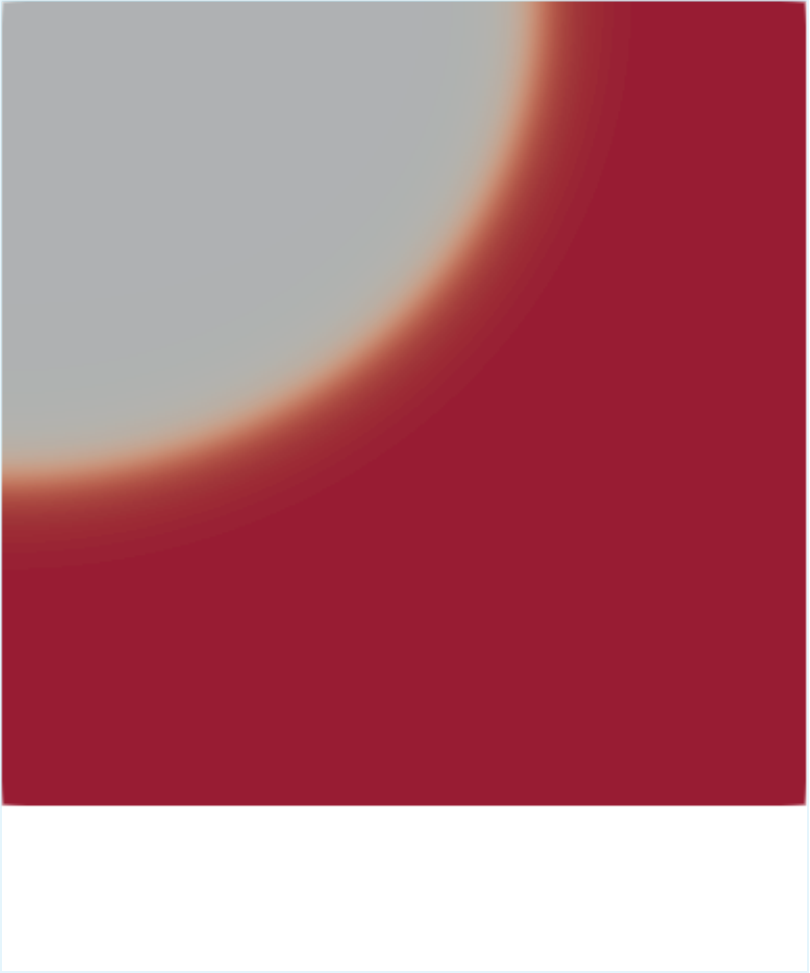}}
    \caption{Initial solid-liquid interface intersecting the domain boundary: evolution of phase indicator $\phi$ and total concentration $c$.}
    \label{phi_diss_irr_2D}
\end{figure}

Next, we consider the case of an initial solid phase that has an irregular shape and once again we provide an example of precipitation and one of dissolution. In the case of precipitation, the initial liquid-solid boundary is defined by the equation
    \begin{equation}
\left(x-\frac{L}{2}\right)^2+\left(y-\frac{L}{2}\right)^2 = R_0^2+\frac{1}{75}\sin(25x)\sin(25y),
\end{equation}
with $R_0=0.25$ (Fig. \ref{phi_prep_irr_2D}).
In the case of dissolution, the initial liquid-solid boundary is defined by
\begin{equation}
y=\frac{2}{5}\left(1+2x\right)+\frac{1}{5}x\sin(10x),
\end{equation}
as shown in Fig. \ref{phi_diss_irr_2D}; unlike in the previous examples, it intersects the boundary of the domain.
The final times (sufficient to reach steady-state conditions) for the precipitation and dissolution simulations are $T=7.5$ and $T=1.5$, respectively. For the dissolution case, unlike in the previous examples, we consider an initial ion concentration in the liquid $c_{0}=0.3$ and a reduced time step $\Delta t=10^{-3}$, whereas all the other data are the same as in Tab. \ref{Dataex1}. The phase indicator and total ion concentration evolutions in the precipitation and dissolution cases are reported in Fig. \ref{phi_prep_irr_2D} and Fig. \ref{phi_diss_irr_2D}, respectively. In the case of precipitation, the solid phase tends to the circular shape (due to the assumption of constant interface diffusivity $\gamma$ and constant interfacial mobility $M_\phi$) and enlarges until the final steady-state conditions are reached. In the case of dissolution, the irregular solid domain again tends to a circular shape and recedes until reaching the steady state.

\subsection{Precipitation and dissolution in $3\mathrm{D}$}
Finally, we report some $3$D results where we simulate the growth of a single spherical precipitate and the dissolution of two spherical precipitates. The domain is given by the cube $V = \left[0, L\right]^3$ and the data used in the simulations are the same as for the $2$D case, listed in Tab. \ref{Dataex1}, except for the initial ion concentration in the liquid for the precipitation example, that is now $c_{0}=0.9$. Using Eq. \eqref{criticalradiusformula3D}, the critical radius of an initial spherical precipitate is $R_{0c}^{3D}=0.208$ in the precipitation case and $R_{0c}^{3D}=0.312$ in the dissolution case. The final simulation times for the precipitation and dissolution cases are $T=0.75$ and $T=0.45$, respectively (the steady state is not reached in the case of precipitation).
\subsubsection{Growth of a single spherical precipitate}
\begin{figure}[htbp]
    \centering
    \subfloat[$\phi=0.5$ at $t=0$]{\includegraphics[width=0.2\textwidth]{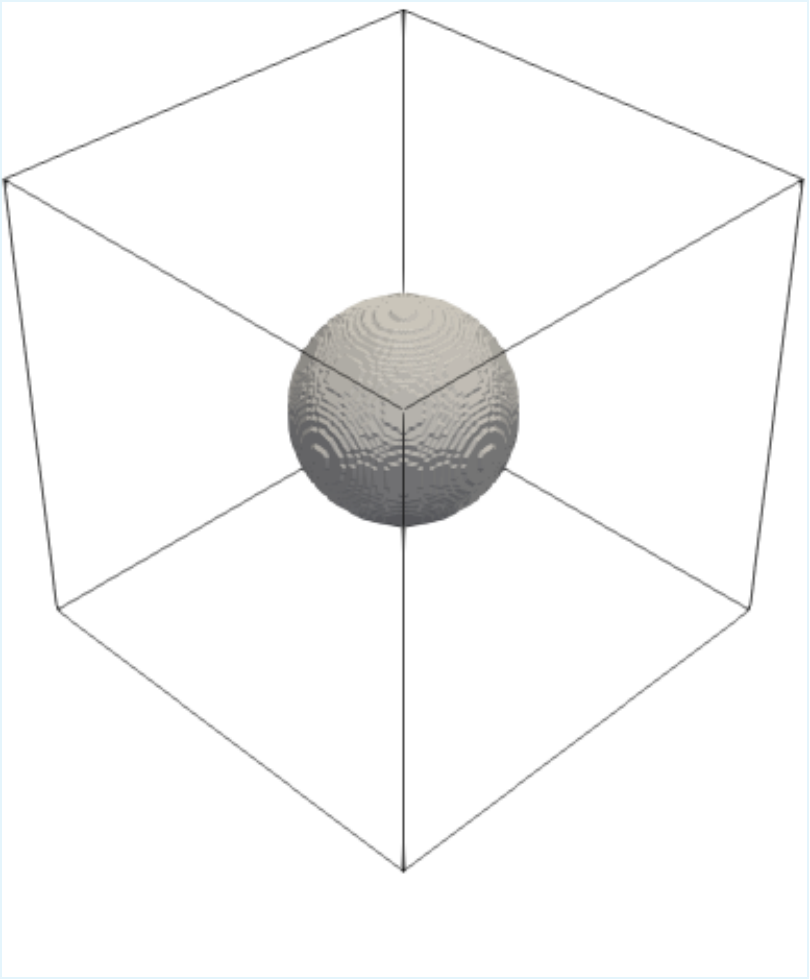}} \quad
    \subfloat[$\phi=0.5$ at $t=0.5$]{\includegraphics[width=0.2\textwidth]{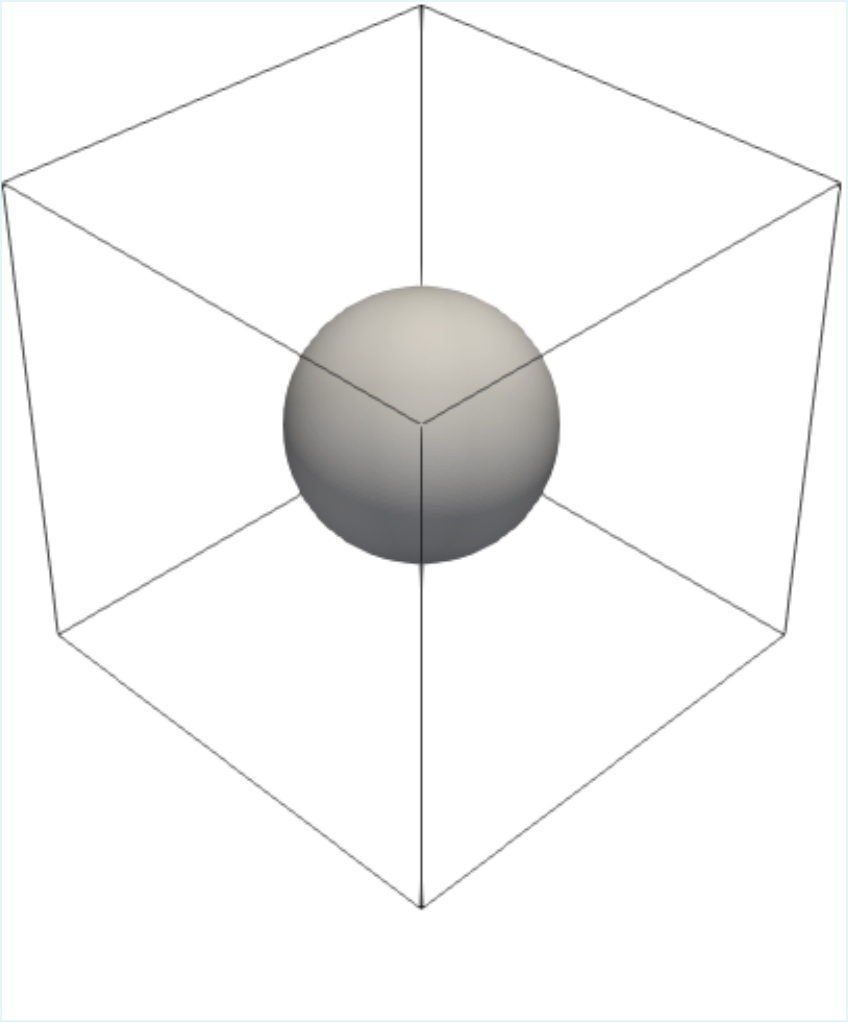}}\quad
    \subfloat[$\phi=0.5$ at $t=0.75$]{\includegraphics[width=0.2\textwidth]{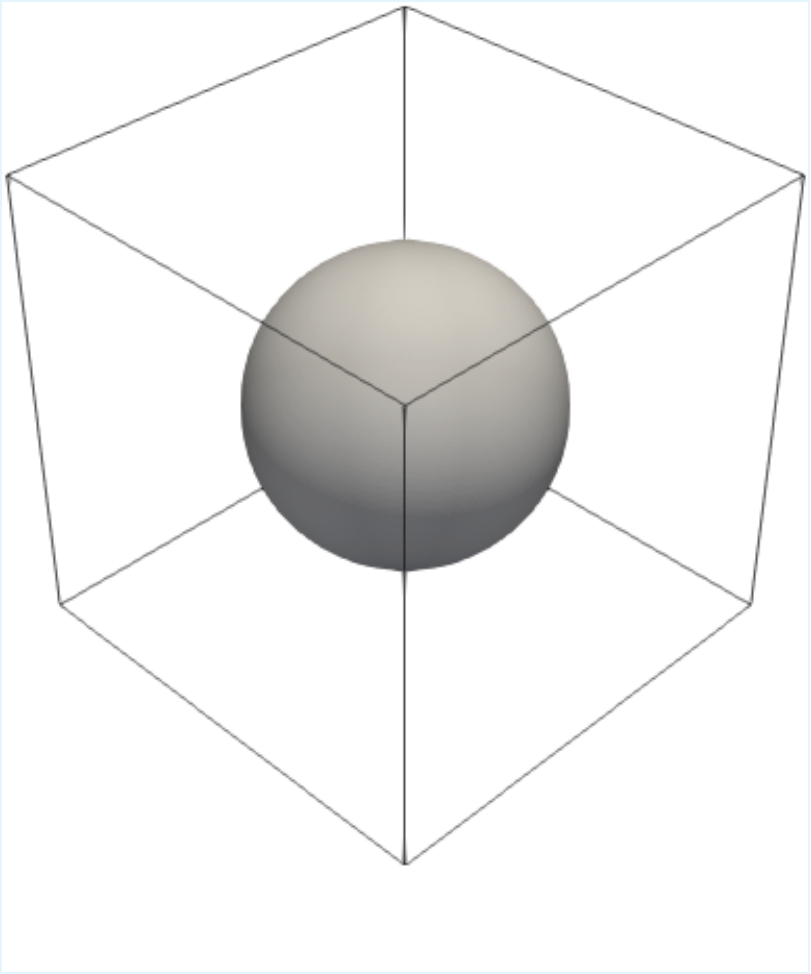}}\\
    \subfloat[$\phi(\mathbf{x},t=0)$]{\includegraphics[width=0.2\textwidth]{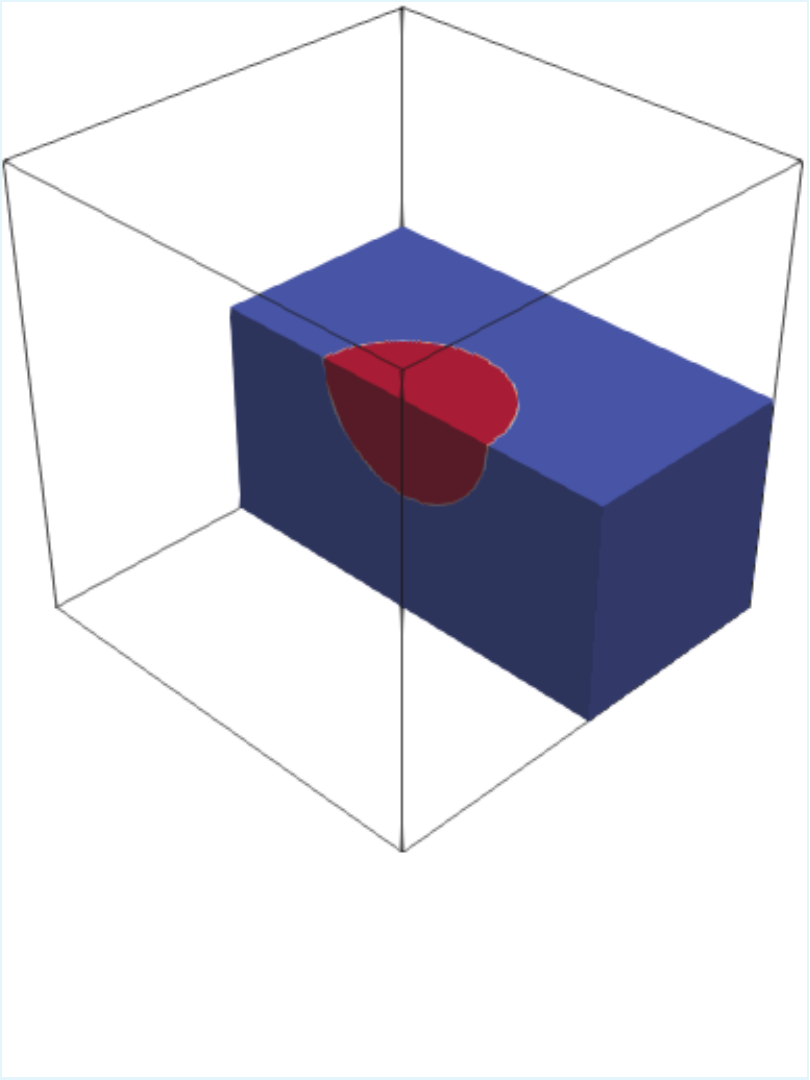}} \quad
    \subfloat[$\phi(\mathbf{x},t=0.5)$]{\includegraphics[width=0.2\textwidth]{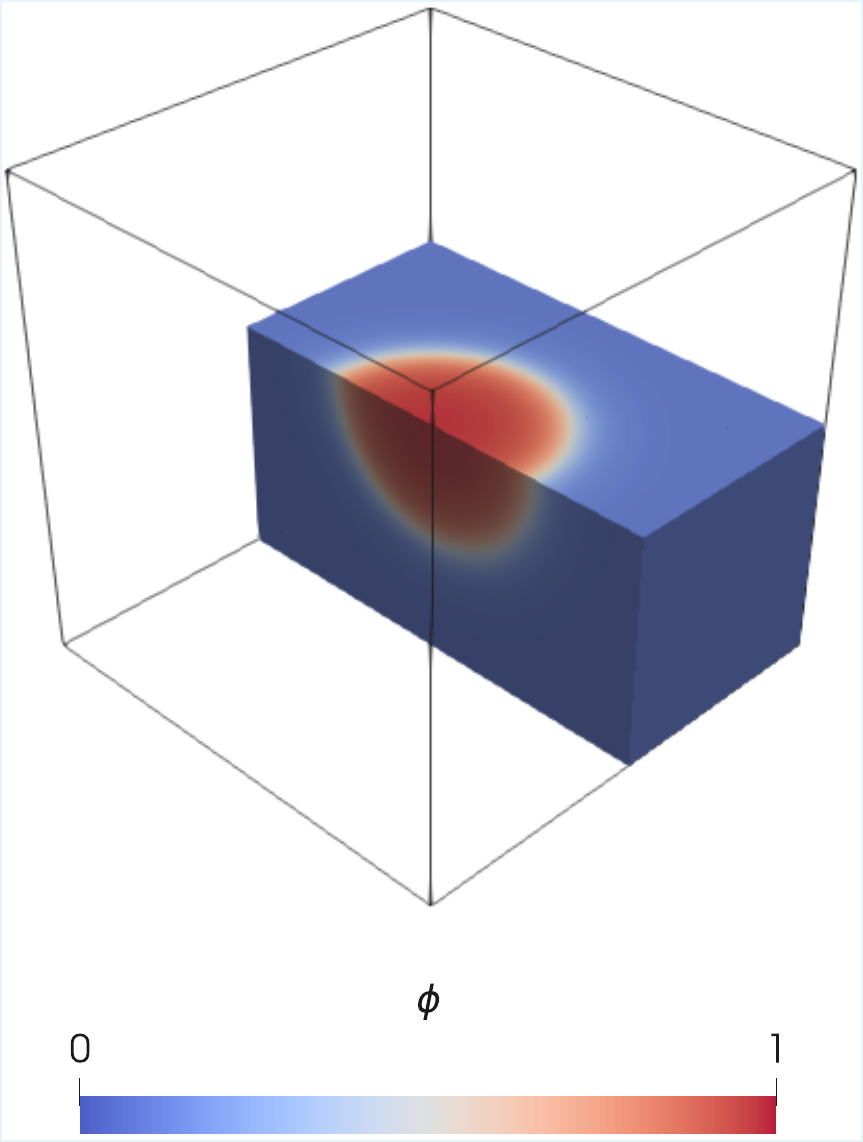}} \quad
    \subfloat[$\phi(\mathbf{x},t=0.75)$]{\includegraphics[width=0.2\textwidth]{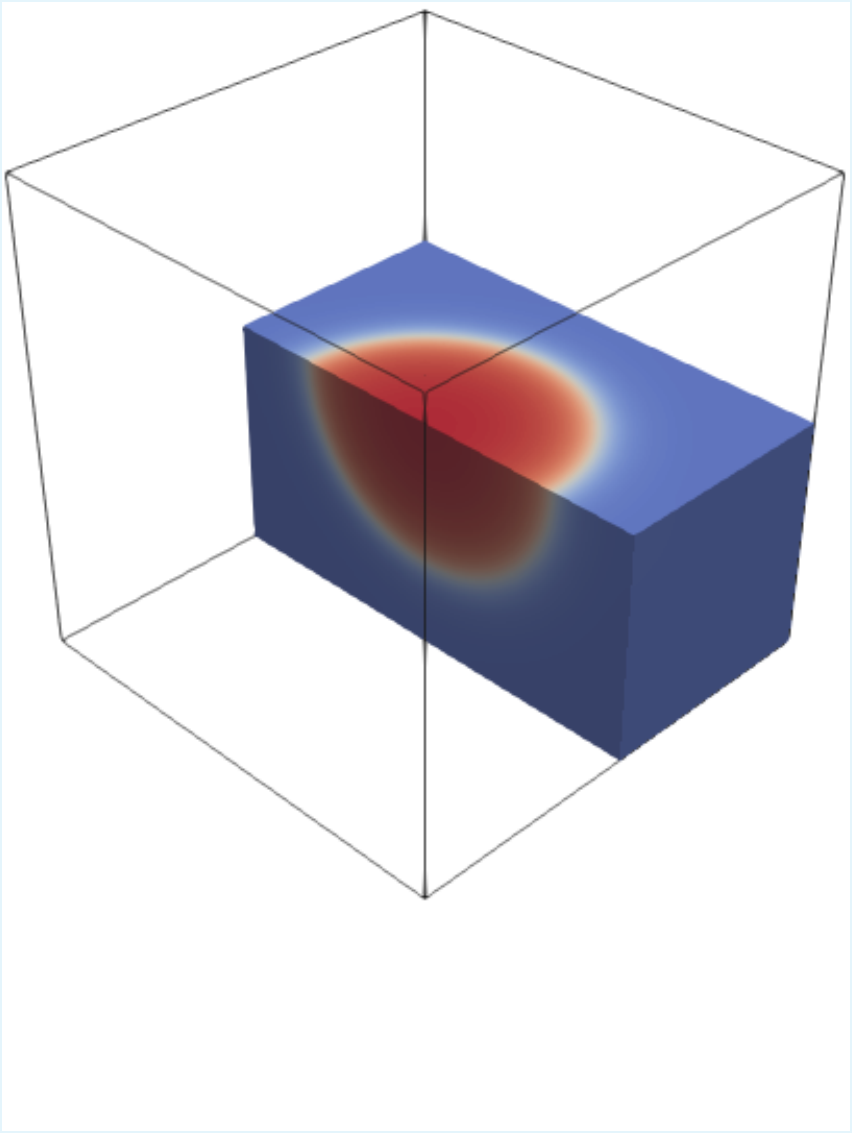}}\\
    \subfloat[$c(\mathbf{x},t=0)$]{\includegraphics[width=0.2\textwidth]{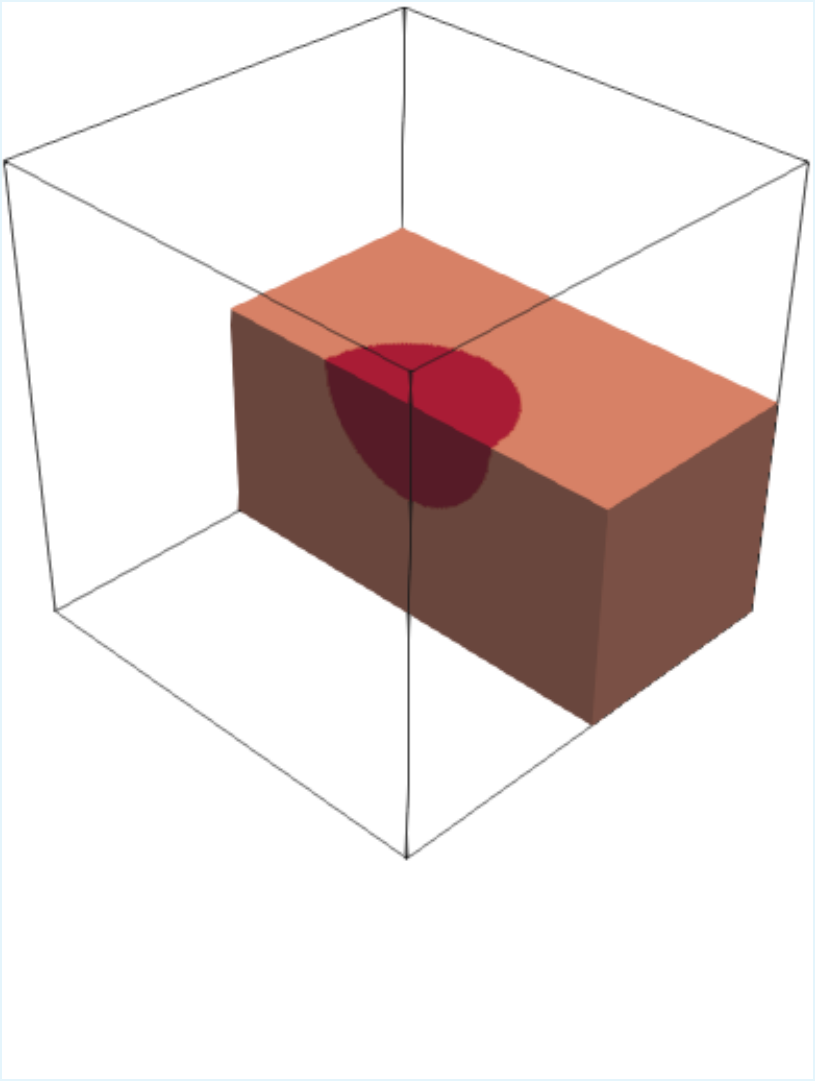}} \quad
    \subfloat[$c(\mathbf{x},t=0.5)$]{\includegraphics[width=0.2\textwidth]{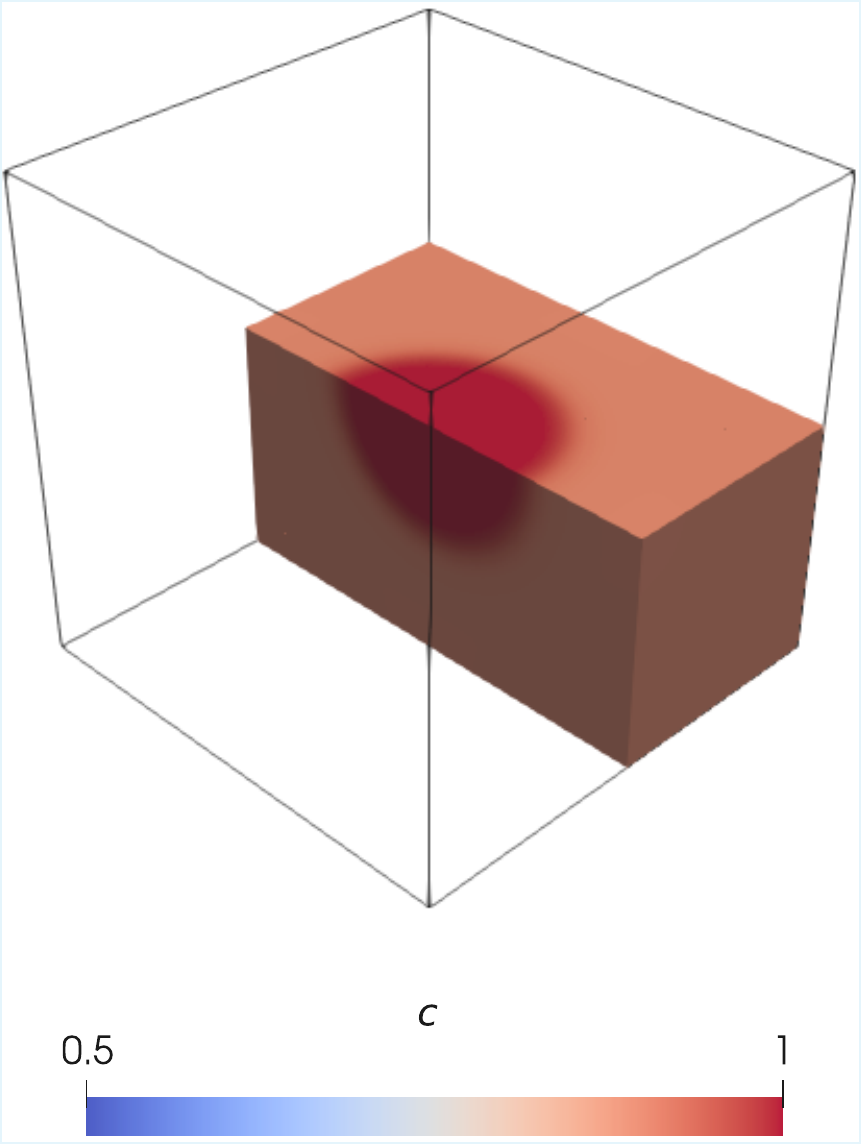}} \quad
    \subfloat[$c(\mathbf{x},t=0.75)$]{\includegraphics[width=0.2\textwidth]{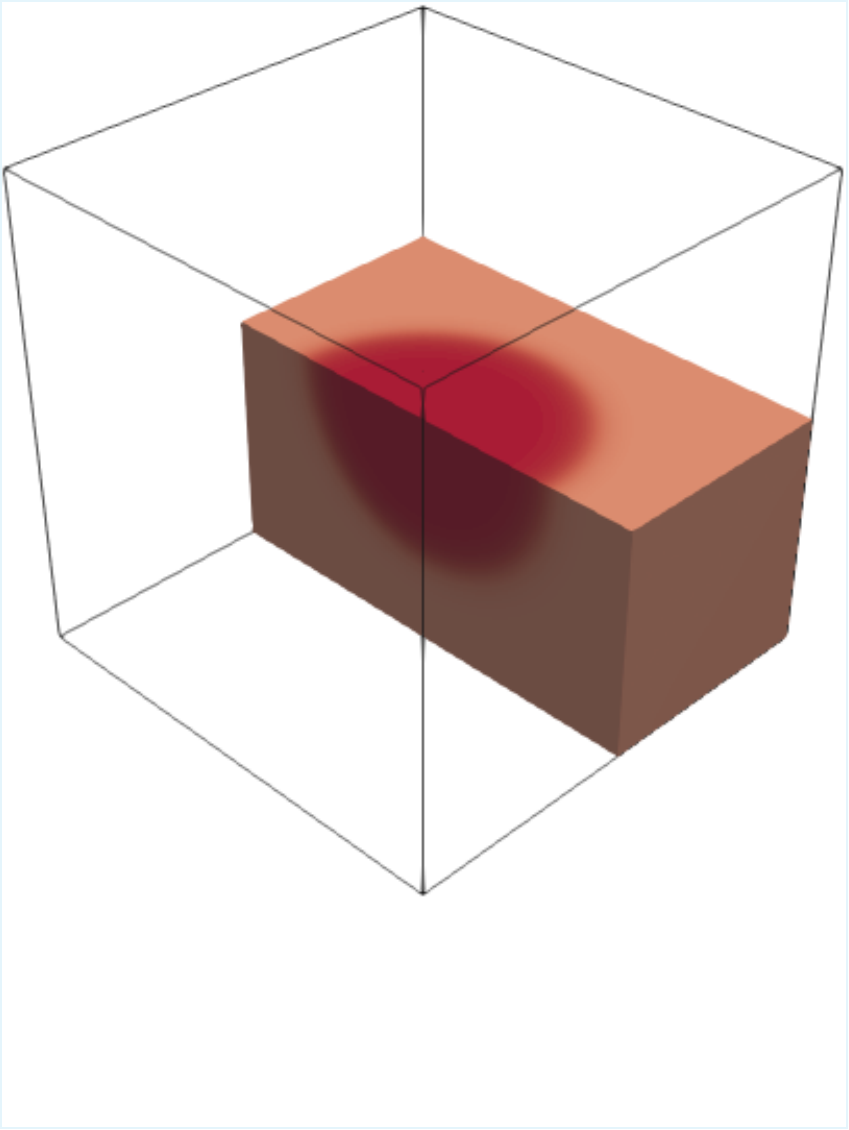}}\\
    \caption{Initial spherical precipitate of radius $R_0=0.22$ (larger than the critical radius): evolution of isosurface $\phi=0.5$, phase indicator $\phi$ and total concentration $c$.}
    \label{isophi_prep3D}
\end{figure}

We consider a single spherical precipitate with initial radius $R_0=0.22>R_{0c}^{3D}=0.208$. As expected, in the numerical simulation the precipitate grows over time. The evolution of the isosurface $\phi=0.5$, of the phase indicator $\phi$ and of the total ion concentration $c$ for a quarter of the domain is reported in Fig. \ref{isophi_prep3D}.

\subsubsection{Dissolution of two spherical precipitates}
\begin{figure}[htbp]
    \centering
    \subfloat[$\phi=0.5$ at $t=0$]{\includegraphics[width=0.2\textwidth]{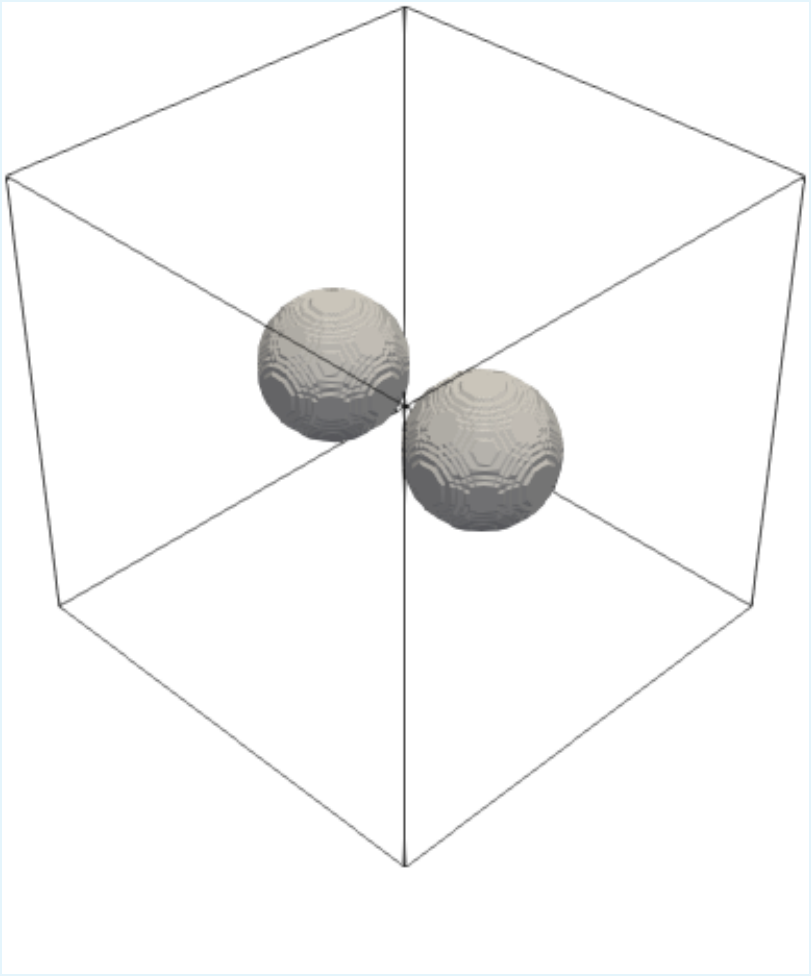}} \quad
    \subfloat[$\phi=0.5$ at $t=0.025$]{\includegraphics[width=0.2\textwidth]{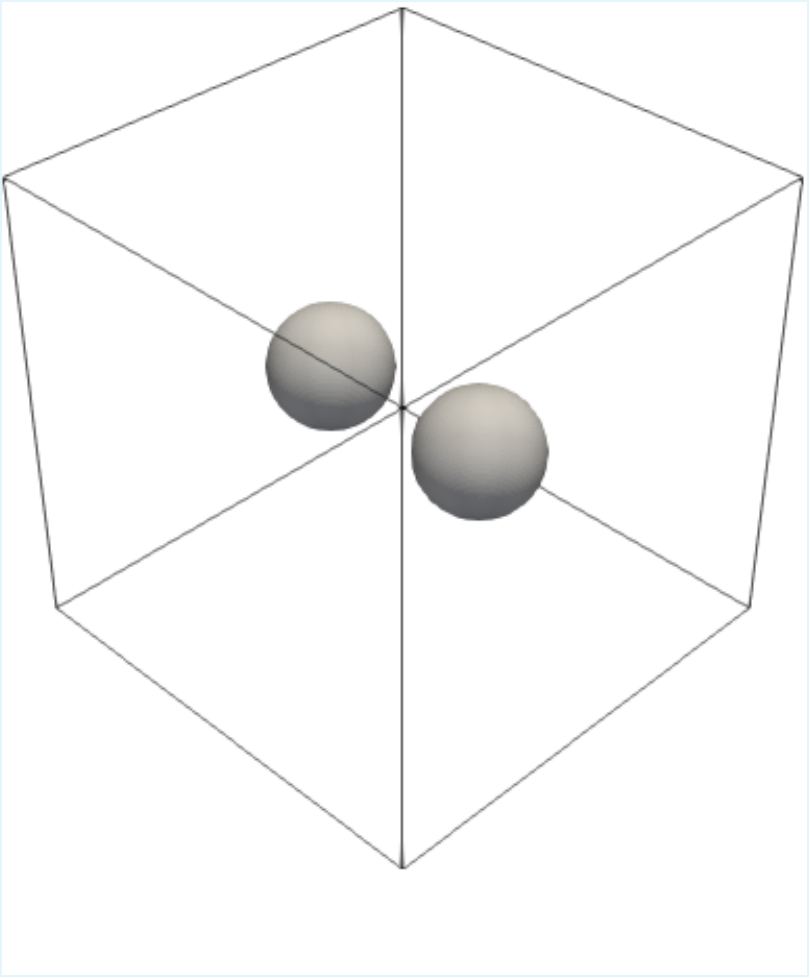}} \quad
    \subfloat[$\phi=0.5$ at $t=0.05$]{\includegraphics[width=0.2\textwidth]{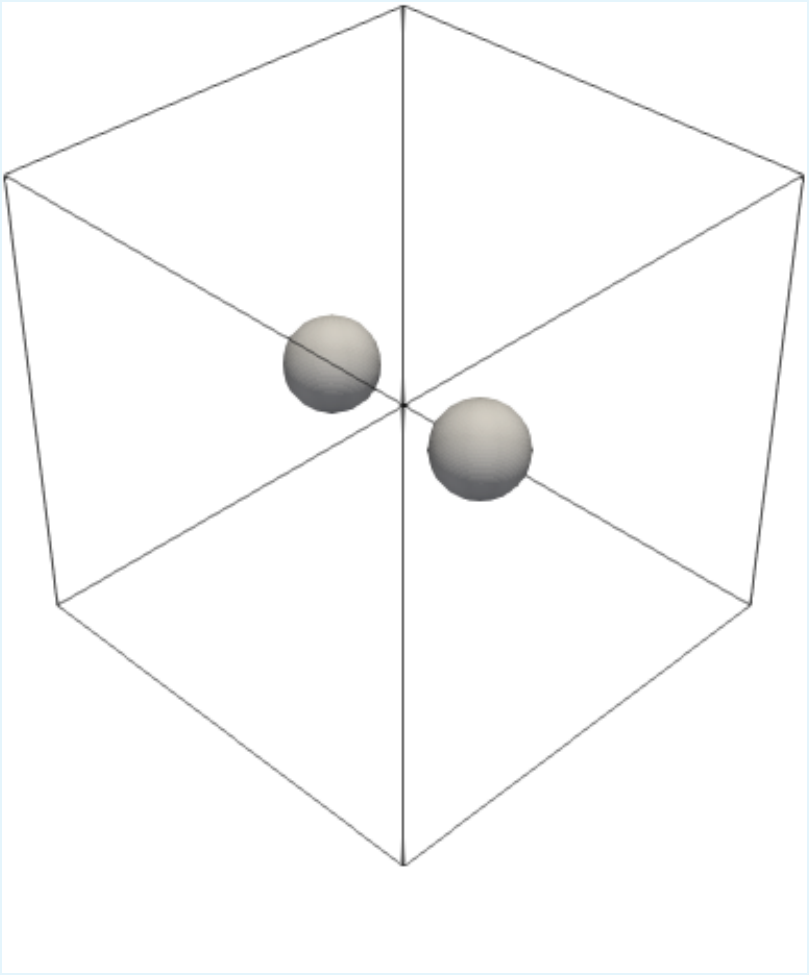}}\\
    \subfloat[$\phi(\mathbf{x},t=0)$]{\includegraphics[width=0.2\textwidth]{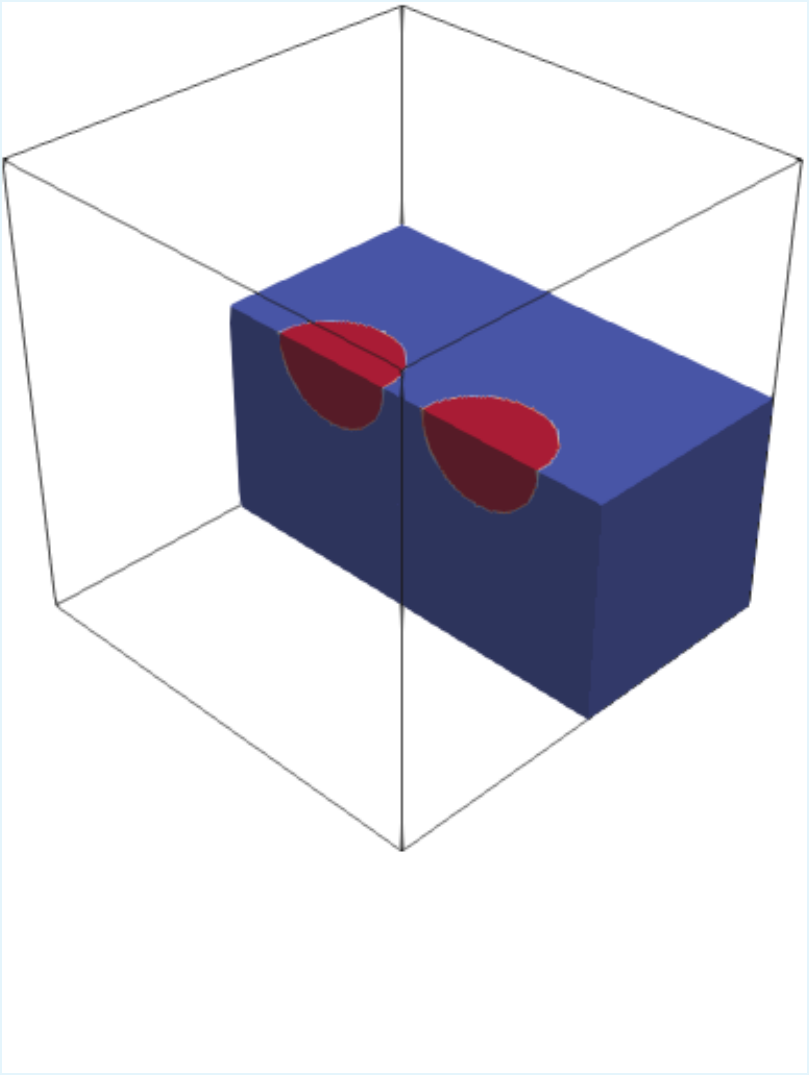}} \quad
    \subfloat[$\phi(\mathbf{x},t=0.025)$]{\includegraphics[width=0.2\textwidth]{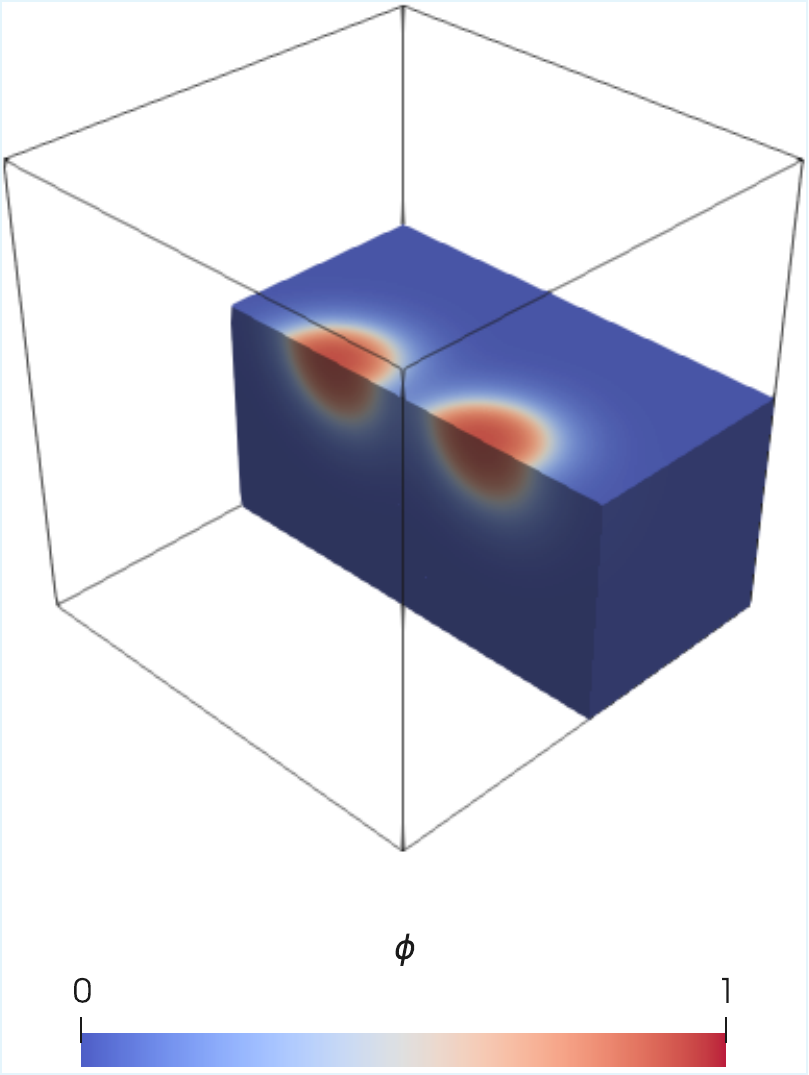}} \quad
    \subfloat[$\phi(\mathbf{x},t=0.05)$]{\includegraphics[width=0.2\textwidth]{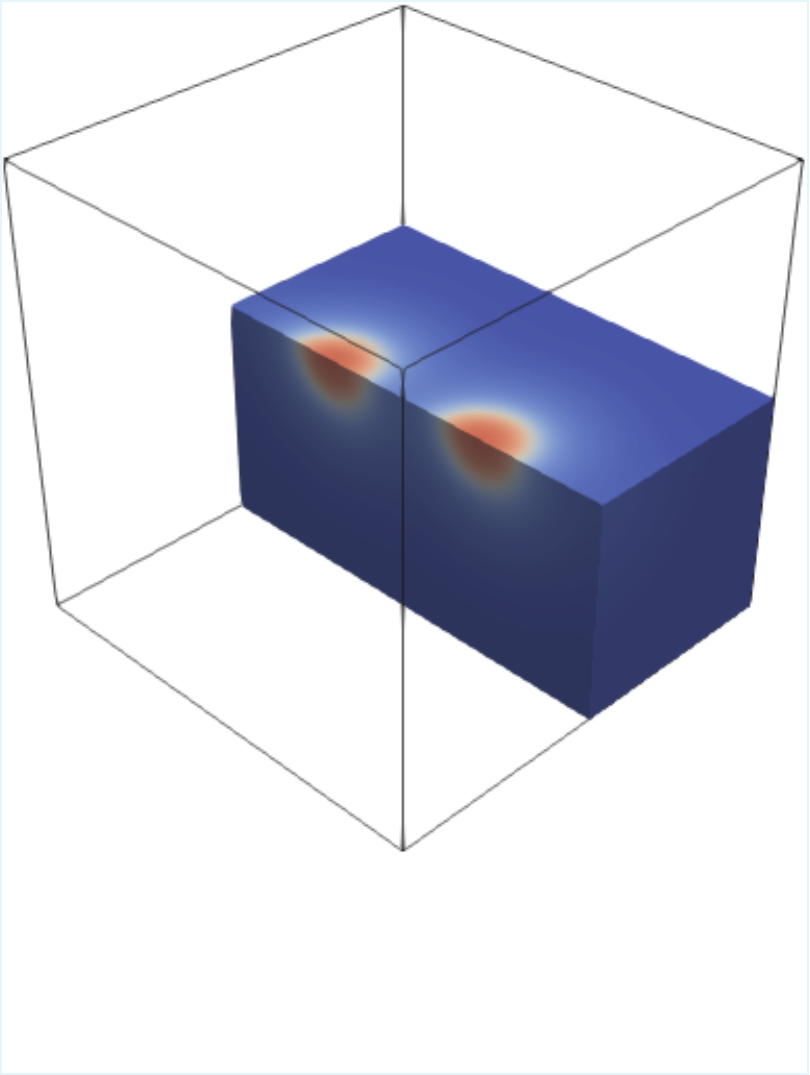}}\\
    \subfloat[$c(\mathbf{x},t=0)$]{\includegraphics[width=0.2\textwidth]{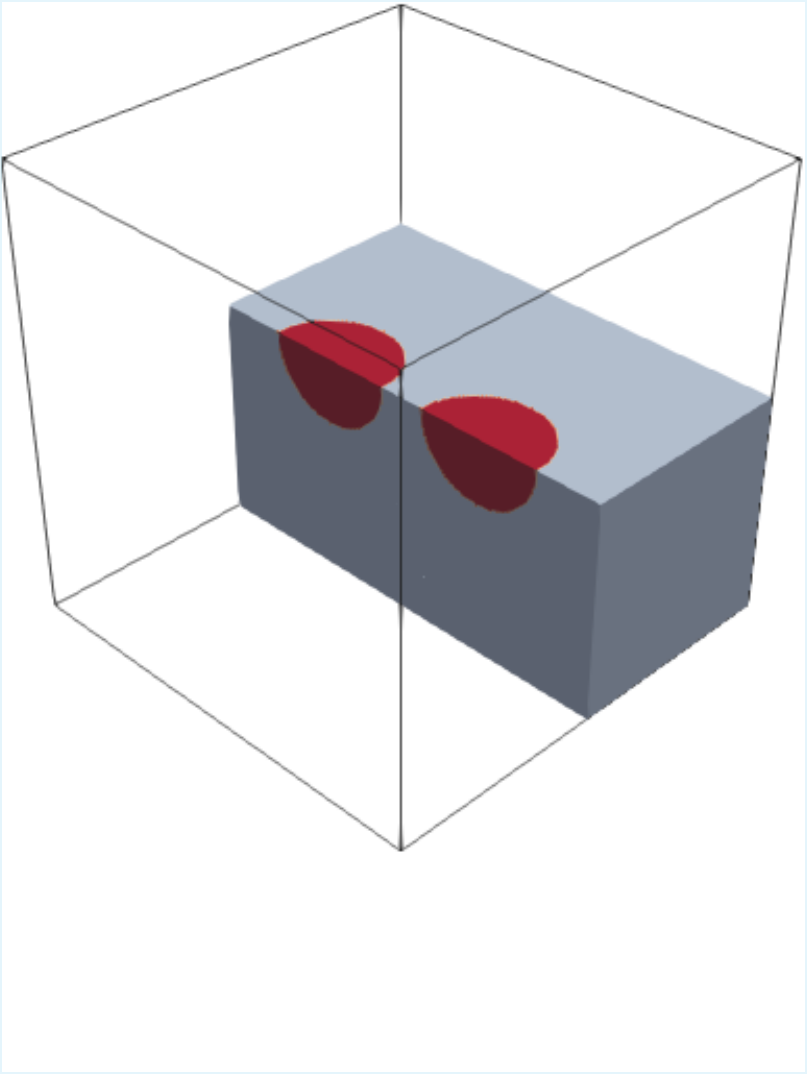}} \quad
    \subfloat[$c(\mathbf{x},t=0.025)$]{\includegraphics[width=0.2\textwidth]{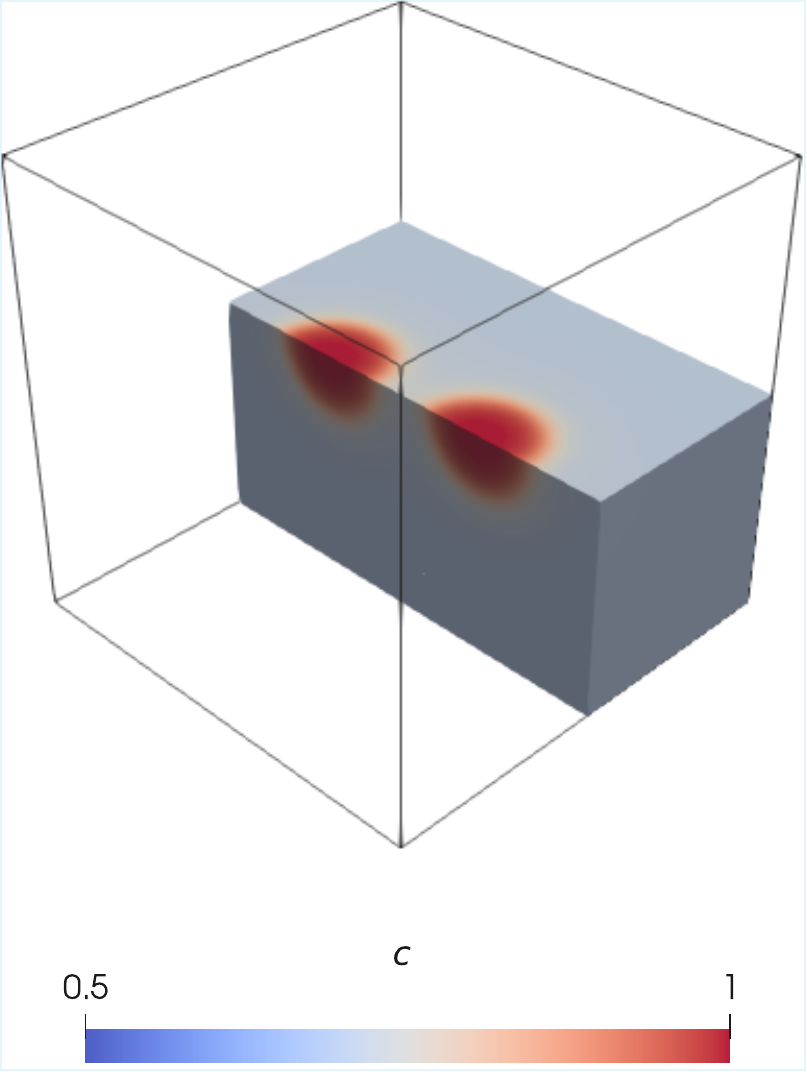}} \quad
    \subfloat[$c(\mathbf{x},t=0.05)$]{\includegraphics[width=0.2\textwidth]{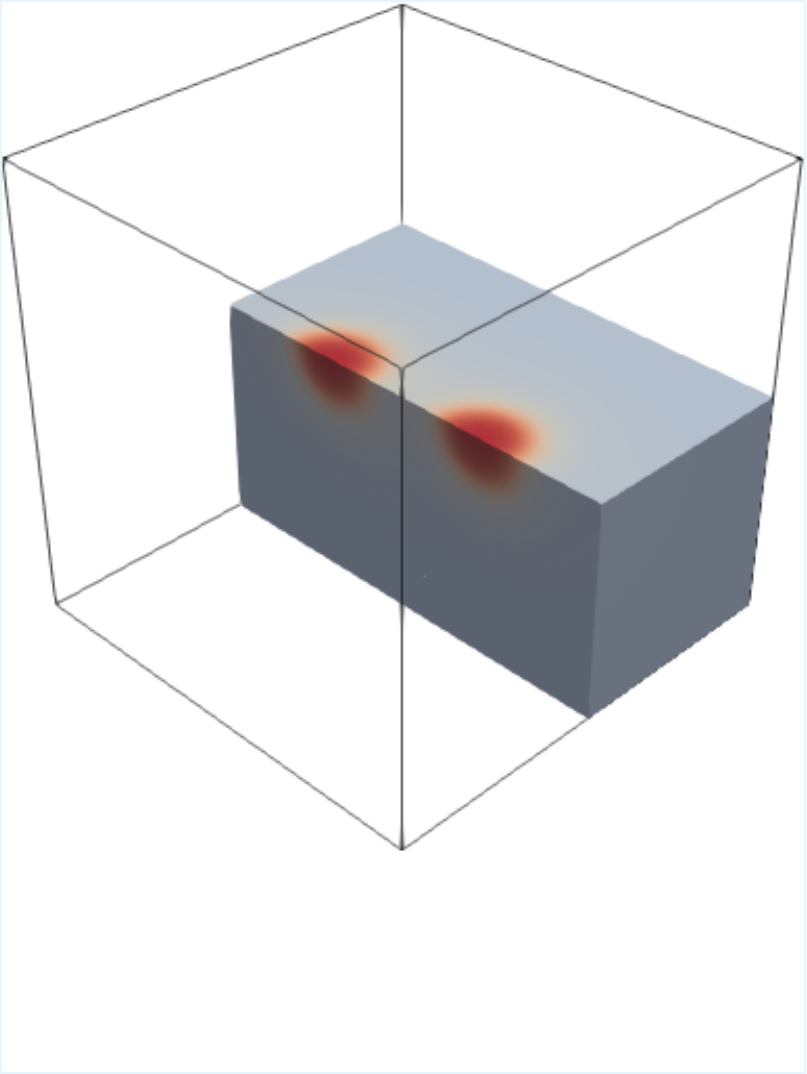}}\\
    \caption{Two initial spherical precipitates of radius $R_0=0.15$ (smaller than the critical radius): evolution of isosurface $\phi=0.5$, phase indicator $\phi$ and total concentration $c$.}
    \label{isophi_diss3D}
\end{figure}

We consider two spherical precipitates with centers in $(3L/10, L/2, L/2)$ and $(7L/10, L/2, L/2)$ and initial radius $R_0=0.15$. Over time, the two precipitates become smaller until they completely dissolve (the final state with a fully liquid domain is not shown).  The evolution of the isosurface $\phi=0.5$, of the phase indicator $\phi$ and of the total ion concentration $c$ for a quarter of the domain is reported in Fig. \ref{isophi_diss3D}.

\section{Conclusions}
\label{conclusions}
We proposed a new phase-field model for solute precipitation and dissolution which, to the best of our knowledge, is the first one in the literature to provably reconcile the two requirements of variational consistency and asymptotic convergence. The model assumes fixed solute concentration in the solid phase, neglects the liquid flow and takes into account the Gibbs-Thomson effect. The basic ideas underlying the construction of the proposed energy functional are: 1) a parabolic approximation of the bulk free energy densities around the equilibrium configuration and an expression of the ion concentration in the liquid as a function of the total ion concentration and the phase indicator, using an appropriate mixture rule; 2) an expression of the interfacial free energy density of Modica-Mortola type.
The obtained governing equation for the total ion concentration field is identical to that proposed in \cite{van2011phase}. However, the other governing equation is different, as the model in \cite{van2011phase} has no underlying variational structure. The convergence of the proposed model to the target sharp-interface model was assessed using the method of matched asymptotic expansions \cite{holmes2012introduction} and a novel expression of the precipitation/dissolution rate was found.
The finite element discretization of the new model was implemented using the deal.II library \cite{bangerth2007deal} and used to run $2$D and $3$D simulations. Results were in agreement with the theoretical predictions on the critical radius of initial circular and spherical precipitates, and demonstrated the flexibility of the model in dealing with precipitates of irregular geometry and different topologies.

While the proposed model, like the previous ones, is not able to predict nucleation of a new solid precipitate in a liquid solution, we anticipate that its variational nature can be useful to incorporate nucleation in future work.

\section*{Acknowledgments}
\par This work was performed within the framework of the ALIVE initiative (Advanced Engineering with Living Materials) and funded by the SFA-AM program (Strategic Focus Area – Advanced Manufacturing) at ETH Zürich.

\section*{Code and data availability}
The deal.II code will be made publicly available upon acceptance of the paper.
  
\begin{appendices}
\section{Governing equations in dimensionless form}  
\label{appx:eqs_nondim} A dimensionless form of the phase-field model is useful for the numerical implementation. To derive it, we introduce the dimensionless coordinate $\bar{\mathbf{x}}=\mathbf{x}/L$ and time $\bar{t}=tD_l/L^2$, where $L$ is a characteristic length of the problem domain. The governing equations in dimensionless form read
    \begin{align}
        \frac{\partial \phi}{\partial \bar{t}} &= -\bar{M}_\phi \left \{h'(\phi)\left[-\bar{\psi}_l(\bar{c}_l(\phi, \bar{c}))+ \frac{\bar{c}-(1+\delta)}{1-h(\phi)+\delta}\frac{\partial \bar{\psi_l}}{\partial \bar{c_l}}(\bar{c}_l(\phi, \bar{c}))\right]+\frac{\bar{B}}{c_w}\left [ \frac{W'(\phi)}{\bar{\varepsilon}}-2\bar{\varepsilon}\bar{\nabla}^2\phi\right]  \right \}\\
        \frac{\partial \bar{c}}{\partial \bar{t}} &= \bar{\nabla} \cdot \left \{  \bar{\nabla} \bar{c}+\frac{\left[ \bar{c}-(1+\delta)\right]h'(\phi)\bar{\nabla}\phi}{1-h(\phi)+\delta} \right \},
    \end{align}
where we defined the following dimensionless quantities
\begin{equation}
   \bar{\varepsilon} = \frac{\varepsilon}{L}, \qquad \bar{\gamma}=\frac{\gamma}{D_l}, \qquad \bar{M}_\phi = \frac{M_\phi A L^2}{D_l}=\frac{\bar{\gamma}c_w}{2\bar{B}\bar{\varepsilon}}, \qquad \bar{B} = \frac{B}{AL}
\end{equation}
\begin{equation}
   \bar{c}_l(\mathbf{x},t) = \frac{c_l(\mathbf{x},t)}{c_s}, \qquad \bar{c}_l^{eq} = \frac{c_l^{eq}}{c_s}, \qquad \bar{c}(\mathbf{x},t) = \frac{c(\mathbf{x},t)}{c_s}\quad 
\end{equation}
\begin{equation}
    \bar{\psi_l}(\bar{c}_l)=\frac{1}{2\bar{c}_l^{eq2}}\left(\bar{c}_l-\bar{c}_l^{eq} \right)^2,  \qquad \frac{\partial \bar{\psi_l}}{\partial \bar{c}_l}(\bar{c}_l)=\frac{1}{\bar{c}_l^{eq2}}\left(\bar{c}_l-\bar{c}_l^{eq} \right)\quad 
\textbf{}\end{equation}
and the dimensionless nabla operator $\bar{\nabla}:=L\nabla$.
\textbf{}
\end{appendices}

\begin{appendices}
\section{Supplementary calculations for matched asymptotic expansions}  
\label{appx:asymptotic}
\subsection{Outer expansions}
\label{B1}
Considering Eq. \eqref{interpolationeq}, we can rewrite a term appearing in Eq. \eqref{eqctilde} 
\begin{equation}\label{asymptoticouter1}
    \frac{h'(\phi^{o})}{1-h(\phi^{o})+\varepsilon}=\frac{6\left(\phi_0^{o}+\varepsilon\phi_1^{o}+\dots\right)\left(1-\phi_0^{o}-\varepsilon\phi_1^{o}-\dots\right)}{1-3\left(\phi_0^{o}+\varepsilon\phi_1^{o}+\dots\right)^2+2\left(\phi_0^{o}+\varepsilon\phi_1^{o}+\dots\right)^3+\varepsilon}.
\end{equation}
In the case $\phi_0^{o}=0$
\begin{equation}\label{asymptoticouter2}
    \frac{h'(\phi^{o})}{1-h(\phi^{o})+\varepsilon}=\frac{6\varepsilon\phi_1^{o}\left(1-\varepsilon \phi_1^{o}\right)+\dots}{1-3\varepsilon^2\phi_1^{out2}+\varepsilon+\dots}\to 0\quad \mathrm{as}\quad \varepsilon\to 0.
\end{equation}
In the case $\phi^{o}_0=1$, the following expansion holds
\begin{equation}\label{asymptoticouter3}
    \frac{h'(\phi^{o})}{1-h(\phi^{o})+\varepsilon}=-6\phi_1^{o}+\mathcal{O}(\varepsilon),
\end{equation}
so that
\begin{equation}\label{asymptoticouter4}
\frac{h'(\phi^{o})}{1-h(\phi^{o})+\varepsilon}\nabla\phi^{o}=\left(-6\phi_1^{o}+\dots\right)\left(\varepsilon \nabla \phi_1^{o}+\dots\right)\to \mathbf{0}\quad \mathrm{as}\quad \varepsilon \to 0
\end{equation}
where $\nabla \phi_0^{o}=\mathbf{0}$ due to $\phi_0^{o}=1$. Eqs. \eqref{asymptoticouter2} and \eqref{asymptoticouter4} lead to Eq. \eqref{diffusioneqouter}.\par
\subsection{Governing equations in curvilinear coordinates and their expansion}
\label{B2}
 As a result of the change of coordinates between $(\mathbf{x},t)$, and $(\xi, \mathbf{s},t)$ and using \eqref{signd}, the differential operators present in the governing equations \eqref{eqphi} and \eqref{eqctilde} transform as follows
 \begin{equation}\label{timeder}
\frac{\partial}{\partial t} = -\frac{v_{n}}{\varepsilon} \frac{\partial}{\partial \xi}+\frac{\partial}{\partial t}+\frac{\partial \mathbf{s}}{\partial t}\cdot \nabla_s 
\end{equation}
\begin{equation}\label{lapl}
\nabla^2 = \frac{1}{\varepsilon^2}\frac{\partial^2}{\partial \xi^2}+\frac{1}{\varepsilon}\nabla^2 d \frac{\partial}{\partial \xi}+\nabla^2 \mathbf{s}\cdot\nabla_s +\sum_{i=1}^{n_d-1} |\nabla s_i|^2 \frac{\partial^2}{\partial s_i^2}.
\end{equation}
Hence, the following relations hold between $\phi(\mathbf{x},t)$, $c(\mathbf{x},t)$ and their counterparts in local coordinates $\phi^{in}(\xi,\mathbf{s},t)$, $c^{in}(\xi,\mathbf{s},t)$
\begin{equation}\label{asymptoticinner1}
\frac{\partial}{\partial t}\phi(\mathbf{x},t) = -\frac{v_{n}}{\varepsilon} \frac{\partial}{\partial \xi}\phi^{in}(\xi,\mathbf{s},t)+\frac{\partial}{\partial t}\phi^{in}(\xi,\mathbf{s},t)+\frac{\partial \mathbf{s}}{\partial t}\cdot \nabla_s \phi^{in}(\xi,\mathbf{s},t)
\end{equation}
\begin{equation}\label{asymptoticinner1b}
\frac{\partial}{\partial t}c(\mathbf{x},t) = -\frac{v_{n}}{\varepsilon} \frac{\partial}{\partial \xi}c^{in}(\xi,\mathbf{s},t)+\frac{\partial}{\partial t}c^{in}(\xi,\mathbf{s},t)+\frac{\partial \mathbf{s}}{\partial t}\cdot \nabla_s c^{in}(\xi,\mathbf{s},t)
\end{equation}
\begin{equation}\label{asymptoticinner2}
\nabla^2 \phi(\mathbf{x},t) = \frac{1}{\varepsilon^2}\frac{\partial^2}{\partial \xi^2}\phi^{in}(\xi,\mathbf{s},t)+\frac{1}{\varepsilon}\nabla^2 d \frac{\partial}{\partial \xi}\phi^{in}(\xi,\mathbf{s},t)+\nabla^2 \mathbf{s}\cdot\nabla_s \phi^{in}(\xi,\mathbf{s},t)+\sum_{i=1}^{n_d-1} |\nabla s_i|^2 \frac{\partial^2}{\partial s_i^2} \phi^{in}(\xi,\mathbf{s},t)
\end{equation}
\begin{equation}\label{asymptoticinner2b}
\nabla^2 c(\mathbf{x},t) = \frac{1}{\varepsilon^2}\frac{\partial^2}{\partial \xi^2}c^{in}(\xi,\mathbf{s},t)+\frac{1}{\varepsilon}\nabla^2 d \frac{\partial}{\partial \xi}c^{in}(\xi,\mathbf{s},t)+\nabla^2 \mathbf{s}\cdot\nabla_s c^{in}(\xi,\mathbf{s},t)+\sum_{i=1}^{n_d-1} |\nabla s_i|^2 \frac{\partial^2}{\partial s_i^2} c^{in}(\xi,\mathbf{s},t),
\end{equation}
where $\nabla^2 \mathbf{s}=\left(\nabla^2 s_1\,...\,\nabla^2 s_{n_d-1}\right)^T$. 
Furthermore, defining
\begin{equation}\label{asymptoticinner3}
\mathcal{K} = \left[c-c_s(1+\varepsilon)\right]\frac{h'(\phi)}{1-h(\phi)+\varepsilon}
\end{equation}
we have
\begin{equation}\label{asymptoticinner4}
\nabla \cdot (\mathcal{K}\nabla \phi(\mathbf{x},t))=\frac{1}{\varepsilon^2}\frac{\partial}{\partial \xi}\left(\mathcal{K} \frac{\partial}{\partial \xi}\phi^{in}(\xi,\mathbf{s},t)\right)+\frac{1}{\varepsilon}\kappa\mathcal{K} \frac{\partial}{\partial \xi}\phi^{in}(\xi,\mathbf{s},t).
\end{equation}

Rewriting \eqref{eqphi} and \eqref{eqctilde} in the new coordinate system $(\xi,\mathbf{s},t)$ with \eqref{asymptoticinner1}-\eqref{asymptoticinner4} and substituting \eqref{innerphi} and \eqref{innerctilde}, we obtain
\begin{equation}\label{eqphi_inner}
\begin{split}
&\frac{2B\varepsilon^2}{\gamma c_w}\left(-\frac{1}{\varepsilon}v_{n0}\frac{\partial \phi_0^{in}}{\partial \xi}+\dots\right) - \frac{2B\varepsilon^2}{c_w}\left(\frac{1}{\varepsilon^2}\frac{\partial^2 \phi_0^{in}}{\partial \xi^2}+\frac{1}{\varepsilon} \frac{\partial ^2\phi_1^{in}}{\partial \xi^2}+\frac{1}{\varepsilon}\kappa_0\frac{\partial \phi_0^{in}}{\partial \xi}+\dots\right)
+ \frac{B}{c_w}\left[ W'\left(\phi_0^{in}\right) + W''\left(\phi_0^{in}\right)\varepsilon\phi_1^{in} +\dots\right]+\\&+
\varepsilon \left[h'\left(\phi_0^{in}\right)+h''\left(\phi_0^{in}\right)\varepsilon\phi_1^{in}+\dots\right] 
\cdot \left\{ -\frac{1}{2}\frac{A}{c_l^{eq2}}\left(c_l^{eq}-c_s\right)^2 + \frac{1}{2}\frac{A}{c_l^{eq2}}\left[\frac{c_0^{in}-c_s}{1-h\left(\phi_0^{in}\right)}\right]^2 + \dots \right\} = 0
\end{split}
\end{equation}
\begin{equation}\label{inner2}
\begin{split}
    &-v_{n0}\frac{\partial c_0^{in}}{\partial \xi} = D_l\frac{\partial ^2 c_1^{in}}{\partial \xi^2}+\kappa_0 D_l\left[\frac{\partial c_0^{in}}{\partial \xi}+\left(c_0^{in}-c_s\right)f\frac{\partial \phi_0^{in}}{\partial \xi}\right]+\\&+D_l\frac{\partial}{\partial \xi}\Bigg \{ \left(c_0^{in}-c_s\right)\left[f\frac{\partial \phi_1^{in}}{\partial \xi}+\left(\frac{h''\phi_1^{in}}{1-h}+f^2 \phi_1^{in}-\frac{f}{1-h} \right)\frac{\partial \phi_0^{in}}{\partial \xi}\right] +\left(c_1^{in}-c_s\right)f\frac{\partial \phi_0^{in}}{\partial \xi} \Bigg \},
\end{split}
\end{equation}
where
\begin{equation}
f\left(\phi_0^{in}\right) := \frac{h'\left(\phi_0^{in}\right)}{1-h\left(\phi_0^{in}\right)} .
\end{equation}

\subsection{Inner expansions}
\label{B4}

\subsubsection{Eq. \eqref{eqphi_inner}, $\mathcal{O}(1)$  terms}
Multiplying both sides of Eq. \eqref{phi0insecond} by $\partial \phi_0^{in}/ \partial \xi$, rearranging the obtained left-hand side and integrating between $-\infty$ and $\xi$ we obtain
\begin{equation}\label{integralw'}
\int_{-\infty}^\xi \frac{\partial}{\partial \hat{\xi}}\left(\frac{\partial \phi_0^{in}}{\partial \hat{\xi}}\right)^2\,\textrm{d}\hat{\xi}=\int_{0}^{\phi_0^{in}} W'\left(\hat{\phi}_0^{in}\right)\,\textrm{d}\hat{\phi}_0^{in},
\end{equation}
where we changed variable to compute the integral on the right-hand side, taking into account that $\textrm{d}\phi_0^{in} = \left(\partial \phi_0^{in}/\partial \xi\right) \,\textrm{d}\xi$ and using Eq. \eqref{phi0=1matching}$_2$.

\subsubsection{Eq. \eqref{inner2}, $\mathcal{O}(1)$  terms}
The term $\frac{h'(\phi^{in})}{1-h(\phi^{in})+\varepsilon}$ admits the following first-order expansion around $\phi_0^{in}$
\begin{equation}\label{h'term}
\frac{h'(\phi^{in})}{1-h(\phi^{in})+\varepsilon}=\frac{h'}{1-h}+\varepsilon \left [ \frac{(1-h)h''\phi_1^{in}+h'^2\phi_1^{in}-h'}{(1-h)^2}\right ]+\dots .
\end{equation}
Setting $\eta=\phi_0^{in}(\xi)$, we can rewrite the previous differential equation as
\begin{equation}
\frac{\partial  c_0^{in}}{\partial \eta}=-\left(c_0^{in}-c_s\right)\frac{h'(\eta)}{1-h(\eta)},
\end{equation}
whose general solution is
\begin{equation}
c_0^{in}(\eta, \mathbf{s},t)=c_s-G(\mathbf{s},t)\left[1-h(\eta)\right],
\end{equation}
with $G(\mathbf{s},t)$ to be determined.
Using the $-\infty$ limit in Eq. \eqref{phi0=1matching}$_3$ we obtain
\begin{equation}
G(\mathbf{s},t) = c_s-c_0^{o}\left(\mathbf{y}_{\varepsilon 0}^-, t\right),
\end{equation}
so that the final solution is the one given by Eq. \eqref{c0in}.

Enforcing the $+\infty$ limit in Eq. \eqref{phi0=1matching}$_3$ we obtain
\begin{equation}\label{csctilde}
\lim_{\eta \to 1} c_0^{in}(\eta, \mathbf{s},t)=c_s= c_0^{o}\left(\mathbf{y}_{\varepsilon 0}^+, t\right) ,
\end{equation}
hence it must be $\lim_{\xi\to +\infty} c_0^{in}(\xi, \mathbf{s},t) = c_s$. Starting from an initial condition $c_0^{o}(\mathbf{x},0)=c_s$ where $\phi_0^{o}=1$ and considering \eqref{csctilde}, we have $c_0^{o}=c_s$ where $\phi_0^{o}=1$ at any time, in agreement with the sharp-interface model \eqref{sharpinterfacemodel}. This implies also $\nabla c_0^{o}\left(\mathbf{y}_{\varepsilon 0}^+, t\right)=\mathbf{0}$.

\subsubsection{Eq. \eqref{eqphi_inner}, $\mathcal{O}(\varepsilon)$  terms}
Considering Eq. \eqref{c0in}, the term $\mathcal{A}\left(\phi_0^{in}\right)$ in Eq. \eqref{rhsfredholm} can be rewritten as
\begin{equation}\label{rhsfredholm2}
\mathcal{A}\left(\phi_0^{in}\right) = \frac{2B}{c_w}\left(\frac{v_{n0}}{\gamma}+\kappa_0\right)\frac{\partial \phi_0^{in}}{\partial \xi}-h'\left(\phi_0^{in}\right) \frac{1}{2}\frac{A}{c_l^{eq2}}\left \{-\left(c_l^{eq}-c_s\right)^2+\left[c_0^{o}\left(\mathbf{y}_{\varepsilon0}^-,t\right)-c_s\right]^2 \right \},
\end{equation}
which can be further expressed as
\begin{equation}\label{rhsfredholm3}
\mathcal{A}\left(\phi_0^{in}\right) = \frac{2B}{c_w}\left(\frac{v_{n0}}{\gamma}+\kappa_0\right)\frac{\partial \phi_0^{in}}{\partial \xi} - h'\left(\phi_0^{in}\right) \frac{1}{2}\frac{A}{c_l^{eq2}}\left \{ - \left[c_0^{o}\left(\mathbf{y}_{\varepsilon0}^-,t\right) - c_l^{eq}\right]^2 + 2\left[c_0^{o}\left(\mathbf{y}_{\varepsilon0}^-,t\right) - c_l^{eq}\right] \left[c_0^{o}\left(\mathbf{y}_{\varepsilon0}^-,t\right) - c_s\right] \right \}.
\end{equation}
Using Eq. \eqref{rhsfredholm3}, Eq. \eqref{solvability} can be conveniently arranged as
\begin{equation}\label{kinetic1}
    \frac{2B}{c_w}\left(\frac{v_{n0}}{\gamma}+\kappa_0\right) \int_{-\infty}^{+\infty} \left(\frac{\partial \phi_0^{in}}{\partial \xi}\right)^2 \,\textrm{d}\xi = \frac{1}{2} \frac{A}{c_l^{eq2}}\int_{-\infty}^{+\infty} \frac{\partial}{\partial \xi}\left \{ h \left \{ -\left[c_0^{o}\left(\mathbf{y}_{\varepsilon0}^-,t\right)-c_l^{eq}\right]^2+2\left[c_0^{o}\left(\mathbf{y}_{\varepsilon0}^-,t\right)-c_l^{eq}\right]\left[ c_0^{o}\left(\mathbf{y}_{\varepsilon0}^-,t\right)-c_s\right]\right \} \right \} \,\textrm{d}\xi.
\end{equation}
Changing variable and using \eqref{2W}, the integral on the left-hand side can be computed as
\begin{equation}
\int_{-\infty}^{+\infty} \left(\frac{\partial \phi_0^{in}}{\partial \xi}\right)^2 \,\textrm{d}\xi = \int_{0}^{1} \frac{\partial \phi_0^{in}}{\partial \xi}  \,\textrm{d}\phi_0^{in} = \int_{0}^{1} \sqrt{W\left(\phi_0^{in}\right)} \,\textrm{d}\phi_0^{in} .
\end{equation}
From the double-well expression in \eqref{W} we have
\begin{equation}
    \int_{0}^{1} \sqrt{W\left(\phi_0^{in}\right)} \,\textrm{d}\phi_0^{in} = \int_0^1 \sqrt{\phi_0^{in4}\left(1-\phi_0^{in}\right)^4}  \,\textrm{d}\phi_0^{in} =\frac{c_w}{2}=\frac{1}{30}.
\end{equation}
The right-hand side can be computed as
\begin{equation}\label{rhsfredholmfinal}
\begin{split}
&\frac{1}{2}\frac{A}{c_l^{eq2}}\int_{-\infty}^{+\infty} \frac{\partial}{\partial \xi} \left \{ h \left \{ -\left[c_0^{o}\left(\mathbf{y}_{\varepsilon0}^-,t\right)-c_l^{eq}\right]^2+2\left[c_0^{o}\left(\mathbf{y}_{\varepsilon0}^-,t\right)-c_l^{eq}\right]\left[c_0^{o}\left(\mathbf{y}_{\varepsilon0}^-,t\right)-c_s\right]\right \} \right \} \,\textrm{d}\xi = \\&=  -\frac{1}{2}\frac{A}{c_l^{eq2}} \left\{\left[c_0^{o}\left(\mathbf{y}_{\varepsilon0}^-,t\right)-c_l^{eq}\right]^2-2\left[c_0^{o}\left(\mathbf{y}_{\varepsilon0}^-,t\right)-c_l^{eq}\right]\left[c_0^{o}\left(\mathbf{y}_{\varepsilon0}^-,t\right)-c_s\right] \right \},
\end{split}
\end{equation}
where we made use of the conditions $h(1)=1$ and $h(0)=0$.\par

\subsubsection{Eq. \eqref{inner2}, $\mathcal{O}(\varepsilon)$  terms}
Eq. \eqref{inner2} has boundary conditions given by Eq. \eqref{bcderivativephi1}$_2$. Considering \eqref{dcodxi} and \eqref{c0in}, Eq. \eqref{inner2} can be rewritten as
\begin{equation}
    -v_{n0}\frac{\partial c_0^{in}}{\partial \xi} = D_l \frac{\partial}{\partial \xi} \left[ \frac{\partial c_1^{in}}{\partial \xi} + \left( c_1^{in} - c_s \right) f \frac{\partial \phi_0^{in}}{\partial \xi} \right] +
    D_l \frac{\partial}{\partial \xi} \left \{ \left[ c_0^{o}\left(\mathbf{y}_{\varepsilon 0}^-, t\right) - c_s \right] (1 - h) \left[ f \frac{\partial \phi_1^{in}}{\partial \xi}  + \left( \frac{h'' \phi_1^{in}}{1 - h} + f^2 \phi_1^{in} - \frac{f}{1 - h} \right) \frac{\partial \phi_0^{in}}{\partial \xi} \right] \right \}.
\end{equation}
Integrating indefinitely with respect to $\xi$ and with some manipulations, we obtain
\begin{equation}
\begin{split}
    &-v_{n0}c_0^{in} = D_l\left(1-h\right)\frac{\partial}{\partial \xi}\left \{\frac{c_1^{in}-c_s}{1-h} + \left[c_0^{o}\left(\mathbf{y}_{\varepsilon 0}^-, t\right)-c_s\right]\frac{h'\phi_1^{in}-1}{1-h}\right \}+ C,
\end{split}
\end{equation}
where $C$ is an integration constant. \par
Considering Eq. \eqref{matchingphi1}$_2$, we have
\begin{equation}\label{limc1}
\lim_{\xi \to - \infty} c_1^{in} = \lim_{\xi \to - \infty} \xi\nabla c_0^{o}\left(\mathbf{y}_{\varepsilon 0}^-,t\right)\cdot  \mathbf{n}_{\varepsilon 0}.
\end{equation}
Eq. \eqref{limc1}, together with \eqref{phi0=1matching}$_2$, \eqref{matchingphi1}$_1$, \eqref{bcderivativephi}$_1$, \eqref{bcderivativephi1}$_1$ and \eqref{bcderivativephi1}$_2$, implies
\begin{equation}
\lim_{\xi \to - \infty} b(\xi) = \lim_{\xi \to - \infty} \left[  \xi\nabla c_0^{o}\left(\mathbf{y}_{\varepsilon 0}^-,t\right)\cdot  \mathbf{n}_{\varepsilon 0} -c_0^{o}\left(\mathbf{y}_{\varepsilon 0}^-, t\right)\right]
\end{equation}
\begin{equation}
\lim_{\xi \to - \infty} \frac{\partial b(\xi)}{\partial \xi} = \nabla c_0^{o}\left(\mathbf{y}_{\varepsilon 0}^-,t\right)\cdot  \mathbf{n}_{\varepsilon 0}.
\end{equation}
Considering the previous limits, we conclude that
\begin{equation}
\lim _{\xi \to -\infty} b(\xi)f\left(\phi_0^{in}(\xi)\right)\frac{\partial \phi_0^{in}}{\partial \xi}=0.
\end{equation}
Now we take the limits as $\xi \to +\infty$. Considering that $\nabla c_0^{o}\left(\mathbf{y}_{\varepsilon 0}^+, t\right)=\mathbf{0}$, it follows
\begin{equation}
\lim_{\xi \to  +\infty} c_1^{in} = c_1^{o}\left(\mathbf{y}_{\varepsilon 0}^+, t\right)
\end{equation}
\begin{equation}
\lim_{\xi \to  +\infty} \frac{\partial c_1^{in}}{\partial \xi} = 0.
\end{equation}
Taking into account also 
\eqref{phi0=1matching}$_1$,
\eqref{matchingphi1}$_1$,
\eqref{bcderivativephi}$_1$
and \eqref{bcderivativephi1}$_1$, we have
\begin{equation}
\lim_{\xi \to  +\infty} b(\xi) =  c_1^{o}\left(\mathbf{y}_{\varepsilon 0}^+, t\right) -c_0^{o}\left(\mathbf{y}_{\varepsilon 0}^-, t\right)
\end{equation}
\begin{equation}
\lim_{\xi \to  +\infty} \frac{\partial b(\xi)}{\partial \xi} = 0.
\end{equation}
Now we want to compute the limit
\begin{equation}
\lim _{\xi \to +\infty} b(\xi)f\left(\phi_0^{in}(\xi)\right)\frac{\partial \phi_0^{in}}{\partial \xi}.
\end{equation}
The situation is more tricky than the case $\xi \to -\infty$ since also $1-h\left(\phi_0^{in}\right)$ tends to $0$ as $\xi \to +\infty$. 
However, it can be shown that
\begin{equation}\label{criticallimit}
\lim _{\xi \to +\infty} b(\xi)f\left(\phi_0^{in}(\xi)\right)\frac{\partial \phi_0^{in}}{\partial \xi}  = \lim _{\xi \to +\infty} b(\xi)\frac{6\phi_0^{in}\left(1-\phi_0^{in}\right)}{1-3\phi_0^{in2}+2\phi_0^{in3}}\sqrt{\phi_0^{in4}\left(1-\phi_0^{in} \right)^4} = 0.
\end{equation}

\end{appendices}

\textbf{}

\bibliographystyle{elsarticle-num}
\bibliography{Precipitation} 

\begin{thebibliography}{10}
\expandafter\ifx\csname url\endcsname\relax
  \def\url#1{\texttt{#1}}\fi
\expandafter\ifx\csname urlprefix\endcsname\relax\def\urlprefix{URL }\fi
\expandafter\ifx\csname href\endcsname\relax
  \def\href#1#2{#2} \def\path#1{#1}\fi

\bibitem{cahn1958free}
J.~W. Cahn, J.~E. Hilliard, Free energy of a nonuniform system. i. interfacial
  free energy, The Journal of chemical physics 28~(2) (1958) 258--267.

\bibitem{cahn1965phase}
J.~W. Cahn, Phase separation by spinodal decomposition in isotropic systems,
  The Journal of chemical physics 42~(1) (1965) 93--99.

\bibitem{allen1979microscopic}
S.~M. Allen, J.~W. Cahn, A microscopic theory for antiphase boundary motion and
  its application to antiphase domain coarsening, Acta metallurgica 27~(6)
  (1979) 1085--1095.

\bibitem{Boettinger2002}
W.~J. Boettinger, J.~A. Warren, C.~Beckermann, A.~Karma, Phase-field simulation
  of solidification, Annual Review of Materials Research 32 (2002) 163--194.
\newblock \href {https://doi.org/10.1146/annurev.matsci.32.101901.155803}
  {\path{doi:10.1146/annurev.matsci.32.101901.155803}}.

\bibitem{Chen2002}
L.-Q. Chen, Phase-field models for microstructure evolution, Annual Review of
  Materials Research 32 (2002) 113--140.
\newblock \href {https://doi.org/10.1146/annurev.matsci.32.112001.132041}
  {\path{doi:10.1146/annurev.matsci.32.112001.132041}}.

\bibitem{Brener2009}
E.~A. Brener, R.~Spatschek, Phase-field modeling of microstructural pattern
  formation, Advances in Physics 58~(3) (2009) 191--314.
\newblock \href {https://doi.org/10.1080/00018730902804194}
  {\path{doi:10.1080/00018730902804194}}.

\bibitem{plapp2015phase}
M.~Plapp, Phase-field models, in: Handbook of Crystal Growth, Elsevier, 2015,
  pp. 631--668.

\bibitem{Elder2016}
K.~R. Elder, N.~Provatas, Phase-field methods in materials science and
  engineering, Physics Today 69~(10) (2016) 53--58.
\newblock \href {https://doi.org/10.1063/PT.3.3326}
  {\path{doi:10.1063/PT.3.3326}}.

\bibitem{xu2008phase}
Z.~Xu, P.~Meakin, Phase-field modeling of solute precipitation and dissolution,
  The Journal of chemical physics 129~(1) (2008).

\bibitem{van2011phase}
T.~Van~Noorden, C.~Eck, Phase field approximation of a kinetic moving-boundary
  problem modelling dissolution and precipitation, Interfaces and Free
  Boundaries 13~(1) (2011) 29--55.

\bibitem{redeker2016upscaling}
M.~Redeker, C.~Rohde, I.~Sorin~Pop, Upscaling of a tri-phase phase-field model
  for precipitation in porous media, IMA Journal of Applied Mathematics 81~(5)
  (2016) 898--939.

\bibitem{bringedal2020phase}
C.~Bringedal, L.~Von~Wolff, I.~S. Pop, Phase field modeling of precipitation
  and dissolution processes in porous media: Upscaling and numerical
  experiments, Multiscale Modeling \& Simulation 18~(2) (2020) 1076--1112.

\bibitem{rohde2021ternary}
C.~Rohde, L.~von Wolff, A ternary cahn--hilliard--navier--stokes model for
  two-phase flow with precipitation and dissolution, Mathematical Models and
  Methods in Applied Sciences 31~(01) (2021) 1--35.

\bibitem{von2021investigation}
L.~von Wolff, F.~Weinhardt, H.~Class, J.~Hommel, C.~Rohde, Investigation of
  crystal growth in enzymatically induced calcite precipitation by
  micro-fluidic experimental methods and comparison with mathematical modeling,
  Transport in Porous Media 137 (2021) 327--343.

\bibitem{amirouche2009phase}
L.~Amirouche, M.~Plapp, Phase-field modeling of the discontinuous precipitation
  reaction, Acta materialia 57~(1) (2009) 237--247.

\bibitem{ji2022phase}
Y.~Ji, L.-Q. Chen, Phase-field model of stoichiometric compounds and solution
  phases, Acta Materialia 234 (2022) 118007.

\bibitem{ORTIZ1999419}
M.~Ortiz, L.~Stainier, The variational formulation of viscoplastic constitutive
  updates, Computer Methods in Applied Mechanics and Engineering 171~(3) (1999)
  419--444.

\bibitem{modica1977esempio}
L.~Modica, Un esempio di $\gamma$-convergenza, Boll. Un. Mat. Ital. B 14 (1977)
  285--299.

\bibitem{modica1987gradient}
L.~Modica, The gradient theory of phase transitions and the minimal interface
  criterion, Archive for Rational Mechanics and Analysis 98 (1987) 123--142.

\bibitem{ambrosio2000variational}
L.~Ambrosio, N.~Dancer, G.~Alberti, Variational models for phase transitions,
  an approach via $\gamma$-convergence, Calculus of variations and partial
  differential equations: topics on geometrical evolution problems and degree
  theory (2000) 95--114.

\bibitem{castro2019microbially}
M.~J. Castro-Alonso, L.~E. Monta{\~n}ez-Hernandez, M.~A. Sanchez-Mu{\~n}oz,
  M.~R. Macias~Franco, R.~Narayanasamy, N.~Balagurusamy, Microbially induced
  calcium carbonate precipitation (micp) and its potential in bioconcrete:
  microbiological and molecular concepts, Frontiers in Materials 6 (2019) 126.

\bibitem{lee2018current}
Y.~S. Lee, W.~Park, Current challenges and future directions for bacterial
  self-healing concrete, Applied microbiology and biotechnology 102 (2018)
  3059--3070.

\bibitem{yang2020review}
S.~Yang, F.~Aldakheel, A.~Caggiano, P.~Wriggers, E.~Koenders, A review on
  cementitious self-healing and the potential of phase-field methods for
  modeling crack-closing and fracture recovery, Materials 13~(22) (2020) 5265.

\bibitem{feng2021microbial}
J.~Feng, B.~Chen, W.~Sun, Y.~Wang, Microbial induced calcium carbonate
  precipitation study using bacillus subtilis with application to self-healing
  concrete preparation and characterization, Construction and Building
  Materials 280 (2021) 122460.

\bibitem{chang2024application}
J.~Chang, D.~Yang, C.~Lu, Z.~Shu, S.~Deng, L.~Tan, S.~Wen, K.~Huang, P.~Duan,
  Application of microbially induced calcium carbonate precipitation (micp)
  process in concrete self-healing and environmental restoration to facilitate
  carbon neutrality: a critical review, Environmental Science and Pollution
  Research (2024) 1--16.

\bibitem{zhang2024application}
Y.~Zhang, Y.~Liu, X.~Sun, W.~Zeng, H.~Xing, J.~Lin, S.~Kang, L.~Yu, Application
  of microbially induced calcium carbonate precipitation (micp) technique in
  concrete crack repair: A review, Construction and Building Materials 411
  (2024) 134313.

\bibitem{fang2024enhancing}
C.~Fang, V.~Achal, Enhancing carbon neutrality: A perspective on the role of
  microbially induced carbonate precipitation (micp), Biogeotechnics 2~(2)
  (2024) 100083.

\bibitem{ma2024biomimetic}
Y.~Ma, S.~Yi, M.~Wang, Biomimetic mineralization for carbon capture and
  sequestration, Carbon Capture Science \& Technology 13 (2024) 100257.

\bibitem{wang2017review}
Z.~Wang, N.~Zhang, G.~Cai, Y.~Jin, N.~Ding, D.~Shen, Review of ground
  improvement using microbial induced carbonate precipitation (micp), Marine
  Georesources \& Geotechnology 35~(8) (2017) 1135--1146.

\bibitem{mujah2017state}
D.~Mujah, M.~A. Shahin, L.~Cheng, State-of-the-art review of biocementation by
  microbially induced calcite precipitation (micp) for soil stabilization,
  Geomicrobiology Journal 34~(6) (2017) 524--537.

\bibitem{fu2023microbially}
T.~Fu, A.~C. Saracho, S.~K. Haigh, Microbially induced carbonate precipitation
  (micp) for soil strengthening: A comprehensive review, Biogeotechnics 1~(1)
  (2023) 100002.

\bibitem{caginalp1994phase}
G.~Caginalp, E.~Socolovsky, Phase field computations of single-needle crystals,
  crystal growth, and motion by mean curvature, SIAM Journal on Scientific
  Computing 15~(1) (1994) 106--126.

\bibitem{xu2012phase}
Z.~Xu, H.~Huang, X.~Li, P.~Meakin, Phase field and level set methods for
  modeling solute precipitation and/or dissolution, Computer Physics
  Communications 183~(1) (2012) 15--19.

\bibitem{kim1999phase}
S.~G. Kim, W.~T. Kim, T.~Suzuki, Phase-field model for binary alloys, Physical
  review e 60~(6) (1999) 7186.

\bibitem{plapp2011unified}
M.~Plapp, Unified derivation of phase-field models for alloy solidification
  from a grand-potential functional, Physical Review E—Statistical,
  Nonlinear, and Soft Matter Physics 84~(3) (2011) 031601.

\bibitem{mai2016phase}
W.~Mai, S.~Soghrati, R.~G. Buchheit, A phase field model for simulating the
  pitting corrosion, Corrosion Science 110 (2016) 157--166.

\bibitem{abubakar2015phase}
A.~A. Abubakar, S.~S. Akhtar, A.~F.~M. Arif, Phase field modeling of v2o5 hot
  corrosion kinetics in thermal barrier coatings, Computational Materials
  Science 99 (2015) 105--116.

\bibitem{gao2020efficient}
H.~Gao, L.~Ju, R.~Duddu, H.~Li, An efficient second-order linear scheme for the
  phase field model of corrosive dissolution, Journal of Computational and
  Applied Mathematics 367 (2020) 112472.

\bibitem{holmes2012introduction}
M.~H. Holmes, Introduction to perturbation methods, Vol.~20, Springer Science
  \& Business Media, 2012.

\bibitem{chen2006rapidly}
X.~Chen, G.~Caginalp, C.~Eck, A rapidly converging phase field model, Discrete
  and continuous dynamical systems 15~(4) (2006) 1017.

\bibitem{bangerth2007deal}
W.~Bangerth, R.~Hartmann, G.~Kanschat, deal. ii—a general-purpose
  object-oriented finite element library, ACM Transactions on Mathematical
  Software (TOMS) 33~(4) (2007) 24--es.

\end{thebibliography}

\end{document}